\pdfoutput=1
\documentclass[a4paper,onecolumn,11pt,unpublished]{quantumarticle}
\pdfoutput=1
\usepackage[T1]{fontenc}
\usepackage[utf8]{inputenc}
\usepackage{mathtools,amssymb,bm,mathrsfs,graphicx,amsthm,booktabs,tabularx}
\usepackage[numbers,sort&compress]{natbib}
\usepackage{float}
\usepackage[colorlinks=true,citecolor=blue,linkcolor=blue,urlcolor=blue]{hyperref}
\newtheorem{theorem}{Theorem}
\newtheorem{lemma}{Lemma}
\numberwithin{equation}{section}
\hypersetup{pdftitle={Geometric limits of quantum recovery across an Ising transition},pdfauthor={Yi Liang and Chen Qian},pdfsubject={},pdfkeywords={}}
\title{Geometric limits of quantum recovery across an Ising transition}
\author{Yi Liang}
\affiliation{Independent Researcher, Beijing, China}
\email{iantrans2042@gmail.com}
\orcid{0009-0006-3168-972X}
\author{Chen Qian}
\affiliation{Beijing Academy of Quantum Information Sciences, Beijing, China}
\email{qianchen@baqis.ac.cn}
\orcid{0000-0002-3851-7713}
\pdftrailerid{}
\newcommand{\phaseproofintro}{Theorem~\ref{thm:phase} relies on the explicit endpoint codes formulated above and their filtered spectral transport through a fixed gapped phase. All spatial estimates below are established on the full physical spin chain, invoking a finite spectral gap solely within the symmetry sector of conserved defect charges.}
\newcommand{\criticalproofintro}{Having established necessary filling, we evaluate the spatial cost by enumerating retained lattice sites, characterize the near-minimal density profile, and track omission counts to deduce the quantitative small-error bound. The conditional extension to other correlation exponents is presented in Sec.~\ref{app:eta}.}
\newcommand{\finiteproofintro}{We derive the fixed-support polar fidelity optimum, establish the step-like performance threshold for connected intervals, and present the complete deterministic recovery protocol based on single-spin operations. The uncompressed quantum capacity of the full access channel is analyzed in Appendix~\ref{app:access-capacity}.}
\newcommand{\taskphasecaption}{Recovery task and leading spatial cost across the reference Ising transition. (a) Schematic of the decoding task: an unknown fusion qubit is entangled with an inert reference $R$ and encoded into $L$ physical spins. The receiver collects an input-independent subsystem $A$ and reconstructs the logical qubit following erasure of the environment $\bar A$. (b) Asymptotic cost envelopes under defect gap proportions $3d,2d,d$. Numerical labels indicate minimum retained-spin counts normalized by $d$ in the thermodynamic-then-separation limit. Shaded regions illustrate individual gap contributions to the leading price at error threshold $\epsilon_d=d^{-p}$ ($p>0$) for fixed finite positive reference field $h$. The zero-error column gives the leading cost when $\epsilon_d=0$ is required at each finite size. Diamonds mark the ordered core coordinates $p_1<p_2<p_3<p_4$. Leading cost bands omit the subleading boundary padding required by quasi-adiabatic transport in the gapped phases.}
\newcommand{\recordscaption}{Complementary environmental records and interval-based exact extraction. Marked test sites define physical spin-pair records: central--exterior omissions generate order coefficients for logical $Z_f$, whereas omissions across both outer gaps generate disorder coefficients for logical $X_f$. These test observables are evaluated on a non-wrapping coordinate window, while candidate retained sets $A$ may be disconnected or wrap the periodic seam. Bulk gaps between defect cores are denoted $G_i=(p_i,p_{i+1})\cap\mathbb Z$, and $E_d$ marks an exterior window of $\lceil(M+1)d\rceil$ sites for retained budgets $|A|\le Md$ with fixed finite $M$. Three-defect segments $J_\pm$ support deterministic unitary extraction. The $3{:}2{:}2$ layout serves as the benchmark for comparing connected and fragmented access; Fig.~\ref{fig:task-phase} adopts $3{:}2{:}1$ instead so that the two gapped phases exhibit distinct inverse-power-error costs while sharing the same leading zero-error cost across phases. The operator dictionaries hold for arbitrary positive gap ratios.}
\newcommand{\protocolcaption}{Single-spin measurement recovery on two disconnected blocks at the critical point $h=1$. (a) Partitioned support geometry with spacing ratios $3{:}2{:}2$, budget coefficient $2<b<4$, and retained strips of length $r_d=\lfloor(b-2)d/2\rfloor$ flanking the macroscopic middle hole. Alice measures single-spin Pauli observables ($Y$ on site $p_2$ and $X$ across the remainder of $A_L$) and transmits the integer sum $s$ via a classical channel (dashed arrow). Bob measures designated helper spins while preserving target output site $p_3$ and the intact flag block $[p_3,p_4]$ prior to helper readout, applying local Clifford and Pauli corrections conditioned on $s$ and his measurement outcomes. Every helper spin counts toward the budget, all outcome branches are retained, and the middle hole is erased before decoding. (b) Entanglement fidelity: the solid curve traces the analytical lower bound of Theorem~\ref{thm:one-way} for the specified protocol, contrasted with the upper ceiling for connected regions (dashed line) and the critical vanishing-error cost $b_0=4$ (dotted line). For a Bell input, fidelity above $1/2$ certifies entanglement between the recovered spin and the inert reference. Open circles delimit the domain $2<b<4$. Discrete benchmark values for $d=2, L=20$ are listed in Table~\ref{tab:protocol} rather than plotted on this asymptotic curve.}

\newcommand{\finiteplotcaption}{Finite-size recovery performance in the $3{:}2{:}2$ defect configuration. (a) Single-hole subregions within $J_+$ at scale $d=2$ with fixed budget $|A|=8$. Data markers (circles and squares) indicate the preserved logical Pauli flag; solid and dashed curves trace optimal fidelities for ring sizes $L=20$ and $L=18$, while dotted curves display complementary-record converse bounds for $L=20$. (b) Optimal fidelities of all single-gap punctured supports for budgets $B\in\{6,7,8\}$ on $L=20$ sites, with degenerate data points horizontally offset for clarity. Dashed and dotted horizontal baselines indicate the connected-support ceiling and the analytical endpoint lower bound, respectively. Data points are floating-point finite-ring evaluations, distinct from the ordered-limit analytical guarantee.}
\newcommand{\authorcontribution}{All authors contributed equally to this work. During the preparation of this manuscript, the authors used large language models for proofreading, language editing, bibliographic searching, and code writing. The authors subsequently reviewed and edited the complete manuscript and take full responsibility for its final content.}

\newcommand{\taskphasefigure}{\begin{figure*}[t]\centering\includegraphics[width=\textwidth]{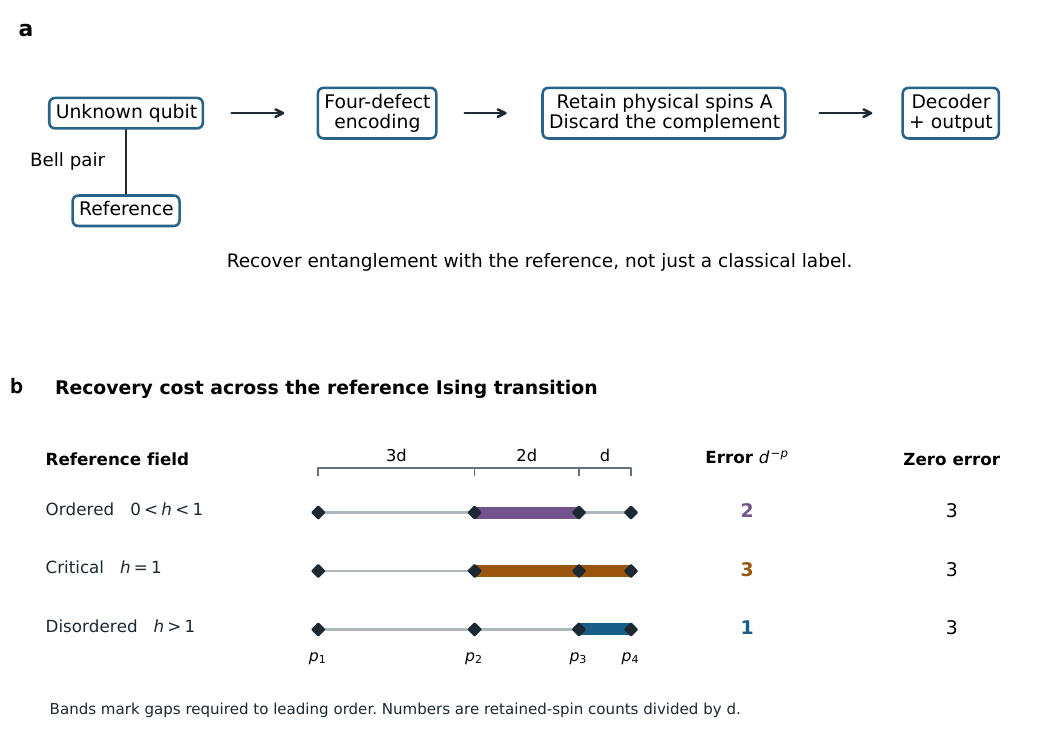}\caption{\taskphasecaption}\label{fig:task-phase}\end{figure*}}
\newcommand{\recordsfigure}{\begin{figure*}[t]\centering\includegraphics[width=0.92\textwidth]{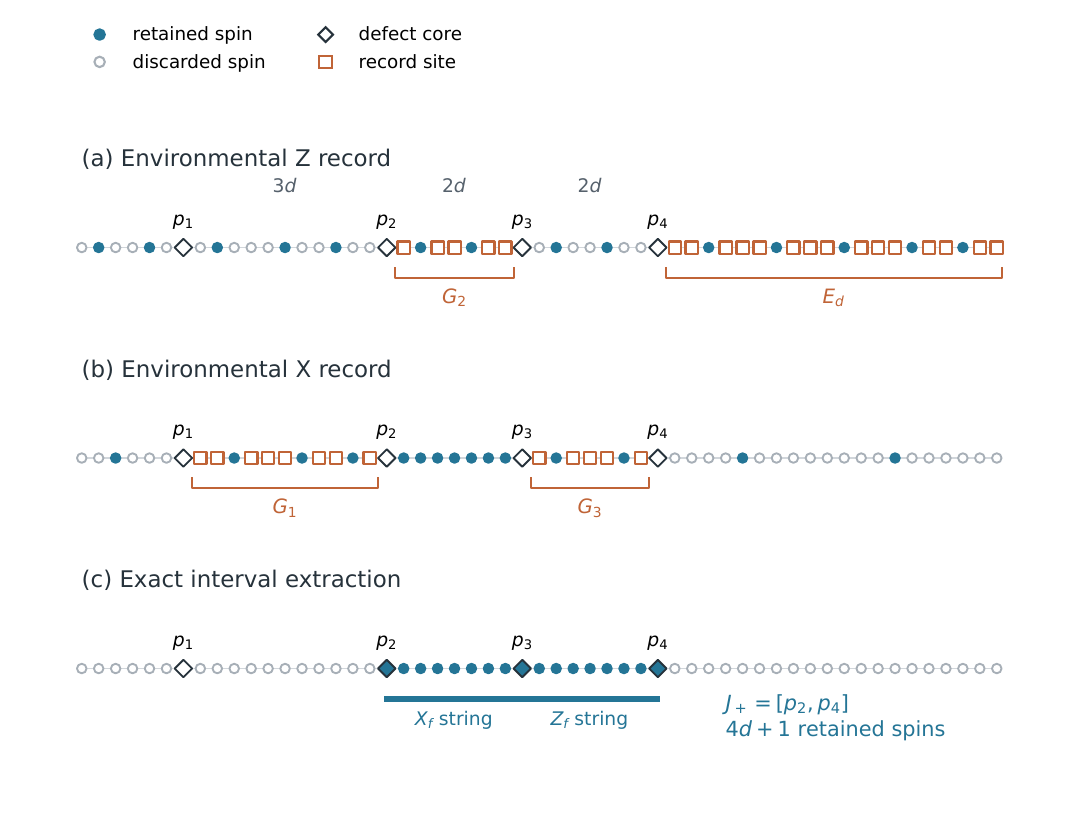}\caption{\recordscaption}\label{fig:records}\end{figure*}}
\newcommand{\protocolfigure}{\begin{figure*}[t]\centering\includegraphics[width=0.94\textwidth]{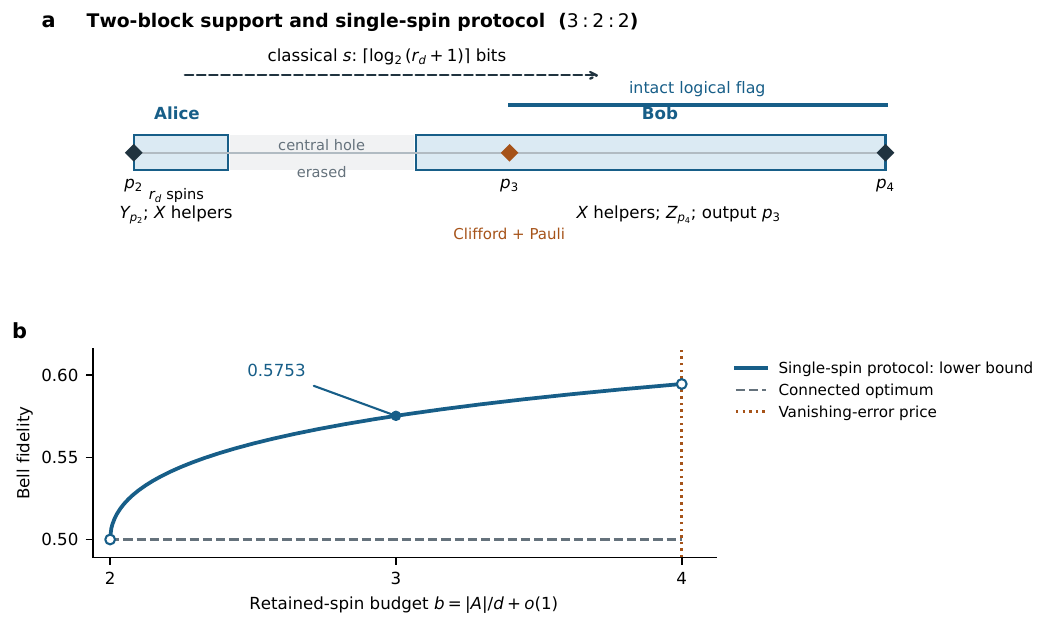}\caption{\protocolcaption}\label{fig:two-block-protocol}\end{figure*}}
\newcommand{\finitescanfigure}{\begin{figure*}[t]\centering\includegraphics[width=\textwidth]{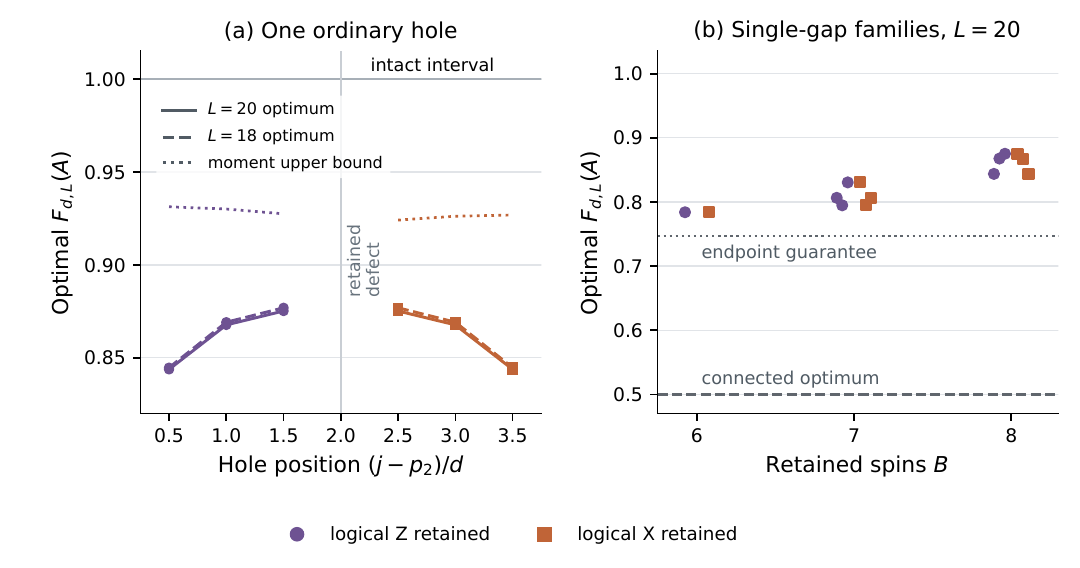}\caption{\finiteplotcaption}\label{fig:finite-recovery}\end{figure*}}

\begin{document}
\maketitle
\hypersetup{pdfauthor={Yi Liang and Chen Qian}}
\begin{abstract}Four Ising duality defects encode an unknown fusion qubit in a periodic physical-spin chain, in the sector of even physical parity and trivial total defect charge. Its basis states share a pure even bulk reference ground state. A matched local Hamiltonian family varies the reference state and encoder at fixed geometry, physical-spin partition and logical algebra. We minimize retained physical-spin count over input-independent supports and arbitrary joint quantum decoders at a prescribed loss in squared Bell entanglement fidelity. At fixed finite positive reference field, the infinite-chain limit precedes large defect separation. Zero error imposed at each finite size gives a common leading cost: the central spacing plus the shorter outer spacing. Allowing any fixed inverse-power error reduces this cost to the central spacing in the ordered phase and the shorter outer spacing in the disordered phase. At criticality, the two-gap cost holds for every vanishing-error sequence, and near-minimal supports achieving such recovery must concentrate in density on leading-shortest three-defect intervals. In a specified critical two-block geometry, an ideal deterministic protocol of single-spin Pauli measurements, one-way classical communication and an output correction maintains a nonvanishing Bell-fidelity margin above $1/2$ in the ordered limits, at budgets where every connected interval is entanglement breaking.

\end{abstract}
\section{Introduction}
\label{sec:introduction}

The physical region needed to recover an encoded qubit can depend on the target accuracy. Standard quantum error correction characterizes recoverable noise through correctable error sets or code distance~\cite{Shor1995QuantumMemory,KnillLaflamme1997QEC}. When physical access is spatially constrained, however, the receiver must select which degrees of freedom to measure or keep. A faint trace of logical information left in unobserved spins strictly prevents exact reconstruction, yet can become negligible if one requires only high-fidelity approximate recovery or the preservation of entanglement with an inert reference. Determining the minimal physical footprint for a specified accuracy therefore requires optimizing the retained spatial region and the decoding operation together.

This spatial optimization connects directly to existing formulations of access and code overhead. Code price quantifies the smallest subsystem that supports a complete logical algebra~\cite{PastawskiPreskill2017CodePrice}, whereas code distance constrains the size of correctable erasures under adversarial loss~\cite{BravyiPoulinTerhal2010Tradeoffs,Arxiv1610_06169}. In holographic and scale-invariant tensor networks, code price admits sparse or fractal reconstructions on disconnected geometries under entanglement-wedge and ideal infinite-scaling-dimension assumptions, respectively~\cite{PastawskiPreskill2017CodePrice,KimKastoryano2017MERAQEC}. Quantum secret sharing similarly classifies access structures into qualified, forbidden, and intermediate share subsets~\cite{CleveGottesmanLo1999QSS,OgawaEtAl2005RampQSS}. An intermediate set in a ramp scheme does not guarantee preserved reference entanglement; it may retain only a classical logical variable. In a physical many-body encoding, high-precision retrieval and nonvanishing entanglement retention below the high-precision price must therefore be treated as distinct operational targets.

Correlated many-body states furnish concrete realizations of both error-correcting codes and recovery diagnostics. Approximate codes in spin-chain eigenstates and error detection in low-energy subspaces link quantum protection directly to Hamiltonian spectral structure~\cite{BrandaoEtAl2019EigenstateCodes,GschwendtnerEtAl2019LowEnergy}. Subsystem variance connects approximate correctability to state complexity and quantum order~\cite{YiYeGottesmanLiu2024AQECOrder}, while conformal-field-theory codes relate dephasing thresholds to scaling dimensions and fusion data under continuum scaling assumptions~\cite{SangHsiehZou2024CFTCodes}. Critical-Ising source--channel codes address small contiguous erasures of thermal sources~\cite{PastawskiEisertWilming2017SourceChannel}, whereas Sachdev--Ye--Kitaev price constructions concern coherent-qubit recovery from fermionic subsystems~\cite{ChandrasekaranLevine2022SYKPrice}. In each case, recovery depends on the accessible operator algebra and the underlying state preparation, rather than on a macroscopic phase designation alone.

Recovery operations likewise diagnose spatial correlations in many-body states. Petz-map reconstructions in conformal field theories and long-range-entangled media probe correlation structures and quantum Markov properties beyond von Neumann entropies~\cite{VardhanWeiZou2024CFTPetz,HuZou2024PetzLongRange}. In Abelian code subspaces of chiral topological edges, approximate recovery can act on the erased region plus a power-law-range buffer while leaving the remaining code region intact~\cite{SongEtAl2026ChiralEdgeCodes}. In the task examined here, by contrast, all spins outside a preselected receiver region are permanently discarded, leaving only that region accessible to an unrestricted joint completely positive, trace-preserving quantum decoder. Access is defined strictly across physical spin tensor factors rather than fermionic modes, so that Jordan--Wigner strings traversing discarded intervals are unavailable~\cite{FagottiCalabrese2010DisjointBlocks,BenyZimborasPastawski2019LocalRecovery}.

We investigate this physical access problem in a periodic Ising spin chain hosting four Kramers--Wannier duality defects [Fig.~\ref{fig:task-phase}(a)]~\cite{FrohlichFuchsRunkelSchweigert2004,AasenMongFendley2016}. Their non-Abelian fusion algebra encodes a logical qubit in the sector of even physical spin parity and trivial total defect charge, where both logical basis states share a pure, even-parity bulk reference ground state. A matched local Hamiltonian family varies the transverse reference field $h$ at fixed defect geometry, fusion algebra, and physical spin factorization, modifying the shared reference state and encoder while leaving the defect operators fixed. Determining the minimal spatial footprint requires bounding logical information leakage across arbitrary discarded configurations uniformly prior to optimizing the receiver support, paired with matching constructive decoders that achieve the leading bounds in the proved accuracy regimes.

\subsection*{Main results and scope}

Let $L$ be the number of spins in the ring, $d$ the defect separation scale, and $\ell_i$ the three consecutive defect spacings, with $\ell_i/d\to\lambda_i>0$ for $i=1,2,3$. Set $m=\min(\lambda_1,\lambda_3)$ as the normalized length of the shorter outer gap. All spatial limits evaluate the infinite-chain limit $L\to\infty$ at fixed $d$ before taking $d\to\infty$. Table~\ref{tab:phase-overview} summarizes the leading-order minimum retained-spin count, optimized over all input-independent physical spin subsets and all joint completely positive, trace-preserving decoders. Here $\epsilon_d$ denotes the allowed loss in optimized squared Bell entanglement fidelity. Demanding exact recovery ($\epsilon_d=0$) at every finite lattice size imposes an identical leading spatial cost of $(\lambda_2+m)d$ across all finite positive fields. For inverse-power error $\epsilon_d=d^{-p}$ with fixed $p>0$, the central spacing sets the ordered-phase coefficient, the shorter outer spacing sets the disordered-phase coefficient, and their sum sets the critical coefficient [Fig.~\ref{fig:task-phase}(b)].

\begin{table}[t]
\centering
\caption{Leading retained-spin count divided by $d$ at fixed reference field. Zero error is imposed before taking either spatial limit. The inverse-power column holds for each fixed $p>0$. The formal optimization and limits are defined in Sec.~\ref{sec:encoding}.}
\label{tab:phase-overview}
\begin{tabular}{lcc}
\toprule
Reference field & Zero error & Error $d^{-p}$\\
\midrule
Ordered: $0<h<1$ & $\lambda_2+m$ & $\lambda_2$\\
Critical: $h=1$ & $\lambda_2+m$ & $\lambda_2+m$\\
Disordered: $h>1$ & $\lambda_2+m$ & $m$\\
\bottomrule
\end{tabular}
\end{table}
For spacings $3d,2d,d$, a fixed Bell fidelity of at least $0.99$ reduces the leading cost coefficient from $3$ to $2$ at $h=1/2$ and to $1$ at $h=2$, whereas at criticality it remains between $2.9964$ and $3$ [Eq.~\eqref{eq:fixed-target-example}].
\taskphasefigure

At criticality ($h=1$), the two-gap coefficient $\lambda_2+m$ applies to every nonnegative infidelity sequence $\epsilon_d\to0$, not merely power laws. Uniform bounds on collective logical records leaking into unobserved spins establish that any successful support of linear budget must fill the central gap and at least one outer gap in density. Consequently, any sequence of near-minimal successful regions has vanishing normalized symmetric difference from a three-defect interval that is shortest to leading order. When the outer ratios tie, the comparison interval may change with system size. This density constraint governs arbitrary irregular or fragmented geometries, including configurations that change with system size. Exact unitary extractors on three consecutive defects supply the matching upper bound.

Below this high-precision threshold, spatial geometry determines whether quantum reference entanglement can survive. For the critical benchmark with exact spacings $3d,2d,2d$, under the finite-ring conditions of Eq.~\eqref{eq:finite-geometry-regime}, each fixed budget coefficient $2<b<4$ admits a two-block fragmented support of $bd+O(1)$ spins that preserves a nonvanishing Bell-fidelity margin above $1/2$. At $b=3$, this support retains $B_d=2d+2\lfloor d/2\rfloor=3d+O(1)$ spins. At this same budget, the optimum over connected cyclic intervals is $1/2$ for $d\ge2$, whereas the specified single-spin protocol attains an ordered lower Bell-fidelity limit of at least $3371/5860\simeq0.575256$. A deterministic four-step instrument realizes this guarantee using individual Pauli measurements, one classical message transmitted from the left block to the right, and a one-spin Clifford and Pauli correction on a designated quantum output. The unobserved central void is permanently erased before decoding, and every measured helper spin counts toward the physical budget. While localizable entanglement and one-way extraction have established precedents in spin chains, stabilizer codes, and helper frameworks~\cite{VerstraetePoppCirac2004LocalizableEntanglement,ShiraishiYamasakiMurao2024LOCCExtraction,ChitambarEtAl2026HelperCodes}---where helper schemes may utilize joint Pauli measurements within a share---this result provides an explicit macroscopic guarantee for a critical spin encoding using strictly individual measurements, rather than asserting an optimum over all fragmented supports or one-way decoders.

The underlying mechanism reflects a dual role for bulk correlations. On the environment side, complementary order and disorder records constrain every possible decoder. At criticality, the conditional mean magnitude and the square root of the raw second moment both scale as $d^{7/4}$ for the positive-density omission sets used in the converse, producing an information-disturbance bound that holds uniformly over their spatial patterns. On the receiver side, retaining an unbroken gap preserves an exact logical flag that isolates encoded populations, while a conjugate collective record across the remaining boundaries transfers quantum coherence. The phase-dependent high-precision price and the low-budget fragmented protocol thus address distinct operational accuracy targets within the same physical access problem.

\begin{samepage}
Section~\ref{sec:encoding} establishes the defect encoding, physical access channel, and complementary moment bound. Sections~\ref{sec:phase} and~\ref{sec:critical} determine the phase-dependent recovery costs and critical recovery geometry, while Section~\ref{sec:quantum-access} presents the fragmented recovery protocol and its operational resource bounds. Finally, Section~\ref{sec:discussion} addresses physical implications, with technical proofs and extensions detailed in the appendices.
\par
\end{samepage}

\section{Encoding and recovery task}
\label{sec:encoding}

The receiver retains an input-independent set of physical spins, with the complement permanently erased before decoding. Changing the common bulk reference state changes the encoding, even though the defect geometry and fusion algebra stay fixed. We therefore minimize the retained-spin count at a prescribed squared Bell entanglement-fidelity target, allowing arbitrary joint completely positive, trace-preserving decoders. An exact extractor on an interval containing three consecutive defect cores gives an upper bound; logical records in the discarded spins give converse bounds that apply to every decoder.

\subsection{A fusion qubit and a common reference state}

The physical Hilbert space consists of a periodic ring of $L$ spins with local Pauli operators $X_j,Y_j,Z_j$, subject to periodic boundary conditions identifying site indices modulo $L$. Four nonadjacent defect cores occupy coordinates $D=\{p_1<p_2<p_3<p_4\}$, positioned away from the boundary seam such that $1<p_1$ and $p_4<L$. At criticality, the physical Hamiltonian takes the form~\cite{AasenMongFendley2016}
\begin{equation}
H_D=-\sum_{j\notin D}X_j-\sum_{j:j+1\notin D}Z_jZ_{j+1}
-\sum_{p\in D}Z_{p-1}Y_p .
\label{eq:hamiltonian}
\end{equation}
At each defect site $p\in D$, the localized term $Z_{p-1}Y_p$ replaces both the local transverse field $X_p$ and the incoming Ising coupling $Z_{p-1}Z_p$.

Under the standard Jordan--Wigner transformation
\begin{equation}
 c_{2j-1}=\left(\prod_{k<j}X_k\right)Z_j,\qquad
 c_{2j}=\left(\prod_{k<j}X_k\right)Y_j,
\label{eq:jw}
\end{equation}
the odd Majorana modes at the defect coordinates, $\gamma_a=c_{2p_a-1}$ for $a=1,\ldots,4$, decouple entirely from the bulk bonds. Although individual Majorana operators do not necessarily commute with the parity-dependent boundary seam~\cite{SeibergShao2024}, their even bilinears commute with the full Hamiltonian $H_D$ and preserve the physical spin parity $P=\prod_j X_j$. The fusion-qubit Pauli operators and the total defect charge are given by
\begin{equation}
 Z_f=i\gamma_1\gamma_2,\quad X_f=i\gamma_2\gamma_3,\quad
 Q=(i\gamma_1\gamma_2)(i\gamma_3\gamma_4).
\label{eq:fusion-pauli}
\end{equation}
Here $Z_f$ and $X_f$ are mutually anticommuting Hermitian involutions that generate the logical qubit algebra, while the central involution $Q$ fixes the total fusion charge sector.

Excluding the four defect modes leaves an ordered sequence of $2n$ active Majorana modes $\alpha_1,\ldots,\alpha_{2n}$, where $n=L-2$. Their active parity $P_b=i^n\alpha_1\cdots\alpha_{2n}$ satisfies the sector relation $P=QP_b$. Restricting to the joint sector $P=Q=+1$, an isometry $W$ maps $\mathbb C_f^2\otimes\mathcal H_{b,+}$ onto the physical subspace, where $\mathcal H_{b,+}$ is the even-parity subspace of $n$ reference spins. We vary the bulk background through the matched parent Hamiltonian
\begin{equation}
\begin{split}
 H_D(h)={}&-ih\sum_{j=1}^n\alpha_{2j-1}\alpha_{2j}
-i\sum_{j=1}^{n-1}\alpha_{2j}\alpha_{2j+1}
+iP\alpha_{2n}\alpha_1 .
\end{split}
\label{eq:matched-parent}
\end{equation}
Every bond term in Eq.~\eqref{eq:matched-parent}, including the parity-dependent seam term, is a local physical spin interaction whose coupling coefficient is determined by its sequential position in the active-mode ordering; the assignment of these weights to physical terms changes at each defect core. In the selected sector, $H_D(h)$ acts as $I_f\otimes H_{b,+}(h)$, corresponding to the even restriction of the periodic reference Ising chain
\begin{equation}
 H_b(h)=-h\sum_{j=1}^n\widetilde X_j
-\sum_{j=1}^n\widetilde Z_j\widetilde Z_{j+1} .
\label{eq:reference-hamiltonian}
\end{equation}
Tildes distinguish the reference spin operators from physical spins. For any $h>0$, the off-diagonal entries of $H_b(h)$ in the computational $\widetilde Z$ basis are nonpositive and connect all basis configurations, so the Perron--Frobenius theorem guarantees a unique, strictly positive ground state $\Omega_{n,h}$, whose global spin-inversion symmetry ensures even parity $P_b=+1$~\cite{Pfeuty1970}. The resulting encoding isometry is defined by
\begin{equation}
 V_{h,d,L}|a\rangle=W\bigl(|a\rangle_f\otimes\Omega_{n,h}\bigr),
 \qquad a=0,1,
\label{eq:encoding}
\end{equation}
where $Z_f|a\rangle_f=(-1)^a|a\rangle_f$. Both logical basis states share the identical pure reference bulk $\Omega_{n,h}$ and differ solely in their fusion state. Tuning the field $h$ modifies this shared bulk state without altering the defect coordinates, the fusion algebra, or the physical spin tensor factorization. The family thus traverses the ordered ($0<h<1$), critical ($h=1$), and disordered ($h>1$) phases, interpolating between the even Greenberger--Horne--Zeilinger (GHZ) reference at $h=0$ and the fully polarized transverse product reference at $h=\infty$.

\subsection{Physical access, accuracy and spatial cost}

Physical access is defined by assigning the receiver an unconstrained subset of physical spins $A\subseteq\{1,\ldots,L\}$, selected independently of the unknown logical input state. Tracing out the unobserved environment on the complementary set $\bar A=\{1,\ldots,L\}\setminus A$ induces the physical erasure channel
\begin{equation}
 \mathcal E_{A,h,d,L}(\rho)=\operatorname{Tr}_{\bar A}
 (V_{h,d,L}\rho V_{h,d,L}^\dagger).
\label{eq:access-channel}
\end{equation}
Crucially, the tensor product partition between $A$ and $\bar A$ acts on the physical spin Hilbert space rather than on fermionic modes. Nonlocal Jordan--Wigner strings threading through the discarded region $\bar A$ cannot be evaluated from physical access to $A$ alone~\cite{FagottiCalabrese2010DisjointBlocks,BenyZimborasPastawski2019LocalRecovery}. The algebraic identification $W$ establishes the code structure internally but does not modify the physical subsystem accessible to the receiver.

To evaluate recovery accuracy for arbitrary unknown logical inputs, the encoded qubit is coupled to an isolated reference in the maximally entangled Bell state $|B\rangle=(|00\rangle+|11\rangle)/\sqrt2$. We optimize the squared Bell entanglement fidelity~\cite{Schumacher1996EntanglementFidelity},
\begin{equation}
 F_{h,d,L}(A)=\max_{\mathcal R_A}
 \langle B|[\operatorname{id}\otimes(\mathcal R_A\mathcal E_{A,h,d,L})]
 (|B\rangle\langle B|)|B\rangle .
\label{eq:fidelity}
\end{equation}
The maximum is taken over all completely positive, trace-preserving (CPTP) decoding maps $\mathcal R_A$ from the physical algebra on $A$ to a single output qubit. The decoding operation permits unrestricted joint quantum operations on $A$ and the inclusion of arbitrary fresh ancillas, with every retained helper spin accounted for in the resource budget $|A|$. The complementary spins $\bar A$ are completely discarded prior to decoding, distinguishing this channel from local-buffer error correction where intact code regions persist~\cite{SongEtAl2026ChiralEdgeCodes}, as well as from monitored code models assisted by classical measurement histories~\cite{LeeYoshida2026MonitoredCodes}. Achieving $F_{h,d,L}(A)>1/2$ certifies nonclassical entanglement between the decoded output and the reference.

For a specified error tolerance $0\le\epsilon<1$, the spatial resource cost is defined by the minimal physical spin cardinality supporting recovery with fidelity at least $1-\epsilon$:
\begin{equation}
 n_{*,L}^{(h)}(d,\epsilon)=\min\{|A|:F_{h,d,L}(A)\ge1-\epsilon\}.
\label{eq:price-definition}
\end{equation}
This formulation corresponds to a code price optimization, which seeks the smallest physical footprint that permits successful recovery~\cite{PastawskiPreskill2017CodePrice}. It stands in operational contrast to the standard code distance, which minimizes the erasure size that destroys correctability via a complementary quantifier over regions, and to approximate logical-support criteria formulated in operator norm~\cite{Arxiv1610_06169}.

To separate macroscopic spatial scaling from microscopic lattice effects, we parameterize the defect layout by an integer scale $d$. At each fixed $d$, the defect coordinates are held constant as $L$ increases, with the inter-defect separations scaling linearly:
\begin{equation}
 \ell_i=p_{i+1}-p_i,\qquad \ell_i/d\longrightarrow\lambda_i>0,
 \qquad i=1,2,3 .
\label{eq:gap-scaling}
\end{equation}
Candidate recovery regions $A$ may vary arbitrarily with $L$ and $d$, cross the boundary cut, or consist of multiple disconnected intervals. For an error tolerance sequence $\epsilon_\bullet=(\epsilon_d)$, the asymptotic leading-order spatial costs are bounded by the lower and upper envelopes
\begin{equation}
\begin{split}
 \underline b_h[\epsilon_\bullet]&=
 \liminf_{d\to\infty}\liminf_{L\to\infty}
 \frac{n_{*,L}^{(h)}(d,\epsilon_d)}d,\\
 \overline b_h[\epsilon_\bullet]&=
 \limsup_{d\to\infty}\limsup_{L\to\infty}
 \frac{n_{*,L}^{(h)}(d,\epsilon_d)}d .
\end{split}
\label{eq:cost-envelopes}
\end{equation}
In these asymptotic envelopes, the thermodynamic limit $L\to\infty$ is evaluated prior to the large-defect limit $d\to\infty$. Exact recovery requires $\epsilon_d=0$ at each finite lattice size, prior to taking either limit.

\subsection{Exact extraction on three-defect intervals}

Physical spin intervals spanning three consecutive defect cores directly support the logical Pauli algebra. Let $X,Z$ denote the input-qubit Pauli matrices, with $X|a\rangle=|1-a\rangle$ and $Z|a\rangle=(-1)^a|a\rangle$ for $a=0,1$. Spatially, for ordered defect labels $1\le a<b\le4$, the Majorana bilinears take the physical spin-string form
\begin{equation}
i\gamma_a\gamma_b
=-Y_{p_a}\left(\prod_{p_a<j<p_b}X_j\right)Z_{p_b}.
\label{eq:physical-logical-string}
\end{equation}
Because the canonical interval $[p_a,p_b]$ avoids the chosen boundary seam, this expression makes its physical support explicit. A receiver region itself need not avoid that seam.

On $J_+=[p_2,p_4]\cap\mathbb Z$, the strings $X_f=i\gamma_2\gamma_3$ and $\widehat Z=i\gamma_3\gamma_4=QZ_f$ are supported entirely in $J_+$ and form anticommuting Hermitian involutions on its full spin Hilbert space. Restricted to the code $Q=+1$, they satisfy $X_fV_{h,d,L}=V_{h,d,L}X$ and $\widehat ZV_{h,d,L}=V_{h,d,L}Z$. Pairing an orthonormal basis of the $+1$ eigenspace of $\widehat Z$ with its image under $X_f$ yields a decoding unitary $U_{J_+}$ acting only on $J_+$~\cite{BenyKempfKribs2007HeisenbergQEC}, such that
\begin{equation}
 (U_{J_+}\otimes I_{\bar J_+})V_{h,d,L}|a\rangle
 =|a\rangle\otimes|\chi_h\rangle,
\label{eq:extractor}
\end{equation}
for every $h>0$, with the same residual state $|\chi_h\rangle$ for both inputs. Tracing out the residual retained degrees of freedom recovers the unknown qubit exactly and preserves its entanglement with an inert reference.

The left interval $J_-=[p_1,p_3]\cap\mathbb Z$ supports the analogous pair $X_f$ and $Z_f=i\gamma_1\gamma_2$, which likewise anticommutes on the full spin Hilbert space of $J_-$ and implements the logical algebra on the code subspace. Retaining the shorter of $J_\pm$ gives the finite-lattice upper bound
\begin{equation}
 n_{*,L}^{(h)}(d,0)\le\ell_2+\min(\ell_1,\ell_3)+1.
\label{eq:exact-upper}
\end{equation}
Appendix~\ref{app:encoding} proves the general extraction lemma and verifies the intertwining for every matched encoding.

\subsection{Logical records and a recovery bound}

Let $G_i=(p_i,p_{i+1})\cap\mathbb Z$ denote the open intervals of ordinary spins between adjacent defect cores. The two record geometries appear in Fig.~\ref{fig:records}. Throughout the fixed sector $P=Q=+1$, two microscopic physical-spin identities connect physical correlators to logical Pauli operators and reference observables:
\begin{equation}
\begin{aligned}
 Z_jZ_k&=-Z_f\widetilde Z_{j-1}\widetilde Z_{k-2},
 &j&\in G_2,\ k>p_4,\\
 Z_jZ_k&=-X_f\prod_{m=j}^{k-2}\widetilde X_m,
 &j&\in G_1,\ k\in G_3 .
\end{aligned}
\label{eq:two-dictionaries}
\end{equation}
These test strings are chosen not to cross the boundary seam. Their reference-spin factors correspond, respectively, to an order-parameter pair and a disorder string. Because multiplying two bilinears of the same pairing cancels the logical Pauli operator—including for coincident indices—any linear combination of such physical records exhibits conditional expectation values of equal magnitude and opposite sign in the corresponding logical eigenbasis, accompanied by an identical raw second moment. As established in Appendix~\ref{app:critical}, these algebraic identities provide the direct interface between the spatial geometry of erasure and the logical information leaked to the environment.

\recordsfigure

\begin{lemma}[Complementary moment bound]
\label{lem:moment}
Suppose a Hermitian observable $O$ on $\bar A$ has conditional means $\pm\mu$ on two opposite logical basis states and a common raw second moment $\nu>0$. Then
\begin{equation}
 F_{h,d,L}(A)\le\frac{1+\sqrt{1-\mu^2/\nu}}2 .
\label{eq:moment-bound}
\end{equation}
\end{lemma}
\begin{proof}
Let $p(x)$ and $q(x)$ denote the complete measurement outcome probability distributions of the observable evaluated on the two logical basis states, corresponding to conditional means $+\mu$ and $-\mu$, respectively. Defining the average distribution $\tau=(p+q)/2$ and the signed likelihood ratio $v=(p-q)/(p+q)$ on the support of $\tau$, direct evaluation yields $\mathbb E_\tau[xv]=\mu$ and $\mathbb E_\tau[x^2]=\nu$. Applying the Cauchy--Schwarz inequality followed by the concavity of the square root yields
\begin{equation}
 \sum_x\sqrt{p(x)q(x)}
 =\mathbb E_\tau\sqrt{1-v^2}\le\sqrt{1-\mu^2/\nu}.
\label{eq:affinity}
\end{equation}
Because quantum operations and measurements cannot decrease root fidelity, this classical affinity bounds the quantum root fidelity between the complementary states on $\bar A$. By Uhlmann's theorem~\cite{Uhlmann1976TransitionProbability}, this fidelity matches the trace norm $\|C_A\|_1$ of the off-diagonal correlation operator on the accessible subsystem, $C_A=\operatorname{Tr}_{\bar A}|\Phi_+\rangle\langle\Phi_-|$, where $|\Phi_\pm\rangle$ are the code states corresponding to the chosen logical basis.

For an arbitrary CPTP decoding channel $\mathcal R_A$, the Bell entanglement fidelity can be expressed in terms of the unital adjoint map $\mathcal R_A^*$. The two diagonal contributions to $4F_{h,d,L}(A)$ are each bounded by unity, while each of the two off-diagonal terms has modulus at most $\|C_A\|_1$ because the unital completely positive map $\mathcal R_A^*$ contracts the operator norm. Summing these terms yields $F_{h,d,L}(A)\le(1+\|C_A\|_1)/2$, which establishes Eq.~\eqref{eq:moment-bound}.
\end{proof}

The normalization in Eq.~\eqref{eq:moment-bound} depends strictly on the raw second moment $\nu=\langle O^2\rangle$ rather than the variance, dispensing with any requirement for an operator-norm bound that remains uniform in system size. Furthermore, because optimal Bell recovery fidelity is invariant under single-qubit rotations of the logical basis, both the $Z_f$ and $X_f$ complementary records constrain the identical recovery task. While general information--disturbance principles dictate that environmental leakage limits reconstructability~\cite{Arxiv0907_5391}, our operational objective is to evaluate these physical records uniformly across all candidate retained configurations $A$.

\section{Optimal spatial cost across the Ising transition}
\label{sec:phase}

With defect separations scaling as $\lambda_i d$, an interval spanning the central defect gap together with the narrower outer gap achieves exact recovery on $b_0 d + o(d)$ physical spins, where $m=\min(\lambda_1,\lambda_3)$ and $b_0=\lambda_2+m$. At any finite positive reference field, this leading spatial footprint is irreducible so long as zero error is demanded. Relaxing the target to an infidelity bounded by an arbitrary fixed power law $d^{-p}$, however, bifurcates the spatial requirement across the quantum phase transition.

\begin{theorem}[Fixed-field recovery costs]
\label{thm:phase}
Fix $0<h<\infty$, $p>0$, and the asymptotic gap ratios in Eq.~\eqref{eq:gap-scaling}, with the code space defined by the even reference state of Eq.~\eqref{eq:encoding}. The optimal asymptotic spatial recovery envelopes in Eq.~\eqref{eq:cost-envelopes}, minimized over all preselected physical-spin supports and arbitrary joint CPTP decoders, satisfy
\begin{equation}
 \underline b_h[0]=\overline b_h[0]=b_0
\label{eq:exact-field-price}
\end{equation}
for exact recovery at every system size. Under the power-law error threshold $\epsilon_d=d^{-p}$, these envelopes evaluate to
\begin{equation}
 \underline b_h[d^{-p}]=\overline b_h[d^{-p}]
 =\begin{cases}
 \lambda_2,&0<h<1,\\
 \lambda_2+m,&h=1,\\
 m,&1<h<\infty .
 \end{cases}
\label{eq:finite-field-price}
\end{equation}
\end{theorem}

The phase costs are illustrated in Fig.~\ref{fig:task-phase}. The proof of Theorem~\ref{thm:phase} divides naturally: this section establishes the exact-recovery price and the gapped phases ($h \ne 1$), while the critical price at $h=1$ is deferred to Theorem~\ref{thm:critical}, where collective rather than single-pair environmental records govern recovery. All statements here apply at fixed $h$, without invoking a scaling crossover $h(d) \to 1$; detailed finite-ring transport proofs and quasi-adiabatic locality estimates appear in Appendix~\ref{supp:finite-field}.

\subsection{Why exact recovery retains both gap contributions}

Because the connected three-defect geometry of Eq.~\eqref{eq:exact-upper} already achieves exact recovery on $s_d+1$ spins, where $s_d=\ell_2+\min(\ell_1,\ell_3)$, optimal supports need never exceed this cardinality. In the thermodynamic regime where the exterior region satisfies $L-p_4 > s_d+1$, any competing support of size at most $s_d+1$ necessarily leaves at least one exterior spin unobserved. If such a candidate support also omits an ordinary spin from the central gap, Eq.~\eqref{eq:two-dictionaries} gives an unobserved physical spin pair $Z_jZ_k$ whose code compression is $-t_Z Z_f$, where $t_Z$ is the corresponding reference order correlator. If both outer gaps contain an omitted ordinary spin, their physical spin pair instead compresses to $-t_X X_f$, where $t_X$ is the reference disorder-string expectation.

At any finite field $h>0$, both reference coefficients $t_Z$ and $t_X$ are strictly positive on the finite ring. In the reference $\widetilde Z$ basis, the Perron--Frobenius ground state has strictly positive amplitudes, which a disorder string merely permutes with positive signs. In the dual $\widetilde X$ basis, nearest-neighbour ferromagnetic couplings connect all even-parity spin configurations via adjacent pair flips, and an order pair similarly permutes the positive components of the even ground state. Because these unobserved environmental operators square to the identity, Lemma~\ref{lem:moment} with $\nu=1$ bounds the attainable entanglement fidelity strictly below unity at each finite size whenever either physical spin-pair record falls into the environment. Zero-error recovery therefore requires retaining every ordinary spin across the central gap, as well as the entirety of at least one outer gap. Tallying these obligatory sites against the upper bound yields
\begin{equation}
 s_d-2\le n_{*,L}^{(h)}(d,0)\le s_d+1
\label{eq:finite-exact-sandwich}
\end{equation}
along the thermodynamic tail. Dividing by $d$ and passing to the limit confirms Eq.~\eqref{eq:exact-field-price}. While this constant-width bracket leaves the precise integer optimum on finite rings undetermined, its asymptotic slope is exact.

\subsection{The phase selects which record persists}

Away from criticality ($h \ne 1$), the long-range behavior of the bulk reference state is governed by the duality parameter $\kappa=\min(h,1/h)<1$. At the ordered reference coupling $\kappa$, the infinite-chain two-point order correlator at large spatial separation approaches the square of the spontaneous magnetization~\cite{Pfeuty1970}
\begin{equation}
 t_h=(1-\kappa^2)^{1/4}>0.
\label{eq:gapped-response}
\end{equation}
Kramers--Wannier duality on the finite ring maps a disorder string at coupling $h$ directly to an order pair at coupling $1/h$. Long-range order therefore preserves a macroscopic central--exterior record in the ordered reference phase ($0<h<1$); its dual disorder response preserves an outer--outer record in the disordered phase ($h>1$).

To establish that these records rule out recovery for arbitrary candidate geometries, it is sufficient to restrict the search to supports satisfying a linear budget $|A| \le M d$, since the upper bound is already $O(d)$. On the thermodynamic tail, an exterior window of length $\lceil(M+1)d\rceil$ sites beyond $p_4$ must contain an unobserved spin. For any candidate support missing the designated bulk spins, the possible separation distances between the unobserved partner spins fall into a finite coordinate band whose width depends only on the defect positions and $M$. As $L \to \infty$ at fixed $d$, the finite-ring correlators converge uniformly across this band to their infinite-chain values. Because the lower edge of this distance window scales linearly with $d$, the minimum environmental expectation over the band subsequently approaches $t_h$ as $d \to \infty$.

By Lemma~\ref{lem:moment}, any support suffering from these omissions is subject to an asymptotic Bell-fidelity ceiling
\begin{equation}
 f_h^{\rm pair}=\frac{1+\sqrt{1-t_h^2}}2<1.
\label{eq:gapped-fidelity-ceiling}
\end{equation}
Achieving an error tolerance $\epsilon_d = d^{-p}$ therefore requires eliminating the persistent record entirely: the retained subsystem must contain the entire central ordinary gap when $0<h<1$, or at least one complete outer ordinary gap when $h>1$. Accounting for these mandatory spins yields the lower leading coefficients $\lambda_2$ and $m=\min(\lambda_1,\lambda_3)$, respectively.

\subsection{Constructive upper bounds from the endpoints}

Exact endpoint decoders exploit cancellations in the physical strings of Eq.~\eqref{eq:physical-logical-string}. Setting $\widehat Z=i\gamma_3\gamma_4=QZ_f$, the two-endpoint physical boundary operators
\begin{equation}
\begin{aligned}
O_X&=-Y_{p_2}Z_{p_3}=X_f\prod_{j\in G_2}X_j,\\
O_Z&=-Y_{p_3}Z_{p_4}=\widehat Z\prod_{j\in G_3}X_j
\end{aligned}
\label{eq:conjugate-endpoints}
\end{equation}
satisfy operator identities on the full physical spin Hilbert space; the identification $\widehat Z=Z_f$ holds on the $Q=+1$ sector, not globally. Under the fixed-sector isometry $W$, the bulk product across $G_2$ maps to the disorder string $\prod_{j\in G_2}\widetilde X_{j-1}$, while the product across $G_3$ telescopes to the reference order pair $\widetilde Z_{p_3-1}\widetilde Z_{p_4-2}$, formed prior to projecting onto the even reference parity subspace. Because the product reference $\Omega_{\rm P}=|+\rangle_X^{\otimes n}$ stabilizes every bulk disorder string and the even-GHZ reference $\Omega_{\rm F}=(|0\cdots0\rangle_Z+|1\cdots1\rangle_Z)/\sqrt2$ stabilizes every order pair, the corresponding pure even-parity encoders $V_{\rm P}$ and $V_{\rm F}$ satisfy the exact intertwining relations $O_XV_{\rm P}=V_{\rm P}X$ and $O_ZV_{\rm F}=V_{\rm F}Z$ with the input Pauli operators. Rather than furnishing short logical representatives simultaneously on a single code, each two-endpoint observable replaces an entire gap string on a distinct reference encoding.

In the even-GHZ encoding ($V_{\rm F}$), retaining $A_0=[p_2,p_3]\cup\{p_4\}$ supports the full central string $X_f$ alongside the conjugate endpoint observable $O_Z$, forming an exact anticommuting logical pair on $\ell_2+2=\lambda_2d+o(d)$ physical spins. For the product encoding ($V_{\rm P}$), both recovery branches share the central two-endpoint observable $O_X$, while the complete outer $Z$ string alternates with the defect geometry: retaining $A_0=\{p_2\}\cup[p_3,p_4]$ when $\ell_3\le\ell_1$ (capturing $\widehat Z$) and $A_0=[p_1,p_2]\cup\{p_3\}$ otherwise (capturing $Z_f$). These product supports require $\min(\ell_1,\ell_3)+2=md+o(d)$ physical spins. Lemma~\ref{lem:exact-extraction} gives exact decoders, with the endpoint recovery bounds proved in Appendix~\ref{supp:reference-endpoints}. At these solvable endpoints, the minimum physical spin budgets required for strict Bell advantage and for exact recovery differ by at most three spins, with the exterior-volume condition $L-p_4>\ell_2+2$ sufficient to ensure an unobserved exterior spin in the even-GHZ comparison; this tight finite-size bracket remains an endpoint property and does not classify intermediate fidelity thresholds at generic finite fields.

To extend these constructions to any fixed gapped coupling $h \ne 1$, one deforms the endpoint code along the local Hamiltonian path: tuning the reference transverse-field coefficient from $0$ to $h$ in the ordered regime, or holding that coefficient at unity while ramping the reference exchange couplings, including the seam, from $0$ to $1/h$ in the disordered regime. Throughout either deformation, the fusion doublet remains separated from excited states in its symmetry sector by a uniform spectral gap of at least $2(1-h)$ or $2(1-1/h)$, respectively. Intersector states, even if degenerate or lower in energy, decouple by parity and defect conservation. Filtered spectral flow~\cite{Arxiv1102_0842} then generates a unitary operator $U_h$ on the full physical spin chain that intertwines the code spaces while commuting with the logical Pauli representatives, thereby preserving the encoded qubit.

Let $A_w$ denote the $w$-neighbourhood of $A_0$ on the spin ring, and let $\mathcal D_0$ be the exact endpoint decoder. With $\iota_X$ denoting extension by identity outside $X$, and $\mathbb E_X$ representing the conditional expectation that traces out the complement of $X$ with normalized trace and reinserts identity, we define the adjoint of the transported decoder by
\begin{equation}
 \iota_{A_w}\mathcal D_{h,w}^{*}(O)=
 \mathbb E_{A_w}\!\left[U_h\iota_{A_0}\mathcal D_0^{*}(O)U_h^\dagger\right].
\label{eq:field-decoder}
\end{equation}
Because the mapping $\mathcal D_{h,w}^{*}$ is unital and completely positive, its predual channel $\mathcal D_{h,w}$ is deterministic (trace preserving) and acts strictly on the retained subsystem $A_w$ via joint operations.

For a support $X$, let $X_w$ denote its $w$-neighbourhood. Quasi-adiabatic decay bounds on the unprojected spin chain guarantee that for any local observable $O_X$ and every integer $N>0$,
\begin{equation}
 \|U_h O_XU_h^\dagger-\mathbb E_{X_w}(U_hO_XU_h^\dagger)\|
 \le C_N|X|(1+w)^{-N}\|O_X\| ,
\label{eq:field-tail}
\end{equation}
where the constants $C_N$ remain uniform across $d$, $L$, and compact subsets of couplings within either gapped phase. Write $\eta_w=C_N|A_0|(1+w)^{-N}$ and let $\Phi$ denote the resulting logical recovery channel. Expanding an arbitrary norm-one test observable on the output qubit and a reference of any dimension into the four qubit matrix units, with reference blocks of norm at most one, gives $\|\Phi-\mathrm{id}\|_\diamond\le4\eta_w$. The squared Bell fidelity loss is therefore at most $2\eta_w$. Selecting a spatial buffer $w=\lceil\sqrt d\rceil$ and an exponent $N>2(p+1)$ yields an error that is $o(d^{-p})$, while adding at most $4w=o(d)$ boundary spins to the endpoint geometry. The leading spatial footprint of $A_w$ thus matches the lower-bound envelopes in Eq.~\eqref{eq:finite-field-price}.

The divergence between exact and polynomial-error spatial costs does not reflect a change in the underlying logical algebra. Rather, while both complementary environmental records are present at every finite positive field, only a single record retains macroscopic magnitude in each gapped phase. At the critical point $h=1$, neither isolated two-point witness survives; as the next section demonstrates, recovery is then governed by collective environmental records that reinstate the full two-gap price $b_0$.

\section{Critical records and near-minimal recovery geometry}
\label{sec:critical}

At the critical point $h=1$, where we abbreviate $F_{d,L}=F_{1,d,L}$ and $n_{*,L}=n_{*,L}^{(1)}$, the algebraic decay of two-point correlations suppresses the response of any isolated spin pair across macroscopic distances. Nonetheless, an extensive ensemble of discarded spins aggregates into a collective logical witness whose expectation scales at the same rate as the square root of its raw second moment. This balance provides the mechanism for a uniform converse that encompasses arbitrary, irregular, and disconnected candidate recovery regions.

For any sequence of physical support regions $A_{d,L}$, we write $u_i=|G_i\setminus A_{d,L}|$ for the number of omitted sites within the ordinary gap $G_i$. Let $\mathfrak J$ denote the candidate three-defect interval determined by the limiting gap ratios: $\mathfrak J=\{J_-\}$ when $\lambda_1<\lambda_3$, $\mathfrak J=\{J_+\}$ when $\lambda_3<\lambda_1$, and $\mathfrak J=\{J_-,J_+\}$ when the outer ratios coincide, with each interval defined at its finite-scale lattice position at scale $d$.

\begin{theorem}[Critical cost and necessary geometry]
\label{thm:critical}
Suppose $|A_{d,L}|\le Md$ for a fixed finite $M$ on the large-size tails and
\begin{equation}
 \lim_{d\to\infty}\liminf_{L\to\infty}F_{d,L}(A_{d,L})=1.
\label{eq:success-family}
\end{equation}
Then
\begin{equation}
\begin{split}
 \lim_{d\to\infty}\limsup_{L\to\infty}\frac{u_2}{d}&=0,\\
 \lim_{d\to\infty}\limsup_{L\to\infty}\frac{\min(u_1,u_3)}d&=0 .
\end{split}
\label{eq:necessary-filling}
\end{equation}
Consequently, for every nonnegative error sequence $\epsilon_d\to0$,
\begin{equation}
 \underline b_1[\epsilon_\bullet]=\overline b_1[\epsilon_\bullet]=b_0 .
\label{eq:sharp-price}
\end{equation}
If the successful family also satisfies
\begin{equation}
 \limsup_{d\to\infty}\limsup_{L\to\infty}\frac{|A_{d,L}|}{d}\le b_0,
\label{eq:near-minimal-budget}
\end{equation}
its near-minimal shape obeys
\begin{equation}
 \lim_{d\to\infty}\limsup_{L\to\infty}
 \frac{\min_{J\in\mathfrak J}|A_{d,L}\triangle J|}{d}=0.
\label{eq:necessary-shape}
\end{equation}
Here $\triangle$ denotes symmetric difference.
\end{theorem}

Theorem~\ref{thm:critical} gives necessary constraints on recovery supports at the level of macroscopic spatial density rather than strict microscopic connectedness, without ruling out sublinear perforations and external helper spins. In the symmetric case $\lambda_1=\lambda_3$, the target interval may alternate between $J_-$ and $J_+$ as either $d$ or $L$ varies. The argument proceeds by establishing the necessary filling conditions through collective environmental records before optimizing cardinality; matching this converse against the independent three-defect extractor of Sec.~\ref{sec:encoding} then confirms the sharp asymptotic price.

\subsection{From arbitrary omissions to two fixed containers}

To constrain an arbitrary candidate support $A$ subject to the linear budget $|A|\le Md$, we introduce an exterior test window $E_d=(p_4,p_4+\lceil Kd\rceil]\cap\mathbb Z$ with $K=M+1$. By cardinality, at least $d$ sites within $E_d$ belong to the unobserved environment $\bar A$. On a sufficiently large ring, the two complementary configurations---the central--exterior pairing $G_2\times E_d$ and the outer--outer pairing $G_1\times G_3$---are supported entirely within the fundamental domain and do not traverse the boundary cut, regardless of whether $A$ itself wraps around the ring. Given residual unobserved subsets $U$ and $V$ within these respective regions, we construct the collective observable
\begin{equation}
 O_{U,V}=\left(\sum_{j\in U}Z_j\right)
 \left(\sum_{k\in V}Z_k\right).
\label{eq:collective-observable}
\end{equation}
Supported entirely on $\bar A$, $O_{U,V}$ acts as an environmental witness. By virtue of the microscopic identities in Eq.~\eqref{eq:two-dictionaries}, $O_{U,V}$ exhibits conditional expectations $\pm\mu$ on the logical basis states and a common raw second moment $\nu$ on both states. At self-duality, an active-Majorana signed cyclic translation maps disorder strings and their arbitrary products directly onto reference order-spin correlators. Consequently, the same ferromagnetic correlation bounds govern both logical axes, including terms with coincident indices.

Let $T_{r,n}=\langle\widetilde Z_x\widetilde Z_{x+r}\rangle_{\Omega_{n,1}}$ denote the reference two-point correlation function on an active chain of $n$ spins. For a contiguous reference interval of $w$ sites, we define the container sum
\begin{equation}
 S_n(w)=w+2\sum_{r=1}^{w-1}(w-r)T_{r,n} .
\label{eq:container-sum}
\end{equation}
Because all reference spin correlations are nonnegative, the double sum over an arbitrary perforated subset of omissions is bounded above by the sum over its enclosing contiguous container. Lebowitz's zero-field four-point inequality~\cite{Lebowitz1974GHS} then yields
\begin{equation}
\begin{split}
 \mu&=\sum_{j\in U,k\in V}T_{k-j-1,n},\\
 \nu&\le S_n(w_L)S_n(w_R)+2\mu^2,
\end{split}
\label{eq:perforated-moments}
\end{equation}
where $w_L$ and $w_R$ denote the lengths of the enclosing reference intervals, and each of the two cross pairings contributes $\mu^2$. As shown in Appendix~\ref{app:critical}, applying this inequality to the unprojected ferromagnetic Gibbs state via Trotterization and taking the zero-temperature limit projects directly onto the unique ground state $\Omega_{n,1}$ without requiring an auxiliary parity projection.

On the infinite chain, the critical correlation kernel is strictly positive, decreases monotonically with distance, and exhibits the asymptotic power law $T_r\sim Cr^{-1/4}$~\cite{Pfeuty1970}. Summing this kernel over a macroscopic interval yields
\begin{equation}
 S(w):=\lim_{n\to\infty}S_n(w)\sim k_0Cw^{7/4},
 \qquad k_0=\frac{32}{21} ,
\label{eq:critical-sum}
\end{equation}
where the prefactor evaluates from $2\int_0^1(1-x)x^{-1/4}\,dx=32/21$. While critical order-parameter statistics are well known in statistical mechanics~\cite{LamacraftFendley2008OrderStatistics}, here they determine the information leakage accessible to the environment across arbitrary recovery geometries.

When the residual sets contain macroscopic omission densities, $|U|\ge ad$ and $|V|\ge cd$ for fixed $a,c>0$, the collective signal satisfies $\mu\ge acd^2m_{d,n}$, where $m_{d,n}$ is the infimum of $T_{r,n}$ across the cross-distance band. Inserting the second-moment bound of Eq.~\eqref{eq:perforated-moments} into the complementary-channel bound of Lemma~\ref{lem:moment} gives
\begin{equation}
\begin{split}
 F_{d,L}(A)&\le\mathcal B\!\left(
 \frac{a^2c^2d^4m_{d,n}^2}{S_n(w_L)S_n(w_R)}\right),\\
 \mathcal B(x)&=\frac{1+\sqrt{(1+x)/(1+2x)}}2 .
\end{split}
\label{eq:pattern-independent}
\end{equation}
Here the coefficients $a$ and $c$ represent omission densities normalized by the defect scale $d$, rather than fractions of container lengths. Taking the thermodynamic limit $L\to\infty$ followed by the large-scale limit $d\to\infty$, with container ratios $w_L/d\to a_L>0$, $w_R/d\to a_R>0$ and maximal normalized separation $R_d/d\to R>0$, the argument of $\mathcal B$ converges to
\begin{equation}
 \frac{a^2c^2}{k_0^2(a_La_R)^{7/4}\sqrt R}>0.
\label{eq:density-constant}
\end{equation}
The nonuniversal amplitude $C$ cancels completely from this ratio. Because both the collective expectation value and the root second moment scale as $d^{7/4}$, macroscopic omissions preserve a nonvanishing environmental record of the logical state.

Crucially, this thermodynamic bound holds uniformly over all candidate supports. At fixed $d$, the fixed test containers define a finite set of correlation distances that remains invariant under the choice of $A$. All corresponding correlator entries converge simultaneously as $L\to\infty$ before the containers are expanded with $d$, ensuring rigorous control over candidate supports that shift with $L$ or possess arbitrary internal fragmentation.

\subsection{Filling, cost and concentration}

The macroscopic filling conditions of Eq.~\eqref{eq:necessary-filling} follow directly from these collective records. If the central omission density $u_2/d$ fails to vanish along an asymptotic sequence of successful supports, pairing these omitted spins with the extensive exterior omissions in $E_d$ produces an argument bounded away from zero in Eq.~\eqref{eq:pattern-independent}. Equations~\eqref{eq:pattern-independent} and \eqref{eq:density-constant} then force the recovery fidelity strictly below unity along a subsequence, violating the premise of faithful recovery in Eq.~\eqref{eq:success-family}. Symmetrically, persistent omission densities across both outer gaps leave an unsuppressed environmental record of $X_f$, yielding an identical contradiction. Asymptotically faithful recovery under a linear budget therefore necessitates vanishing omission densities in the central gap and in at least one outer gap.

This filling requirement immediately fixes the asymptotic spatial cost. Suppose a candidate support is restricted to a sub-minimal budget $|A|\le\lfloor(b_0-\delta)d\rfloor$ for a fixed deficit $0<\delta<b_0$. For all sufficiently large $d$, such a support must either omit at least $\delta d/2$ spins in the central gap $G_2$ or, by total cardinality conservation, leave at least $\delta d/4$ omissions in both outer gaps $G_1$ and $G_3$. Applying the uniform bound of Eq.~\eqref{eq:pattern-independent} to both alternatives produces a geometry-independent fidelity gap $\gamma_\delta>0$ such that
\begin{equation}
 \limsup_{d\to\infty}\limsup_{L\to\infty}
 \sup_{|A|\le\lfloor(b_0-\delta)d\rfloor}F_{d,L}(A)
 \le1-\gamma_\delta .
\label{eq:uniform-price-gap}
\end{equation}
Any error sequence vanishing as $\epsilon_d\to0$ must eventually fall below $\gamma_\delta/2$, ruling out all sub-minimal budgets on the corresponding ordered tails. Because the local three-defect extractor of Sec.~\ref{sec:encoding} achieves exact recovery within $b_0 d + o(d)$ sites, the asymptotic price envelope is sharply determined by Eq.~\eqref{eq:sharp-price} without imposing any structural ansatz on the optimal support.

When the resource allowance is restricted to a near-minimal budget [Eq.~\eqref{eq:near-minimal-budget}], necessary filling severely constrains the spatial layout of the support. Asymptotically filling the central gap requires $\lambda_2 d - o(d)$ spins, leaving at most $m d + o(d)$ sites for the remainder of the system. Whenever $\lambda_1\ne\lambda_3$, accommodating the longer outer gap is impossible, compelling the support to concentrate its density on the shorter three-defect interval. Omissions within this designated interval and extraneous helper spins outside it must each scale as $o(d)$, establishing the concentration condition of Eq.~\eqref{eq:necessary-shape}. If the outer gap ratios coincide ($\lambda_1=\lambda_3$), either three-defect span is admissible, and the concentrating branch may alternate as either $d$ or $L$ varies without violating Eq.~\eqref{eq:necessary-shape}. In contrast to holographic and tensor-network codes engineered to permit disconnected reconstruction~\cite{PastawskiPreskill2017CodePrice,KimKastoryano2017MERAQEC}, near-minimal successful supports here must concentrate on a three-defect span in density, as required by the underlying operator dictionary. Appendix~\ref{app:cost} provides the detailed subsequence and cardinality proofs.

\subsection{Fixed error and a common-accuracy comparison}

Holding the target error at a fixed value $\epsilon>0$ defines the constant sequence $[\epsilon]$ with $\epsilon_d=\epsilon$ in the asymptotic envelopes of Eq.~\eqref{eq:cost-envelopes}. Rather than bounding the raw second moment of the unobserved witness by the full container width, one sharpens the converse by bounding the internal correlation sums directly with the number of omitted spins via the count-adapted moment envelope. In the collective observable, the squared signal divided by the product of internal correlation sums exhibits a fourth-root dependence on the product of omission densities. Inverting the fidelity ceiling $\mathcal B(q)=1-\epsilon$ along its small-error branch specifies the error parameter
\begin{equation}
 q(\epsilon)=\frac{4\epsilon(1-\epsilon)}{1-8\epsilon+8\epsilon^2},\qquad
 0\le\epsilon<\frac{1-1/\sqrt2}{2}.
\label{eq:error-function}
\end{equation}
Appendix~\ref{app:count-moments} and Appendix~\ref{app:cost} establish the underlying finite-band construction, uniformity across nonconvergent support families, and this algebraic inversion.

The exact three-defect extractor puts every minimum-cardinality fixed-error support under $|A|\le Md$ for every fixed $M>b_0$ along the large-size tails. Let $\Lambda=\lambda_1+\lambda_2+\lambda_3$ and $\lambda_{\max}=\max(\lambda_1,\lambda_3)$. Balancing the exterior test window against the cross-distance band and evaluating the infimum $M\downarrow b_0$ yields the lower price envelope
\begin{equation}
\begin{split}
 \underline b_1[\epsilon]&\ge b_0-C_\lambda q(\epsilon)^4,\\
 C_\lambda&=k_0^8\left[4(b_0+\lambda_2+\lambda_3)
 +\frac{\Lambda^2}{\lambda_{\max}}\right].
\end{split}
\label{eq:finite-error-cost}
\end{equation}
Expanding $q(\epsilon)=4\epsilon+O(\epsilon^2)$ reveals that small fixed errors can reduce the leading cost coefficient below $b_0$ by at most $O(\epsilon^4)$. No support-and-decoder construction attaining a leading-order cost saving of this magnitude is given here.

The underlying count estimates also constrain shape at the exact-extractor budget. For each fixed positive ratio triple $\lambda=(\lambda_1,\lambda_2,\lambda_3)$, there are $\epsilon_0(\lambda)>0$ and $C^{\rm sh}_\lambda<\infty$ such that a family with fixed $0<\epsilon<\epsilon_0(\lambda)$, $\liminf_{d\to\infty}\liminf_{L\to\infty}F_{d,L}(A_{d,L})\ge1-\epsilon$, and $|A_{d,L}|\le s_d+1$ on the ordered large-size tails satisfies
\begin{equation}
\limsup_{d\to\infty}\limsup_{L\to\infty}
\frac{\min_{J\in\mathfrak J}|A_{d,L}\triangle J|}{d}
\le C^{\rm sh}_\lambda\epsilon^4.
\label{eq:fixed-error-shape}
\end{equation}
Here $s_d=\ell_2+\min(\ell_1,\ell_3)$; the exact extractor puts minimum-cardinality successful supports within the budget $s_d+1$. The comparison set $\mathfrak J$ depends on the limiting outer ratios, with a minimizing branch that may switch with $d,L$ when they tie. The shape prefactor $C^{\rm sh}_\lambda$ is distinct from the cost coefficient $C_\lambda$. Appendix~\ref{app:cost} constructs the geometry-dependent error threshold and the strict branch margin for unequal outer ratios. The bound is necessary localization, not a decoding criterion for arbitrary perforations.

In the benchmark $3{:}2{:}1$ geometry, where $b_0=3$ and $C_\lambda=36(32/21)^8$, an error tolerance of $\epsilon=0.01$ gives $q=99/2302$, constraining the asymptotic cost envelope to
\begin{equation}
 2.996420\le\underline b_1[0.01]
 \le\overline b_1[0.01]\le3 .
\label{eq:fixed-target-example}
\end{equation}
At this identical target accuracy ($\epsilon=0.01$), the gapped encodings at $h=1/2$ and $h=2$ exhibit leading costs of exactly $2$ and $1$, respectively: isolated spin-pair records enforce a fidelity ceiling of approximately $0.6830<0.99$ via Eq.~\eqref{eq:gapped-fidelity-ceiling}, yielding the converse, while transported decoders achieve the corresponding integer coefficients. This comparison isolates the critical price while holding defect geometry and target fidelity fixed.

High precision at criticality thus leaves negligible room for spatial savings under small fixed errors. This rigidity motivates the operational inquiry pursued in Section~\ref{sec:quantum-access}: below the vanishing-error budget $b_0 d$, can a disconnected physical geometry still preserve quantum information?

\section{Fragmented access and single-spin recovery}
\label{sec:quantum-access}

Preserving reference entanglement requires fewer physical spins than vanishing-error recovery in this encoding. In the benchmark defect geometry of the critical Ising chain ($h=1$),
\begin{equation}
 (p_1,p_2,p_3,p_4)=(2,2+3d,2+5d,2+7d),\qquad
 L\ge\max(9d,7d+3),
\label{eq:finite-geometry-regime}
\end{equation}
vanishing-error recovery requires $4d+o(d)$ spins, and Eq.~\eqref{eq:finite-error-cost}, evaluated at these gap ratios in Appendix~\ref{app:cost}, bounds the leading cost coefficient at fidelity $0.99$ from below by $3.99519369$. Every connected interval of at most $4d$ spins has an entanglement-breaking access channel, whereas for each fixed $2<b<4$, a specified two-block support of $bd+O(1)$ spins admits a deterministic single-spin protocol with a scale-stable Bell-fidelity advantage.

\subsection{Connected regions and a retained Pauli flag}

Let $F_{\rm conn}^*(B)$ denote the supremum of Bell recovery fidelity over all connected cyclic subsets of $B$ physical spins. The defect count enclosed within an interval determines which physical Pauli strings are retained by the receiver and which fall into the environment, yielding the exact three-tier staircase
\begin{equation}
 F_{\rm conn}^*(B)=\begin{cases}
 1/4,&1\le B\le2d,\\
 1/2,&2d+1\le B\le4d,\\
 1,&4d+1\le B\le L .
 \end{cases}
\label{eq:connected-budget-staircase}
\end{equation}
Each tier corresponds to a distinct algebraic structure of the accessible subsystem and its complement:
\begin{enumerate}
 \item For $1\le B\le2d$, any connected interval contains at most a single defect core. Its unobserved environment $\bar A$ encloses at least three defects and consequently supports two anticommuting physical Pauli operators; tracing over $\bar A$ twirls the logical state completely, reducing the accessible channel to a constant map of fidelity $1/4$.
 \item For $2d+1\le B\le4d$, the interval encompasses at most two defects, ensuring that $\bar A$ encloses at least two defects. The environment therefore supports an exact physical representation of a logical Pauli operator. Tracing out $\bar A$ eliminates all off-diagonal code matrix units in the corresponding logical eigenbasis, rendering the physical access channel $\mathcal E_A$ measure-and-prepare and hence entanglement breaking~\cite{HorodeckiShorRuskai2003EB}. Bell fidelity is consequently bounded by $1/2$. A contiguous block holding two defects (such as $[p_3,p_4]$) retains the same logical Pauli, whose local projective readout and conditional state preparation saturates this $1/2$ ceiling.
 \item For $B\ge4d+1$, one may choose a segment containing $J_+$ and hence at least three consecutive defect cores, retaining both logical generators and permitting exact unitary extraction with fidelity $1$.
\end{enumerate}
Appendix~\ref{app:finite-recovery} verifies Eq.~\eqref{eq:connected-budget-staircase} across all cyclic intervals, including arcs traversing the Jordan--Wigner boundary seam.

When an accessible support $A$ retains an unbroken physical representative of a single logical Pauli operator, this observable serves as a logical flag; exceeding the entanglement-breaking ceiling also requires nonzero conjugate coherence. Because the eigenstates of this flag isolate the encoded basis states into mutually orthogonal subspaces of $\mathcal H_A$, a decoder can reconstruct diagonal logical populations without error. The remaining recovery task reduces to transferring the off-diagonal coherence operator $C_A=\operatorname{Tr}_{\bar A}|\Phi_+\rangle\langle\Phi_-|$ between the two flag sectors. Aligning the singular vectors of $C_A$ via polar decomposition attains the optimal Bell recovery fidelity for the designated support,
\begin{equation}
 F_{d,L}(A)=\frac{1+\|C_A\|_1}{2} .
\label{eq:flag-optimum}
\end{equation}
While parity-resolved polar decoders have been formulated for average fidelity in uncoupled Majorana wires under parity-preserving noise~\cite{IppolitiRizziGiovannettiMazza2016MajoranaMemory}, Eq.~\eqref{eq:flag-optimum} optimizes squared Bell entanglement fidelity directly on the physical-spin tensor product.

For single-gap perforations $A=J_+\setminus H$, where $H\ne\emptyset$ lies entirely in $G_2$ or $G_3$, the unperforated gap preserves an exact logical flag, while the retained defect endpoints $(p_2,p_3)$ or $(p_3,p_4)$ of the perforated gap support the conjugate observable $O_X$ or $O_Z$ of Eq.~\eqref{eq:conjugate-endpoints}. Under the critical encoding isometry $V=V_{1,d,L}$, this boundary observable compresses on the code space to $V^\dagger O_XV=T_{2d-1,L-2}X$ or $V^\dagger O_ZV=T_{2d-1,L-2}Z$, respectively, providing a nonzero compressed response rather than exact Pauli intertwining. Trace-norm duality against this boundary operator yields $\|C_A\|_1\ge T_{2d-1,L-2}>0$, bypassing the connected entanglement-breaking ceiling; together with environmental leakage from unobserved spins that enforces $F_{d,L}(A)<1$, this establishes $1/2<F_{d,L}(A)<1$ throughout $2d+2\le|A|\le4d$ (Appendix~\ref{app:finite-recovery}). Although this endpoint bound certifies a margin that decays as $d\to\infty$, it does not bound the optimal margin from above. To obtain a scale-stable guarantee across a macroscopic hole, we now use collective receiver records.

\subsection{A macroscopic two-block guarantee}

For the geometry in Fig.~\ref{fig:two-block-protocol} and any chosen budget ratio $2<b<4$ and sufficiently large $d$ so that $r_d\ge1$, define the discrete two-component support
\begin{equation}
\begin{split}
 r_d&=\left\lfloor\frac{(b-2)d}{2}\right\rfloor,\\
 A_L&=[p_2,p_2+r_d-1],\\
 A_R&=[p_3-r_d+1,p_4],\qquad A_d=A_L\cup A_R ,
\end{split}
\label{eq:macroscopic-hole-family}
\end{equation}
where each interval contains its integer lattice sites. At scale $d$, the combined allocation $|A_d|=2d+2r_d=bd+O(1)$ leaves an unobserved macroscopic gap of $(4-b)d+O(1)$ sites inside $G_2$. The right block $A_R$ contains the intact interval $[p_3,p_4]$, preserving an exact logical $Z_f$ flag. Flanking segments across the macroscopic void assemble a collective conjugate $X_f$ record.

\begin{theorem}[Two-block recovery by single-spin operations]
\label{thm:one-way}
For the critical encoding and supports in Eq.~\eqref{eq:macroscopic-hole-family}, a deterministic protocol using single-spin Pauli measurements, one-way classical communication from $A_L$ to $A_R$, and a one-spin output correction attains
\begin{equation}
\begin{split}
 \liminf_{d\to\infty}\liminf_{L\to\infty}F_{\rm meas}(A_d)
 &\ge\frac{1+c_b}{2},\\
 c_b&=\frac{x_b}{1+2x_b},\qquad
 x_b=\frac{\sqrt{b-2}}{2k_0^2}>0 .
\end{split}
\label{eq:constant-advantage}
\end{equation}
Here $F_{\rm meas}$ is the Bell fidelity of the specified protocol, not the optimum over all one-way protocols. The constant $k_0=32/21$ is defined in Eq.~\eqref{eq:critical-sum}.
\end{theorem}
\protocolfigure

Evaluating these expressions at budget parameter $b=3$ yields $x_b=441/2048$ and $c_b=441/2930$, establishing an analytical fidelity floor of $(1+c_b)/2\simeq0.575256$. This guarantee certifies that the constructive single-spin protocol on fragmented supports maintains reference entanglement at spatial allocations where every connected region remains entanglement breaking.

The strict margin $0.575256>0.55$ separates two distinct accuracy targets within this encoding. A support of $3d+O(1)$ spins eventually guarantees fidelity $0.55$, whereas every vanishing-error sequence requires a leading footprint of $4d$. This establishes a sufficient fixed-target footprint rather than an optimized price; Appendix~\ref{app:fixed-target-recovery} provides the ordered-envelope argument.

\subsection{Physical measurements and all-outcome decoding}

With $r=r_d$, take $j\in[p_2,p_2+r-1]$ and $k\in[p_3-r+1,p_3]$. In the physical operator $O_{jk}=X_f\prod_{j<m<k}X_m$, the intervening $X$ factors cancel those of the full $X_f$ string in Eq.~\eqref{eq:physical-logical-string}, leaving two tails on the retained flanks. Their sum factorizes as
\begin{equation}
\begin{aligned}
 O_A&=L_A\otimes(K_RZ_{p_3}),\\
 L_A&=-Y_{p_2}\sum_{j=p_2}^{p_2+r-1}\prod_{p_2<m\le j}X_m,\\
 K_R&=\sum_{k=p_3-r+1}^{p_3}\prod_{k\le m<p_3}X_m .
\end{aligned}
\label{eq:one-way-product-record}
\end{equation}
Under the fixed-sector dictionary, $O_{jk}$ carries $X_f$ times the reference disorder string $\prod_{m=j}^{k-2}\widetilde X_m$. Critical self-duality gives the compression $V^\dagger O_{jk}V=T_{k-j-1,n}X$, with $V=V_{1,d,L}$ and $n=L-2$. Thus $O_A$ has conditional means $\pm\mu$ and common raw second moment $\nu$ on logical $X_f$ eigenstates (Appendix~\ref{app:receiver-record}). These are receiver moments, not environmental records. The operator is a moment witness, not an additional runtime readout: the protocol forms $(s,t,w)$ from helper measurements, never measures the erased hole or $Z_{p_3}$, and keeps spin $p_3$ as the quantum output.

The recovery protocol proceeds through local physical operations:
\begin{enumerate}
 \item Alice measures $Y_{p_2}$ alongside $X$ on each of the remaining $r-1$ sites of $A_L$. Because these single-spin Pauli observables commute, their products directly evaluate each summand of $L_A$, yielding its integer eigenvalue $s\in\{-r,-r+2,\ldots,r\}$. Alice transmits $s$ to Bob using $\lceil\log_2(r+1)\rceil$ classical bits.
 \item Bob measures $X$ on $[p_3-r+1,p_3-1]$ and $G_3$, and measures $Z$ on defect site $p_4$, reserving site $p_3$ as the single-qubit quantum output. His readouts determine the eigenvalue $t$ of $K_R$ and the collective parity $w=(\prod_{m\in G_3}x_m)z_{p_4}$, where $x_m,z_{p_4}\in\{\pm1\}$.
 \item Conditioned on Bob's outcomes, the exact logical flag on $[p_3,p_4]$ reduces to $-wY_{p_3}$. Bob applies the local Clifford unitary $U_w$ satisfying
\begin{equation}
 U_w(-wY)U_w^\dagger=Z,\qquad U_wZU_w^\dagger=X
\label{eq:output-axes}
\end{equation}
 to rotate this conditioned flag into the computational $Z$ basis of the output spin (see Appendix~\ref{app:one-way-recovery} for its explicit eigenbasis).
 \item For each coarse branch $\lambda=(s,t,w)$, let $a_\lambda\in\mathbb R$ denote the unnormalized real coherence transfer coefficient with which this branch, after $U_w$, maps logical $|0\rangle\langle1|$ to output $|0\rangle\langle1|$. Bob applies $Z$ when $\varepsilon_\lambda=-1$ and the identity otherwise, using the fixed offline sign table $\varepsilon_\lambda=\operatorname{sgn}a_\lambda$ with convention $+1$ at $a_\lambda=0$. This ideal table defines the deterministic protocol in Theorem~\ref{thm:one-way}; finite-sample estimates are treated separately below.
\end{enumerate}

Write $V=V_{1,d,L}$ for the critical encoding isometry. Each sequence of fine readouts defines a coarse branch $\lambda=(s,t,w)$ represented by projector $\Pi_\lambda$. Because every measured observable commutes with $X_f$ and the retained physical representative of $Z_f$ on $[p_3,p_4]$, the code-space projection is scalar: $V^\dagger\Pi_\lambda V=p_\lambda I$. The branch probabilities $p_\lambda$ are therefore strictly independent of the encoded logical state. After rotation by $U_w$, branch $\lambda$ maps logical populations as $|a\rangle\langle a|\mapsto p_\lambda|a\rangle\langle a|$ and coherence as $|0\rangle\langle1|\mapsto a_\lambda|0\rangle\langle1|$.

The transfer coefficients $a_\lambda$ are constrained to be real by the antiunitary symmetry $\Theta=(\prod_{a=1}^4 Z_{p_a})\mathsf K$, where $\mathsf K$ denotes complex conjugation in the physical computational basis. Because $\Theta$ commutes with the Hamiltonian, sector projectors, the logical generators $X_f$ and $Z_f$, readout projectors, and output operator $Z_{p_3}$, adopting a $\Theta$-invariant logical basis guarantees $a_\lambda\in\mathbb R$. This symmetry ensures that the real sign table in step 4 suffices to align the branch coherences.

Summing over all corrected branches yields a deterministic, trace-preserving symmetric dephasing channel whose transferred coherence $c_{\rm meas}$, normalization, and record moments satisfy
\begin{equation}
\begin{aligned}
 c_{\rm meas}&=\sum_\lambda|a_\lambda|,&\sum_\lambda p_\lambda&=1,\\
 \mu&=\sum_\lambda st\,a_\lambda,&
 \nu&=\sum_\lambda s^2t^2p_\lambda .
\end{aligned}
\label{eq:one-way-branch-moments}
\end{equation}
Complete positivity of each branch map implies $|a_\lambda|\le p_\lambda$. Applying Cauchy--Schwarz directly certifies
\begin{equation}
 \mu^2\le\left(\sum_\lambda|a_\lambda|\right)
 \left(\sum_\lambda s^2t^2|a_\lambda|\right)
 \le c_{\rm meas}\nu,
\label{eq:one-way-moment-certificate}
\end{equation}
which establishes $F_{\rm meas}=(1+c_{\rm meas})/2\ge1/2+\mu^2/(2\nu)$. Because tracing out Alice's subsystem eliminates off-diagonal cross-terms between orthogonal readout projectors, coarse-graining by $s$ preserves these moments without postselection.

Evaluating the collective record through the physical dictionary yields $\mu\ge r^2m_{d,n}$ and raw second moment $\nu\le S_n(r)^2+2\mu^2$. Applying the critical container estimates of Sec.~\ref{sec:critical} in the ordered asymptotic limit ($L\to\infty$ followed by $d\to\infty$) gives an ordered lower limit of at least $c_b$ for $\mu^2/\nu$. Appendix~\ref{app:receiver-record} supplies the receiver moment estimates, and Appendix~\ref{app:one-way-recovery} proves that the stated physical instrument realizes the branch identities used here. Together they establish Theorem~\ref{thm:one-way}. The exact flag converts the statistical correlation of the conjugate record into usable quantum coherence. Entanglement localization by helper measurements~\cite{VerstraetePoppCirac2004LocalizableEntanglement,PoppEtAl2005LocalizableEntanglement} and one-way extraction in stabilizer codes and specified stabilizer helper schemes~\cite{ShiraishiYamasakiMurao2024LOCCExtraction,ChitambarEtAl2026HelperCodes} are established precedents. The helper schemes may use joint Pauli measurements within a share. Here the bound is realized by individual physical-spin measurements after the hole is erased, with every helper included in the spatial budget.

On the connected intervals $J_+$ and $J_-$, single-spin measurements and classical feedforward likewise achieve exact recovery with unit fidelity (Appendix~\ref{app:one-way-recovery}). At criticality, restricting decoder operations to single-spin measurements and feedforward therefore achieves the sharp leading high-precision cost $b_0 d$.

Table~\ref{tab:protocol} presents finite-ring calculations at $d=2,L=20$ for the six-spin support $A=\{8\}\cup[12,16]$, corresponding to $b=3,r=1$. Evaluating all four logical matrix units across every outcome branch confirms the operational hierarchy: Bob alone is confined to the entanglement-breaking ceiling $0.500000$; Alice's single-bit message allows the single-spin protocol to attain $0.747468$; conditional polar decoding yields $0.782024$; and unrestricted joint decoding reaches the fixed-support optimum $0.783987$.
\begin{table}[htbp]
\centering
\caption{Bell fidelity on $A=\{8\}\cup[12,16]$ at $d=2,L=20$. Conditional polar recovery preserves Bob's whole quantum block after Alice's readout; joint polar recovery optimizes the same support without that restriction. The printed values are rounded finite-state computations.}
\label{tab:protocol}
\begin{tabular}{lr}
\toprule
Recovery model & Fidelity\\
\midrule
Bob alone & $0.500000$\\
Single-spin measurements and correction & $0.747468$\\
Conditional polar & $0.782024$\\
Joint optimum & $0.783987$\\
\bottomrule
\end{tabular}
\end{table}

\subsection{Resources and a nonzero transmission rate}

The branch correction signs can be calibrated offline prior to unknown-input transmission. From $N$ independent copies of the known logical state $|+\rangle$, each unnormalized coefficient $a_\lambda$ is estimated by averaging the branch indicator weighted by the output $X$ readout following $U_w$. Because this measures the unnormalized response directly, it avoids division by small branch probabilities. For $0<\delta<1$ and at most $M_{\rm br}\le2(r+1)^2$ distinct branch labels, the bounded-differences inequality~\cite{McDiarmid1989BoundedDifferences,McDiarmid1998Concentration} guarantees that with confidence at least $1-\delta$,
\begin{equation}
 \|\widehat a-a\|_1\le
 \frac{\sqrt{M_{\rm br}}+\sqrt{2\ln(1/\delta)}}{\sqrt N} .
\label{eq:calibration-bound}
\end{equation}
Assigning correction signs according to $\widehat a_\lambda$, with sign $+1$ at a zero estimate, degrades the achieved Bell fidelity by at most $\|\widehat a-a\|_1$. At fixed target precision and confidence, this calibration uses $O(d^2)$ known-state copies and $O(d^3)$ helper-spin readouts (Appendix~\ref{supp:finite-calibration}).

\begin{table}[htbp]
\centering
\caption{Resources of the two-block protocol. Calibration bounds hold at fixed accuracy and confidence. The last row assumes ideal projections, independent classical report flips of probability $p_{\rm flip}<1/2$, and repeated quantum nondemolition (QND) readout of the same eigenvalue. Transmission and output gates are ideal in that model; $\eta$ is the target half-diamond channel error.}
\label{tab:resources}
\begin{tabular}{l>{\raggedright\arraybackslash}p{0.43\columnwidth}}
\toprule
Resource & Sufficient cost\\
\midrule
Retained spins & $bd+O(1)$, including all helpers\\
Alice-to-Bob message & $\lceil\log_2(r+1)\rceil$ bits per use\\
Calibration & $O(d^2)$ known-state copies; $O(d^3)$ helper readouts\\
Repeated reports & $O\!\left(d\log(d/\eta)/(1-2p_{\rm flip})^2\right)$ readouts\\
\bottomrule
\end{tabular}
\end{table}

Table~\ref{tab:resources} compiles the operational resources required by the protocol. In the presence of imperfect measurement reports—modeled by ideal projections followed by independent classical bit flips of probability $p_{\rm flip}<1/2$—repeated quantum nondemolition readouts of the same underlying eigenvalue followed by local majority voting exponentially suppress report errors. Coupling the noisy and ideal trajectories makes their output states agree whenever every majority vote succeeds, including when processing reference-entangled inputs; choosing the repetition count so that the probability of any wrong majority vote is at most $\eta$ then bounds the half-diamond norm distance between the averaged decoded channels by $\eta$. Alice executes majority voting locally before transmitting $s$, requiring no repeated interblock communication. Appendix~\ref{supp:report-errors} derives this bound and analyzes the associated $O(\log(d/\eta)/(1-2p_{\rm flip})^2)$ parallel measurement depth under ideal classical transmission and logic gates.

Channel coding over independently prepared encoding blocks at fixed geometry—evaluated before taking the thermodynamic ($L\to\infty$) and defect scaling ($d\to\infty$) limits—yields an unassisted quantum transmission rate. Under ideal sign calibration, the protocol generates a symmetric dephasing channel achieving rate
\begin{equation}
 R_{\rm meas}(A_d)=1-H_2\!\left(\frac{1-c_{\rm meas}}2\right)
\label{eq:measurement-rate}
\end{equation}
qubits per block, expressed in terms of the binary entropy function $H_2(x)=-x\log_2x-(1-x)\log_2(1-x)$. Because this single-spin protocol constitutes a completely positive, trace-preserving decoding of the physical spins, the unassisted quantum capacity $Q(\mathcal N_A)$ of the full access channel $\mathcal N_A=\mathcal E_{A,1,d,L}$ satisfies $Q(\mathcal N_A)\ge R_{\rm meas}(A_d)$ for $A=A_d$. In the ordered macroscopic limit ($L\to\infty$ followed by $d\to\infty$), the nonvanishing coherence parameter $c_b$ certifies strictly positive asymptotic rates,
\begin{equation}
\begin{split}
 \liminf_{d\to\infty}\liminf_{L\to\infty}R_{\rm meas}(A_d)
 &\ge1-H_2\!\left(\frac{1-c_b}2\right)>0,\\
 \liminf_{d\to\infty}\liminf_{L\to\infty}Q(\mathcal N_{A_d})
 &\ge1-H_2\!\left(\frac{1-c_b}2\right)>0 .
\end{split}
\label{eq:constant-capacity}
\end{equation}
At $b=3$, this bound certifies an asymptotic floor of approximately $0.016404$ logical qubits per independent encoding block, rather than per host physical spin. By contrast, on the same benchmark geometry, every connected interval of at most $4d$ spins induces an entanglement-breaking access channel and therefore has zero unassisted quantum capacity. The fragmented support $A_d$ instead guarantees the strictly positive rate certified in Eq.~\eqref{eq:constant-capacity}.

When branch signs are assigned from empirical estimates $\widehat a_\lambda$, adopting the tie-breaking convention $\operatorname{sgn}(0)=+1$, the realized channel coherence is $c_{\rm est}=\sum_\lambda\operatorname{sgn}(\widehat a_\lambda)a_\lambda$. Relative to the ideal benchmark $c_{\rm meas}$, this coherence satisfies the deterministic loss bound $c_{\rm est}\ge c_{\rm meas}-2\|\widehat a-a\|_1$ for every estimated table (Appendix~\ref{supp:finite-calibration}). At each geometry, whenever calibration draws enough samples to make the error bound in Eq.~\eqref{eq:calibration-bound} at most $\tau_0$, the resulting confidence event certifies $c_{\rm est}\ge c_{\rm meas}-2\tau_0$. For fixed $2<b<4$, choosing the tolerance $0<\tau_0<c_b/2$ establishes rate certificates with a strictly positive ordered lower limit. This confidence guarantee applies per geometry rather than simultaneously for the random family across all scales; Appendix~\ref{supp:finite-calibration} provides the joint-family conditions and details the distinction between certified bounds and actual channel performance.

The full flagged access channel belongs to the degradable Hadamard class: measuring the flag operator on $A$ simulates the complementary channel $\mathcal N_{\bar A}$ via a classical-quantum map~\cite{BradlerHaydenTouchetteWilde2010Hadamard}. Spatial factorization of the physical Pauli operators endows the channel with covariance, ensuring via concavity of coherent information that the maximally mixed input achieves the single-letter capacity~\cite{DevetakShor2005CQ,YardHaydenDevetak2008MAC}. While polar decoding maximizes single-block Bell fidelity on a fixed support, its one-qubit output channel $\mathcal D_c$, with $c=\|C_A\|_1$, can discard residual coherent information, so that $Q(\mathcal D_c)\le Q(\mathcal N_A)$. Appendix~\ref{app:access-capacity} establishes these capacity properties; Table~\ref{tab:access-capacity} compares the full-access capacity $Q(\mathcal N_A)$ against the polar-decoded capacity $Q(\mathcal D_c)$ across benchmark geometries, illustrating the capacity loss of single-qubit polar compression.

\section{Discussion}
\label{sec:discussion}

The spatial cost of recovering a qubit depends on the requested accuracy as well as the correlations of its host. The matched Ising family separates these ingredients: varying the common pure, even-parity bulk reference changes the encoding while leaving the defect geometry, fusion algebra and physical spin partition fixed. At each fixed finite positive field, both complementary logical records have nonzero response. Imposing zero error at finite size, then taking the infinite-chain limit before increasing defect separation, gives the same leading cost across the transition. Approximate recovery can tolerate small responses, but attainability still requires a decoder. In the gapped phases, persistent order or disorder records give different lower bounds, and transported endpoint decoders attain them with sublinear padding. The phase dependence is therefore an operational property of the reference-dependent encoding, not a change in the fusion algebra.

At criticality, the decay of individual distant correlations does not remove collective environmental information. For positive-density central--exterior and outer--outer omission sets, the conditional mean magnitude of the product of two physical spin sums and the square root of its raw second moment both grow as $d^{7/4}$. Their ratio remains nonzero even for irregular omissions. The physical spin dictionary, ferromagnetic positivity and the four-point bound make this statement uniform before optimizing the support. This supplies a concrete form of the information--disturbance mechanism~\cite{KretschmannSchlingemannWerner2008InformationDisturbance,Arxiv0907_5391,BarnumKnill2002NearOptimalReversal}. High-precision recovery under a linear budget forces density filling, not exact microscopic connectedness: these necessary conditions do not exclude sublinear holes and auxiliary spins in near-minimal supports.

Within the receiver, bulk correlations serve a complementary function: an intact two-defect segment supplies an exact logical flag that preserves diagonal populations, while a conjugate collective record across the retained flanking segments sustains coherence across the erased hole. On the specified two-block family, the single-spin instrument translates this collective record into a scale-stable quantum advantage, with every helper spin charged to the retained budget and the erased hole left untouched. Criticality is not a generic prerequisite for measurement-assisted transmission~\cite{VerstraetePoppCirac2004LocalizableEntanglement,PoppEtAl2005LocalizableEntanglement}: Affleck--Kennedy--Lieb--Tasaki (AKLT) chains display a divergent localizable-entanglement length~\cite{VerstraeteMartinDelgadoCirac2004GappedEntanglement}, and symmetry-protected phases support measurement-based identity wires~\cite{ElseEtAl2012SPTQuantumWire}. In the present critical system, bulk correlations act through a nonvanishing collective signal-to-moment ratio that sustains coherence across a macroscopic discarded region, at spatial budgets where every connected interval is entanglement breaking.

Spatial access and decoder control constitute distinct operational resources. While the recovery cost theorems admit arbitrary joint decoders, the low-budget protocol relies strictly on single-spin Pauli measurements, one-way classical feedforward, and a local output correction. Runtime feedforward excludes known-state calibration; for each fixed $2<b<4$, sufficiently accurate finite calibration certifies a positive, possibly reduced margin on a per-geometry confidence event. Under otherwise ideal operations, repeated QND readouts and majority voting suppress independent classical report flips with $p_{\rm flip}<1/2$. Positive quantum rates follow by coding the actual corrected dephasing channel over independently prepared encoding blocks at fixed geometry, evaluated before the spatial limits. These transmission rates are normalized per block, and Bell-optimal one-qubit compression can discard coherent information present in the full access channel.

The price also depends on the logical subspace within an unchanged host. On two uncoupled critical chains $\sigma,\tau$ with product bulks, localized and repetition embeddings have vanishing-error costs $\lambda_2+m$ and $\lambda_2+2m$, respectively (Appendix~\ref{app:two-chain}). These are distinct code subspaces, not basis rotations. In the repetition subspace $X_\sigma X_\tau=+1$, either layer's fusion $X_c$ represents logical $X$, whereas logical $Z$ requires the physical product $Z_\sigma Z_\tau$; each single-layer $Z_c$ compresses to zero. Thus vanishing error forces an outer gap to be filled in density on each layer. The matching extractor uses a complete three-defect interval on one layer and an outer two-defect interval on the other, paying only one central spacing. The extra leading cost $m$ comes from the companion outer gap. This dictionary realizes embedding-dependent access structures studied in quantum secret sharing~\cite{GheorghiuLooiGriffiths2010InformationLocation,Gottesman2000SecretSharingTheory}. Fixed-prefix local indistinguishability in the single-defect problem~\cite{LiangIsing2026} instead keeps the observed set fixed as separation grows.

The matched Ashkin--Teller ladder uses the localized embedding: the qubit occupies the $\sigma$ fusion factor and the auxiliary $\tau$ fusion factor is fixed in a pure state. The two-layer bulk is prepared in its prescribed pure interacting reference ground state, even on each layer. For nonnegative coupling $g\ge0$, exact interval extraction and the finite-ring connected--perforated separation persist in the benchmark geometry (Appendix~\ref{app:interacting}). The macroscopic price at genuinely interacting critical couplings still requires uniform lattice two-point lower bounds and raw fourth-moment upper bounds. Continuum correlators guide scaling and contact estimates but do not supply the integrated lattice matching. The $g=0$ limit remains the exact Ising control.

The completed single-chain results provide an accuracy-resolved spatial access structure for a fusion qubit. Exact physical operator identities identify candidate decoders and records; at criticality, uniform collective estimates exclude vanishing-error recovery below the optimal leading cost, regardless of support geometry. Together with the explicit measurement instrument, they explain how the same bulk correlations can obstruct recovery in the complement yet support coherence within a fragmented receiver. This connection between an erasure algebra, its collective records and a constructive decoder is the central mechanism behind the cost and geometry results.

Two questions remain. For the critical Ising encoding, allowing a small fixed error can save at most $O(\epsilon^4)$ in the leading cost coefficient; whether any support and decoder attain a saving of this order is open. More generally, the optimal footprint at intermediate fidelity over all physical supports, particularly those without an exact flag, remains unknown. A second problem is a joint scaling limit $h(d)\to1$. Resolving that crossover requires uniform control of dressed-decoder tails and collective environmental records, which are governed by distinct scales at fixed field.

\newcommand{\availabilitystatement}{The archived physical-spin code states, finite-support results, and recovery and capacity calculations are available as data and code version 1.1.0~\cite{Liang2026SpatialRecoveryDataCode}. The fixed-error examples follow by substituting the stated gap ratios and target fidelity into the analytical coefficients of Appendix~\ref{app:cost}.}

\section*{Author contributions}\authorcontribution
\section*{Data and code availability}\availabilitystatement
\section*{Competing interests}The authors declare no competing interests.
\section*{Funding}This research received no external funding.
\appendix
\section*{Appendices}
Appendices~A--D establish the core encoding and recovery proofs, covering the operator-algebra tools, endpoint transport, uniform critical estimates, and the measurement instrument, with Appendix~C developing a separate conditional extension to other correlation exponents. Operational requirements are addressed in Appendices~E and~F, which analyze calibration, classical report errors, and channel capacity. Finally, Appendices~G and~H investigate extensions to two-chain embeddings and an interacting ladder, with Appendix~H separating exact finite-ring recovery from conditional macroscopic bounds.

\section{Physical encoding and exact interval extraction}
\label{app:encoding}

The periodic seam identifies how physical parity and total defect charge enter the encoding. At criticality we abbreviate $V_{d,L}=V_{1,d,L}$ and $\Omega_n=\Omega_{n,1}$. We derive the common-bulk factorization and the interval extractor used in Sec.~\ref{sec:encoding}.

\subsection{Active Majoranas and the periodic seam}

Applying the Jordan--Wigner transformation in Eq.~\eqref{eq:jw} maps the physical spin operators to Majorana fermions satisfying
\begin{equation}
\begin{aligned}
ic_{2j-1}c_{2j}&=X_j,&
ic_{2j}c_{2j+1}&=Z_jZ_{j+1},\\
Z_LZ_1&=-iPc_{2L}c_1,&
Z_{p-1}Y_p&=ic_{2p-2}c_{2p}.
\end{aligned}
\label{eq:jw-identities}
\end{equation}
Excluding the four defect operators $\gamma_a=c_{2p_a-1}$ leaves an ordered sequence of $2n$ active Majorana modes $\alpha_1,\ldots,\alpha_{2n}$, where $n=L-2$. Each defect core removes one odd-indexed Majorana from the active set and couples its adjacent active modes directly, resulting in the Hamiltonian
\begin{equation}
H_D=-i\sum_{r=1}^{2n-1}\alpha_r\alpha_{r+1}
+iP\alpha_{2n}\alpha_1.
\label{eq:active-hamiltonian}
\end{equation}
The boundary interaction depends explicitly on the physical spin parity $P$. As a consequence, although any even bilinear of defect Majoranas commutes with $H_D$, an isolated defect Majorana anticommutes with $P$ and fails to commute with the boundary term.

The active bulk parity is defined by $P_b=i^n\alpha_1\cdots\alpha_{2n}$. Permuting the four defect modes to the front of the full Majorana sequence produces a sign $(-1)^{\sum_{a=1}^4(2p_a-1-a)}=+1$, establishing the parity relation
\begin{equation}
P=QP_b.
\label{eq:parity-dictionary}
\end{equation}
An inverse Jordan--Wigner transformation on the active chain maps it to a reference spin chain with an Ising boundary seam $t$, generating the boundary coupling $itP_b\alpha_{2n}\alpha_1$. Matching this interaction with Eq.~\eqref{eq:active-hamiltonian} uniquely identifies the boundary seam as $t=Q$.

Local defect fusion provides a physical derivation of this boundary seam. Away from the Jordan--Wigner boundary cut, the unitary operator
\begin{equation}
\mathcal U_r=e^{-i\pi Z_rZ_{r+1}/4}e^{-i\pi X_r/4}
\label{eq:move-core}
\end{equation}
translates a defect core from site $r$ to the adjacent site $r+1$ while commuting with $P$. After bringing two defect cores to adjacent positions, define the fusion observable $a=ic_{2r-1}c_{2r+1}=-Y_rZ_{r+1}$ and the effective spin operators $\widetilde Z=Z_{r+1}$ and $\widetilde X=Z_rY_{r+1}$. These operators commute with $a$, square to identity, and mutually anticommute. In an eigenspace of $a$, the localized defect interactions reduce to
\begin{equation}
-Z_{r-1}Y_r-Z_rY_{r+1}
=aZ_{r-1}\widetilde Z-\widetilde X,
\qquad X_rX_{r+1}=-a\widetilde X.
\label{eq:pair-fusion}
\end{equation}
Each fused pair of defect cores introduces an effective seam $-a$ alongside the corresponding transformation of physical parity. Fusing the consecutive pairs $(1,2)$ and $(3,4)$ with respective bilinear eigenvalues $a$ and $b$ yields the net seam $(-a)(-b)=ab=Q$. Choosing $Q=+1$ selects the identity fusion channel while maintaining the physical spin parity $P$ as an independent symmetry label.

\subsection{The common bulk state}

In the sector $P=Q=+1$, the active parity is fixed to $P_b=+1$. The even defect algebra at fixed $Q$ constitutes the full matrix algebra $M_2$, while the active even algebra at fixed $P_b$ forms $M_{2^{n-1}}$. Because these two algebras commute, and the product of their irreducible representation dimensions $2\cdot2^{n-1}=2^{L-2}$ equals the full dimension of the $P=Q=+1$ subspace, the state space and Hamiltonian factorize as described in Sec.~\ref{sec:encoding}. The bulk dynamics are governed by the standard periodic transverse-field Ising Hamiltonian
\begin{equation}
H_b=-\sum_{j=1}^n\widetilde X_j
-\sum_{j=1}^{n-1}\widetilde Z_j\widetilde Z_{j+1}
-\widetilde Z_n\widetilde Z_1.
\label{eq:bulk-hamiltonian}
\end{equation}
Expressed in the computational $\widetilde Z$ basis, all off-diagonal matrix elements of $H_b$ are nonpositive single spin flips connecting all basis configurations into an irreducible graph. The Perron--Frobenius theorem ensures that the bulk ground state $\Omega_n$ is unique and strictly positive. Global spin-inversion symmetry forces $\Omega_n$ into the even sector $P_b=+1$. As a result, the two encoded basis states share the identical bulk ground state $\Omega_n$ and differ only in their internal fusion state, defining a degenerate ground doublet within the specified symmetry sector.

\subsection{Exact-record obstruction and exact-extraction lemmas}

\begin{lemma}[Exact-record obstruction]\label{lem:exact-record}
Let $V \colon \mathbb{C}^2 \to \mathcal{H}_A\otimes\mathcal H_E$ be an encoding isometry. Suppose a Hermitian unitary $R_E$ supported entirely within the unobserved environment $E=A^c$ satisfies
\begin{equation}
R_E V = V W
\end{equation}
for some logical Pauli operator $W \in \{X,Y,Z\}$. Then the accessible channel $\mathcal{E}_A(\rho) = \mathrm{Tr}_E(V \rho V^\dagger)$ satisfies
\begin{equation}
\mathcal{E}_A(W \rho W) = \mathcal{E}_A(\rho), \qquad \mathcal{E}_A = \mathcal{E}_A \circ \Delta_W,
\end{equation}
where $\Delta_W(\rho) = \frac{1}{2}(\rho + W \rho W)$ is complete dephasing in the eigenbasis of $W$. Consequently, $\mathcal{E}_A$ is entanglement breaking, and for any completely positive trace-preserving decoder $\mathcal D$, the Bell-input squared entanglement fidelity satisfies
\begin{equation}
F(\mathcal{D} \circ \mathcal{E}_A) \le \frac{1}{2}.
\end{equation}
\end{lemma}

\begin{proof}
The intertwining $R_E V=VW$ and unitary invariance of the partial trace over $E$ give $\mathcal{E}_A(W\rho W) = \mathcal{E}_A(\rho)$, hence $\mathcal{E}_A = \mathcal{E}_A \circ \Delta_W$. Complete qubit dephasing $\Delta_W$ is a rank-one measurement followed by preparation, so this factorization makes $\mathcal{E}_A$ and any CPTP postprocessing entanglement breaking. The resulting Bell-input output is separable and has overlap at most $1/2$ with the maximally entangled target.
\end{proof}

\begin{lemma}[Exact extraction via supported anticommuting pairs]\label{lem:exact-extraction}
Suppose $A$ supports Hermitian unitaries $\bar X_A$ and $\bar Z_A$ satisfying $\{\bar X_A,\bar Z_A\}=0$ and
\begin{equation}
\bar X_A V = V X,\qquad \bar Z_A V = V Z.
\end{equation}
There is a unitary change of basis supported on $A$, identifying $\mathcal H_A$ with $\mathbb C^2_{\rm out}\otimes\mathcal K_A$, such that
\begin{equation}
(U_A\otimes I_E)V|a\rangle
=|a\rangle_{\rm out}\otimes|\chi\rangle_{\mathcal K_A E},
\qquad a=0,1,
\end{equation}
with the same residual vector $|\chi\rangle$ for both logical indices. Therefore a fixed completely positive trace-preserving decoder acting only on $A$ recovers the logical qubit exactly, including its entanglement with an inert reference.
\end{lemma}

\begin{proof}
An orthonormal basis $\{e_\mu^+\}$ of the $+1$ eigenspace of $\bar Z_A$ pairs with $e_\mu^-=\bar X_Ae_\mu^+$ in its $-1$ eigenspace. Define the unitary
\begin{equation}
U_Ae_\mu^+=|0\rangle_{\rm out}\otimes|\mu\rangle,\qquad
U_Ae_\mu^-=|1\rangle_{\rm out}\otimes|\mu\rangle.
\end{equation}
Then $U_A\bar Z_AU_A^\dagger=Z\otimes I$ and $U_A\bar X_AU_A^\dagger=X\otimes I$. The $\bar Z_A$ intertwining fixes the output of $V|0\rangle$ to $|0\rangle$; the relation $V|1\rangle=\bar X_AV|0\rangle$ gives output $|1\rangle$ with the same residual vector. Linearity extends this factorization to arbitrary logical inputs and an inert reference. Tracing out the auxiliary factor gives the CPTP decoder
\begin{equation}
\mathcal D_A(\omega)=\mathrm{Tr}_{\mathcal K_A}(U_A\omega U_A^\dagger).
\end{equation}
\end{proof}

\subsection{Extraction supported on a spin interval}

On $J_+=[p_2,p_4]$, Eq.~\eqref{eq:physical-logical-string} gives $\widehat X=i\gamma_2\gamma_3=X_f$ and $\widehat Z=i\gamma_3\gamma_4$, supported on $[p_2,p_3]$ and $[p_3,p_4]$, respectively. They are anticommuting Hermitian involutions on the full physical spin Hilbert space of $J_+$. For $Q=+1$ and every matched $h>0$, they satisfy $\widehat XV_{h,d,L}=V_{h,d,L}X$ and $\widehat ZV_{h,d,L}=V_{h,d,L}Z$. Lemma~\ref{lem:exact-extraction} yields Eq.~\eqref{eq:extractor}, with a residual depending on $h$ but not on the logical input. The left interval $J_-=[p_1,p_3]$ supports the analogous pair $i\gamma_2\gamma_3$ and $i\gamma_1\gamma_2$.

\section{Endpoint codes and finite-field transport}\label{supp:reference-endpoints}

\subsection{Endpoint reference states and local operators}

For any fixed normalized even reference state $\Omega \in \mathcal{H}_{b,+}$, an encoding map $V_\Omega \colon \mathbb{C}^2 \to \mathcal{H}_{P=Q=+1}$ is defined by
\begin{equation}
V_\Omega |a\rangle = W(|a\rangle \otimes \Omega).
\end{equation}
At $h = 0$, only the even active edges and the boundary seam remain. The product endpoint is the normalized limit $H_D(h)/h$ as $h \to \infty$: the odd-edge coefficients remain finite, while the even-edge and seam coefficients vanish.

The bulk reference ground states at these two solvable endpoints are selected to be
\begin{equation}
\Omega_{\rm P} = |+\rangle_X^{\otimes n}, \qquad \Omega_{\rm F} = \frac{1}{\sqrt{2}} \left( |0\cdots 0\rangle_Z + |1\cdots 1\rangle_Z \right).
\end{equation}
The ferromagnetic reference $\Omega_{\rm F}$ is the unique even-parity combination of the two ferromagnetic ground states, not a symmetry-breaking product state. Write $V_{\rm S}=V_{\Omega_{\rm S}}$ for $\mathrm S=\mathrm P,\mathrm F$, and let $n_{*,L}^{\rm S}$ be the minimum retained-spin count for that encoder under the same Bell criterion. All support derivations are formulated on finite rings before considering spatial scaling limits.

The physical endpoint operators and their fixed-sector bulk factors are given in Eq.~\eqref{eq:conjugate-endpoints}.

\subsection{Product reference: outer-gap price}

For the product reference $\Omega_{\rm P}=|+\rangle_X^{\otimes n}$, every bulk disorder string stabilizes the reference:
\begin{equation}
\widetilde X_r\Omega_{\rm P}=\Omega_{\rm P}
\quad\Longrightarrow\quad
\left(\prod_{r=j}^{k-2}\widetilde X_r\right)\Omega_{\rm P}
=\Omega_{\rm P}.
\end{equation}
Consequently, for ordinary sites $j\in G_1$ and $k\in G_3$,
\begin{equation}
Z_jZ_kV_{\rm P}=-X_fV_{\rm P}=-V_{\rm P}X.
\end{equation}
If both sites are outside $A$, then $-Z_jZ_k$ is an exact environmental logical $X$ record. Lemma~\ref{lem:exact-record} makes the accessible channel entanglement breaking. Thus $F(A)>1/2$ requires
\begin{equation}
G_1\subset A\quad\text{or}\quad G_3\subset A.
\end{equation}
This condition is necessary for arbitrary supports; it is not sufficient for arbitrary choices of the remaining retained spins.

For the upper bound, $O_X=X_f\prod_{j\in G_2}\widetilde X_{j-1}$ acts on the product code as $O_XV_{\rm P}=V_{\rm P}X$. Two explicit supports give exact recovery:
\begin{enumerate}
\item If $\ell_3\le \ell_1$, choose
\begin{equation}
A_{{\rm P},+}=\{p_2\}\cup[p_3,p_4].
\end{equation}
It supports $O_X=-Y_{p_2}Z_{p_3}$ and the full defect string
\begin{equation}
\widehat Z=-Y_{p_3}\left(\prod_{j\in G_3}X_j\right)Z_{p_4}.
\end{equation}
These are anticommuting Hermitian involutions representing logical $X$ and $Z$ exactly. The support has $\ell_3+2$ spins.
\item If $\ell_1<\ell_3$, choose
\begin{equation}
A_{{\rm P},-}=[p_1,p_2]\cup\{p_3\}.
\end{equation}
It supports $O_X$ and
\begin{equation}
Z_f=i\gamma_1\gamma_2
=-Y_{p_1}\left(\prod_{j\in G_1}X_j\right)Z_{p_2},
\end{equation}
again an exact anticommuting logical pair. This support has $\ell_1+2$ spins.
\end{enumerate}
Lemma~\ref{lem:exact-extraction} gives deterministic exact recovery on both supports. Counting the ordinary spins in a necessarily retained outer gap gives, for every $0\le\epsilon<1/2$,
\begin{equation}
\min(\ell_1,\ell_3)-1
\le n_{*,L}^{\rm P}(d,\epsilon)
\le \min(\ell_1,\ell_3)+2.
\label{eq:product-endpoint-bracket}
\end{equation}
The brackets agree up to an additive constant; no exact finite minimum is asserted.

\subsection{Even GHZ reference: central-gap price}

Every bulk order pair stabilizes the even GHZ reference:
\begin{equation}
\widetilde Z_a\widetilde Z_b\,\Omega_{\rm F}=\Omega_{\rm F}.
\end{equation}
The central--exterior dictionary therefore gives, for $j\in G_2$ and $k>p_4$,
\begin{equation}
Z_jZ_kV_{\rm F}=-Z_fV_{\rm F}=-V_{\rm F}Z.
\end{equation}
If a central ordinary site and an exterior site are both discarded, $-Z_jZ_k$ is an exact environmental logical $Z$ record. Hence $F(A)\le1/2$ by Lemma~\ref{lem:exact-record}. Whenever at least one exterior site remains unobserved, $F(A)>1/2$ therefore requires $G_2\subset A$.

For a linear budget $|A|\le Md$, an exterior window of $\lceil(M+1)d\rceil$ sites guarantees an unobserved exterior site once $L$ is sufficiently large. This is the role of the thermodynamic-first limit. The necessity statement does not extend without qualification to finite rings lacking sufficient exterior volume.

For achievability, the string in $O_Z=\widehat Z\prod_{j\in G_3}X_j$ telescopes to
\begin{equation}
\prod_{j\in G_3}X_j
=\widetilde Z_{p_3-1}\widetilde Z_{p_4-2},
\end{equation}
which stabilizes $\Omega_{\rm F}$. Thus $O_ZV_{\rm F}=V_{\rm F}Z$. The set
\begin{equation}
A_{\rm F}=[p_2,p_3]\cup\{p_4\}
\end{equation}
supports the exact anticommuting pair
\begin{equation}
X_f=-Y_{p_2}\left(\prod_{j\in G_2}X_j\right)Z_{p_3},
\qquad O_Z=-Y_{p_3}Z_{p_4}.
\end{equation}
Its size is $\ell_2+2$, so Lemma~\ref{lem:exact-extraction} yields exact recovery.

In particular, if $L-p_4>\ell_2+2$, then for every $0\le\epsilon<1/2$,
\begin{equation}
\ell_2-1\le n_{*,L}^{\rm F}(d,\epsilon)\le \ell_2+2.
\label{eq:ghz-endpoint-bracket}
\end{equation}
To prove the lower bound, a support with fewer than $\ell_2-1$ spins necessarily misses a central ordinary site and cannot cover the exterior. The exact-record obstruction then applies. These are constant-width bounds, not an exact finite minimum or a sufficient classification of arbitrary supports.

\subsection{Explicit single-spin measurement extraction}

The supported operator pairs also give explicit single-spin extraction schemes. In each case the helper measurements commute with the supported logical pair.
\begin{itemize}
\item \textbf{Product reference on $A_{{\rm P},+}$:} Measure $Y_{p_2}$, $X$ on every site of $G_3$, and $Z_{p_4}$, leaving $p_3$ as output. Substituting the measurement outcomes into $O_X$ and $\widehat Z$ gives the logical generators $w_{\rm par} Z_{p_3}$ and $-wY_{p_3}$, where
\begin{equation}
w_{\rm par}=-y_{p_2},\qquad
w=\left(\prod_{j\in G_3}x_j\right)z_{p_4}.
\end{equation}
\item \textbf{GHZ reference on $A_{\rm F}$:} Measure $Y_{p_2}$, $X$ on every site of $G_2$, and $Z_{p_4}$, again leaving $p_3$ as output. The residual generators are $w_{\rm par} Z_{p_3}$ and $-wY_{p_3}$, with
\begin{equation}
w_{\rm par}=-y_{p_2}\prod_{j\in G_2}x_j,\qquad
w=z_{p_4}.
\end{equation}
\end{itemize}
In both cases, the Clifford rotation $U_w$ maps $-wY$ to $Z$ and $Z$ to $X$. Applying it and then a Pauli $Z$ correction when $w_{\rm par}=-1$ restores the logical state in every nonzero-probability branch. All helpers belong to the designated support. This is an ideal measurement implementation; it does not establish preparation complexity or tolerance to noisy control.

\subsection{Ordered asymptotic envelopes}

Let $\ell_i/d\to\lambda_i>0$. At each $d$, keep the defect coordinates fixed while taking $L\to\infty$, before taking $d\to\infty$. For any nonnegative error tolerances $\epsilon_d\to0$, the constant-width finite-ring bounds imply that the lower and upper thermodynamic envelopes have the same large-$d$ limit:
\begin{equation}
\begin{aligned}
\lim_{d\to\infty}\liminf_{L\to\infty}
\frac{n_{*,L}^{\rm S}(d,\epsilon_d)}{d}
&=
\lim_{d\to\infty}\limsup_{L\to\infty}
\frac{n_{*,L}^{\rm S}(d,\epsilon_d)}{d}
=b_{\rm S},
\qquad {\rm S}\in\{{\rm P},{\rm F}\},\\
b_{\rm P}&=\min(\lambda_1,\lambda_3),\qquad
b_{\rm F}=\lambda_2.
\end{aligned}
\end{equation}
The critical encoder has the corresponding price
\begin{equation}
b_{\rm crit}=\lambda_2+\min(\lambda_1,\lambda_3).
\end{equation}
For $(\lambda_1,\lambda_2,\lambda_3)=(3,2,2)$, the endpoint prices are $b_{\rm P}=b_{\rm F}=2$, whereas $b_{\rm crit}=4$.

These endpoint costs compare different encoders in the same physical partition. Appendix~\ref{supp:finite-field} treats the intervening finite fields.

\subsection{Precision-target costs}
\phantomsection\label{supp:precision-costs}
For each reference state $\mathcal S\in\{\mathrm P,\mathrm F,\mathrm{crit}\}$, define $m_{\mathcal S}(q;d,L)$ as the minimum number of retained physical spins achieving optimized Bell fidelity at least $q$, where $1/2<q\le1$. The strict Bell-advantage cost is $m_{\mathcal S,+}(d,L)=\min\{|A|:F(A)>1/2\}$. Thus $m_{\mathcal S}(q;d,L)=n^{\mathcal S}_{*,L}(d,1-q)$, and $m_{\mathcal S}(1;d,L)$ is the exact-recovery cost. The support is chosen independently of the logical input, and recovery ranges over all completely positive trace-preserving maps on its physical spin tensor factors. Every measured helper counts in $|A|$.

For the product and even-GHZ endpoints, put $\ell_{\rm P}=\min(\ell_1,\ell_3)$ and $\ell_{\rm F}=\ell_2$, where $\ell_i=p_{i+1}-p_i$ and $|G_i|=\ell_i-1$. In the GHZ case assume the sufficient finite-volume condition $L-p_4>\ell_2+2$. Subtracting the outer bounds of Eqs.~\eqref{eq:product-endpoint-bracket} and \eqref{eq:ghz-endpoint-bracket} gives
\begin{equation}
0\le m_{\mathcal S}(1;d,L)-m_{\mathcal S,+}(d,L)\le3.
\label{eq:precision-endpoint-gap}
\end{equation}
This compares optimized cardinalities, not equal finite costs, identical minimizers or exact recovery on an unaugmented advantage-achieving support. At fixed $d$, the GHZ volume condition holds on the large-$L$ tail; linear budgets likewise ensure an exterior omission there. Taking $L\to\infty$ first, then $d\to\infty$ with $\ell_i/d\to\lambda_i>0$, both thermodynamic envelopes give the same leading coefficient for strict advantage, every fixed $q>1/2$ and exact recovery: $\min(\lambda_1,\lambda_3)$ for product and $\lambda_2$ for even GHZ.

\subsection{Transport at a fixed finite field}\label{supp:finite-field}
\phaseproofintro
\subsubsection{Transport of the endpoint code space}
Write the ordered active Majoranas as $a_j=\alpha_{2j-1}$ and $b_j=\alpha_{2j}$. On the selected sector $a_{n+1}=-a_1$. With $e_j$ the standard basis of $\mathbb C^n$, let $T e_j=e_{j+1}$ for $j<n$ and $T e_n=-e_1$; the quadratic form is $-i a^{\mathsf T}(hI-T)b$. Its mode energies are $2|h-e^{ik}|$, with $k=(2q+1)\pi/n$ and $q=0,\ldots,n-1$. The vacuum is even by continuity from the product state since $\det(hI-T)=h^n+1$ stays positive, including for odd $n$. In the ordered phase follow the path of fields $sh$, $0\le s\le1$, starting with the even GHZ state. In the disordered phase use the normalized path with field coefficient one and bond coefficient $s/h$, starting with the product state. The seam has the same coefficient as the other reference bonds. The gap above the fusion doublet within its fixed charge sector is bounded below by $2(1-h)$ or $2(1-1/h)$, respectively. Other sectors may contain states arbitrarily close in energy.

The path Hamiltonian $H(s)$ and its derivative commute with both charges and the physical representatives of the logical Pauli operators. In the full physical spin ring they are sums of finite-range terms with uniformly bounded local strengths. We use filtered spectral transport~\cite{Arxiv1102_0842}. Choose $\gamma>0$ below the same-sector gap and a real even Schwartz function $\varphi$ with Fourier transform supported in $(-\gamma,\gamma)$ and equal to one near zero. Define $W_\gamma(t)=\int_t^\infty\varphi(v)\,dv$ for $t>0$ and extend it oddly. For the Fourier convention $e^{it\omega}$, its distributional derivative is $\delta_0-\varphi$, with $\delta_0$ the unit Dirac mass at zero, and its transform is $i/\omega$ outside the filter window. Put
\begin{equation}
D(s)=\int_{\mathbb R} W_\gamma(t)e^{itH(s)}\dot H(s)e^{-itH(s)}\,dt,\qquad
\partial_s U=iD(s)U,\quad U(0)=I .
\label{eq:field-generator}
\end{equation}
The code projector $\Pi(s)$ is the product of the fusion identity with the reference ground projector in the selected sector and zero in the other sectors. For a code vector $g$ and an excited eigenvector $e$ in the same sector, $iD_{eg}=-\dot H_{eg}/(E_e-E_g)=\dot\Pi_{eg}$. The adjoint gives the reverse matrix elements. Differentiating $\Pi^2=\Pi$ gives $\Pi\dot\Pi\Pi=(I-\Pi)\dot\Pi(I-\Pi)=0$. Intersector matrix elements vanish by charge conservation. It follows that $\partial_s\Pi=i[D,\Pi]$. The full-spin unitary $U(1)$ transports the code projector. Its commutation with both logical axes fixes the action on the encoded qubit up to a common phase.

\subsubsection{Localization of a deterministic decoder}
The spatial estimate is on the unprojected physical ring. Fix positive Lieb--Robinson constants $C,a,v$, with $v$ denoting the temporal exponential rate. Let $\mathbb E_X$ trace out $X^c$ with normalized trace and reinsert the identity. For a finite-range term $\dot\Phi_Y$ of $\dot H$, the usual Lieb--Robinson bound on the physical time evolution, summed over sites outside the $R_{\rm loc}$-neighbourhood of $Y$, controls truncation there by $C\|\dot\Phi_Y\|e^{-aR_{\rm loc}/2}$ for $|t|\le aR_{\rm loc}/(2v)$. At larger times use $2\|\dot\Phi_Y\|$ and the superpolynomial time tail of $W_\gamma$. Differences between consecutive truncated filtered terms then decompose $D$ into interactions $\Psi_Z$ supported on these neighbourhoods. To cover sites distance $r$ apart requires $2R_{\rm loc}+\mathrm{diam}(Y)\ge r$, with at most $C(1+R_{\rm loc})$ possible centres at each radius. Hence for every $k>1$ the sum over terms containing both sites of $\sup_s\|\Psi_Z(s)\|$ is bounded by $C_k(1+r)^{-k}$ uniformly in $L,d$. On a cycle this decay function has uniformly bounded sum and convolution constant. Applying the time-dependent Lieb--Robinson estimate~\cite{Arxiv1102_0842} to this interaction, followed by single-site unitary averages outside $X_w$, gives
\begin{equation*}
\|U O_X U^\dagger-\mathbb E_{X_w}(U O_X U^\dagger)\|
\le C_N |X|(1+w)^{-N}\|O_X\|,\qquad U=U(1),
\end{equation*}
for every $N>0$, by choosing $k>N+1$ and summing the spatial tail. These estimates apply in either transport direction. The constants can be chosen uniformly for target fields in a compact subset of either phase. The spectral separation was used only to intertwine the selected code projector. Related quasi-adiabatic constructions for perturbed Majorana memories address storage under Hamiltonian evolution~\cite{BravyiKonig2012MajoranaStorage}.

Let $A_0$ and $\mathcal D_0$ be the endpoint support and exact decoder used in Eq.~\eqref{eq:field-decoder}, and set $U_h=U(1)$ and $V_h=V_{h,d,L}$. Spectral transport carries the endpoint encoding to $V_h$ up to a common phase that cancels under conjugation, so that each input-qubit observable $O$ satisfies
\begin{equation*}
V_h^\dagger U_h\iota_{A_0}\mathcal D_0^*(O)U_h^\dagger V_h=O.
\end{equation*}
Because $\mathcal D_0^*$ is unital and completely positive, it contracts the operator norm as $\|\mathcal D_0^*(O)\|\le\|O\|$. Applying Eq.~\eqref{eq:field-tail} with $X=A_0$ and compressing to the target code bounds the reconstructed logical channel $\Phi$ by
\begin{equation*}
\|(\Phi^*-\mathrm{id})(O)\|
\le\eta_w\|O\|,\qquad
\eta_w=C_N|A_0|(1+w)^{-N}.
\end{equation*}
Expanding an arbitrary reference-extended test observable into the four qubit matrix units yields the stated diamond-norm and Bell-fidelity bounds. Because $A_0$ has at most two connected components, its ring neighbourhood adds at most $4w$ spins, so the designated buffer choice yields the inverse-power target error with sublinear overhead.

\subsubsection{Phase-dependent lower bound}
On the antiperiodic active ring the even signed cyclic shift of the Majoranas interchanges $ia_jb_j$ and $ib_ja_{j+1}$ up to a translation. It sends $H(h)$ to $hH(1/h)$, preserves parity and maps each nonwrapping disorder string of length $r$ to an order pair at separation $r$. Its expectation at $h$ is therefore $T_{r,n}(1/h)$ where $T_{r,n}$ denotes the order correlator. At fixed $r$ and $h\ne1$ the Wick expansion of the order pair uses finitely many covariance entries. These are Fourier sums of the continuous phase symbol of $hI-T$ on the antiperiodic grid and converge to their infinite-chain values. In Pauli units the ordered reference has $\lim_{r\to\infty}T_r(\kappa)=(1-\kappa^2)^{1/4}$ for $0<\kappa<1$~\cite{Pfeuty1970}.

Take $\kappa=\min(h,1/h)$ and a fixed linear budget $Md$ larger than the leading exact cost. In the ordered phase an omitted central spin can be paired with an omitted spin in the exterior window of $\lceil(M+1)d\rceil$ sites immediately to the right of $p_4$. In the disordered phase take one omission in each outer gap. The relevant reference separation in both cases is $r=k-j-1$ for physical sites $j<k$. At fixed $d$ the distances in the relevant containers form a finite band independent of the support, so the minimum of $T_{r,n}(\kappa)$ over the band converges to the minimum of $T_r(\kappa)$. The lower edge grows at least as $\ell_3+1$ in the first case or $\ell_2+1$ in the second. Thus the band minima tend to $t_h=(1-\kappa^2)^{1/4}$ as $d$ increases. The ordered upper fidelity envelope over all supports with the specified omissions is at most $(1+\sqrt{1-t_h^2})/2<1$. A fixed positive margin therefore excludes every such support at target $1-d^{-p}$ on successive large-$d$ and large-$L$ tails. Ordered-phase supports must retain the ordinary central gap, and disordered-phase supports must retain one entire ordinary outer gap. The respective lower counts $\ell_2-1$ and $\min(\ell_1,\ell_3)-1$ match the leading counts of the supports constructed after Eq.~\eqref{eq:field-tail}. This proves the two polynomial-accuracy prices for fixed field.

For a fixed target fidelity strictly between $(1+\sqrt{1-t_h^2})/2$ and one, the same margin gives the corresponding gapped-phase price. The ceiling describes the resolving power of the single-pair witness.

\section{Uniform critical estimates, cost and shape}
\label{app:critical}

The operator dictionaries make the complementary records accessible to ferromagnetic correlation estimates. We derive them on the finite reference ring, where finite-band envelopes give bounds uniform over the omitted sets.

\subsection{The two dictionaries and disorder products}

For any pair of ordinary lattice sites $j<k$ whose connecting interval avoids the boundary cut, the Jordan--Wigner transformation expresses the two-spin product as a string of Majoranas
\begin{equation}
Z_jZ_k=i^{k-j}c_{2j}c_{2j+1}\cdots c_{2k-1}.
\label{eq:spin-word}
\end{equation}
When the spatial interval $[j,k]$ encompasses two defect sites $p_a<p_b$, the corresponding defect modes $\gamma_a=c_{2p_a-1}$ and $\gamma_b=c_{2p_b-1}$ appear within the string at relative positions $2(p_a-j)$ and $2(p_b-j)$. Permuting these two fermionic modes to the front generates a factor of $(-1)$, while their product relates to the defect bilinear via $\gamma_a\gamma_b=-i(i\gamma_a\gamma_b)$. The remaining active Majoranas reconstruct the reference spin operators, yielding the operator identities displayed in Eq.~\eqref{eq:two-dictionaries}. Specifically, for the middle--exterior pairing ($j\in G_2, k\in E_d$), the intercepted defects are $\gamma_3$ and $\gamma_4$, leaving active reference spins at sites $j-1$ and $k-2$. For the outer--outer pairing ($j\in G_1, k\in G_3$), the intercepted defects are $\gamma_2$ and $\gamma_3$, leaving the disorder string $D_{j,k}=\prod_{m=j}^{k-2}\widetilde X_m$ on $[j,k-2]$. Because these relations hold as exact operator identities across the full $P=Q=+1$ subspace prior to taking expectations, they remain valid under operator multiplication and encompass coincident spin indices.

To evaluate expectations involving the disorder strings $D_{j,k}$, we introduce an even unitary operator $\mathcal W$ acting on the active antiperiodic chain according to the signed cyclic shift
\begin{equation}
\alpha_r\mapsto\alpha_{r+1}\ (r<2n),
\qquad \alpha_{2n}\mapsto-\alpha_1.
\label{eq:signed-shift}
\end{equation}
This orthogonal transformation has determinant $+1$ and admits an explicit representation as the ordered product $\mathcal W=\prod_{r=1}^{2n-1}\exp[-\pi\alpha_r\alpha_{r+1}/4]$, with smaller $r$ to the left. Each factor sends $\alpha_r\mapsto\alpha_{r+1}$ and $\alpha_{r+1}\mapsto-\alpha_r$ under conjugation, and applying the factors from right to left gives the stated signed shift. The shift preserves the antiperiodic active Hamiltonian and $P_b$, so it leaves the unique even ground state $\Omega_n$ invariant up to phase. On a nonwrapping interval, conjugating by $\mathcal W$ translates the active disorder string into a product of order spins
\begin{equation}
\mathcal W D_{j,k}\mathcal W^\dagger
=\widetilde Z_j\widetilde Z_{k-1}.
\label{eq:disorder-transfer}
\end{equation}
Because $\mathcal W$ acts multiplicatively across products of disorder operators, conjugating any product of disorder strings maps their joint expectation value directly to a standard multi-point correlation function of reference order spins. This mapping reflects an exact correlation symmetry of the reference ground state, establishing that the statistical fluctuations of disorder strings are governed by the same ferromagnetic bounds that control order spins.

\subsection{Ferromagnetic moments and their scaling}

In the following relation, the four subscripts label arbitrary reference insertions and $T_{ab}=\langle\widetilde Z_a\widetilde Z_b\rangle_{\Omega_n}$ is their two-point correlator. The reference ground state $\Omega_n$ inherits nonnegative two-point correlations and the classical zero-field four-point inequality
\begin{equation}
\langle\widetilde Z_1\widetilde Z_2\widetilde Z_3\widetilde Z_4\rangle
\le T_{12}T_{34}+T_{13}T_{24}+T_{14}T_{23}
\label{eq:lebowitz}
\end{equation}
established by Lebowitz for ferromagnetic spin systems~\cite{Lebowitz1974GHS}. To establish this inequality rigorously for the quantum ground state, we begin with the unprojected thermal Gibbs state of the finite reference chain at temperature $\beta^{-1}>0$. A standard Lie--Trotter--Suzuki decomposition with discrete time step $\delta>0$ maps the quantum partition function to an anisotropic two-dimensional classical Ising model with ferromagnetic spatial couplings $K_s=\delta$ and temporal couplings $K_t=\tfrac12\log\coth\delta>0$. Because the classical system has zero longitudinal magnetic field, equal-time spin observables satisfy Eq.~\eqref{eq:lebowitz}. Taking the continuous-time limit $\delta\to 0$ followed by the zero-temperature limit $\beta\to\infty$ at fixed system size $n$ projects directly onto the unique even ground state $\Omega_n$ without requiring a separate parity projection, thereby preserving the inequality on the finite ring and in the subsequent thermodynamic limit $n\to\infty$. Furthermore, coincident spin indices are encompassed naturally because $\widetilde Z_j^2=I$ and all surviving two-point expectations remain nonnegative.

Because all two-point correlations $T_{r,n}$ are nonnegative, the double sum over any perforated subset contained within an interval of length $w$ is bounded above by the complete container sum $S_n(w)$. Summing Lebowitz's inequality [Eq.~\eqref{eq:lebowitz}] over indices $j,j'\in U$ and $k,k'\in V$ immediately yields Eq.~\eqref{eq:perforated-moments}, where the two cross-pairing channels each evaluate to $\mu^2$.

On the critical infinite chain, the longitudinal ground-state correlation function of spin-$1/2$ operators along the ordering axis, $\rho_r^x = \langle 0 | S_j^x S_{j+r}^x | 0 \rangle$, was derived by Pfeuty~\cite{Pfeuty1970} in Eqs.~(3.4), (3.6), and (3.7a). Converting to Pauli operators yields the infinite-chain order kernel $T_r = 4\rho_r^x$, where the spatial separation $r$ between the two spins is distinguished from the chain length $n$:
\begin{equation}
T_r=\left(\frac2\pi\right)^r
\prod_{j=1}^{r-1}\left(1-\frac1{4j^2}\right)^{j-r},
\qquad T_r\sim Cr^{-1/4}.
\label{eq:kernel-product}
\end{equation}
An elementary analysis of the Wallis product reveals that $T_r$ is strictly positive and monotonically decreasing with separation $r$. Inserting this power-law form into the container sum and bounding the integrable short-distance endpoint via $T_r\le C'(1+r)^{-1/4}$ yields the asymptotic integral
\begin{equation}
S(w)\sim2Cw^{7/4}\int_0^1(1-x)x^{-1/4}\,dx
=\frac{32C}{21}w^{7/4}.
\label{eq:sum-integral}
\end{equation}
Because any finite short-distance correction contributes at most $O(w)$, it is subleading compared to the dominant $w^{7/4}$ power law and does not alter the asymptotic prefactor.

\subsection{Finite distance bands and uniformity}

For the $Z_f$ complementary record, the reference container dimensions and the corresponding cross-distance band are given by
\begin{equation}
(w_L,w_R)=(\ell_2-1,\lceil Kd\rceil),
\quad \ell_3+1\le r\le \ell_2+\ell_3+\lceil Kd\rceil-2.
\label{eq:z-band}
\end{equation}
Similarly, for the $X_f$ record, the enclosing containers and cross-distance band are
\begin{equation}
(w_L,w_R)=(\ell_1-1,\ell_3-1),
\quad \ell_2+1\le r\le \ell_1+\ell_2+\ell_3-3.
\label{eq:x-band}
\end{equation}
At any fixed defect scale $d$, each cross-distance band spans a finite set of integer separations that remains strictly invariant under the choice of candidate support $A$. On a finite ring, taking the thermodynamic limit $n\to\infty$ causes the band infimum $m_{d,n}$ to converge to $T_{R_d}$, where $R_d$ is the maximum separation within the band. Combining this limit with the container asymptotics of Eq.~\eqref{eq:sum-integral} establishes Eq.~\eqref{eq:density-constant}.

For $u_2\ge ad$, the exterior residual contains at least $d$ spins, so the density parameters are $(a,c)=(a,1)$. If $u_1,u_3\ge ad$, they are $(a,c)=(a,a)$. These parameters count residual spins per $d$, rather than per container length. In the asymptotic limit $d\to\infty$, the arguments of the fidelity function $\mathcal B$ for the two records converge to
\begin{equation}
\begin{aligned}
\kappa_Z(a,M)&=\frac{a^2}{k_0^2(\lambda_2K)^{7/4}\sqrt{\lambda_2+\lambda_3+K}},\\
\kappa_X(a)&=\frac{a^4}{k_0^2(\lambda_1\lambda_3)^{7/4}\sqrt{\lambda_1+\lambda_2+\lambda_3}}.
\end{aligned}
\label{eq:axis-constants}
\end{equation}
For any fixed positive omission densities, linear budget $M$, and gap ratios $\lambda_i$, the constants $\kappa_Z$ and $\kappa_X$ are strictly positive. If either of the filling conditions in Eq.~\eqref{eq:necessary-filling} failed to hold along a successful support sequence, a strictly positive omission density would recur across an infinite subsequence of scales $d$ and for arbitrarily large $L$. The finite-ring bound would then force the fidelity along that subsequence below a uniform ceiling strictly less than unity, in direct contradiction with the premise of faithful recovery [Eq.~\eqref{eq:success-family}]. This completes the proof of necessary filling prior to any cardinality optimization.

\subsection{Moment bounds adapted to omission counts}
\label{app:count-moments}

At fixed defect scale $d$, let $D_d^{\mathrm{band}}$ be an integer at least as large as every internal and cross distance entering the two test records, determined solely by the test containers and independent of $A$ or $L$. For $L$ sufficiently large to contain the windows, define the upper envelope only for positive indices not exceeding the band endpoint, and its corresponding count sum by
\begin{equation}
\begin{aligned}
\widehat T_{s,d,n}&=\max_{s\le r\le D_d^{\mathrm{band}}}T_{r,n},\qquad s\ge1,\\
\widehat S_{u,d,n}&=u+2\sum_{s=1}^{u-1}(u-s)\widehat T_{s,d,n},
\qquad \widehat S_{0,d,n}=0.
\end{aligned}
\label{eq:finite-count-envelope}
\end{equation}
Only counts $u\le D_d^{\mathrm{band}}+1$ are needed. For any ordered residual reference set $U=\{x_1<\cdots<x_u\}$, the spacing inequality $x_{i+s}-x_i\ge s$ ensures that
\begin{equation}
\sum_{x,x'\in U}T_{|x-x'|,n}\le\widehat S_{u,d,n}.
\label{eq:ordered-count-sum}
\end{equation}
The diagonal terms use $T_{0,n}=1$. This relation holds as a finite-ring inequality even when the finite-ring kernel is not monotone.

Within each reference region, the operator dictionary preserves the underlying integer coordinates. For the middle--exterior record, these coordinates evaluate to $j-1$ and $k-2$, whereas following the signed shift, the outer record employs coordinates $j$ and $k-1$. Both pairings exhibit cross distance $k-j-1$ while leaving each internal distance unchanged. Because the signed shift acts multiplicatively on disorder products, the same order four-point inequality applies, including all repeated indices. Summing its three pairings and applying Eq.~\eqref{eq:ordered-count-sum} proves Eq.~\eqref{eq:count-moments}.
\begin{equation}
 \nu\le\widehat S_{u,d,n}\widehat S_{v,d,n}+2\mu^2,
\label{eq:count-moments}
\end{equation}
where $u=|U|$ and $v=|V|$. For nonempty residual sets, the band minimum $m_{d,n}$ also yields
\begin{equation}
F_{d,L}(A)\le\mathcal B\!\left(
\frac{u^2v^2m_{d,n}^2}
{\widehat S_{u,d,n}\widehat S_{v,d,n}}
\right).
\label{eq:finite-count-ceiling}
\end{equation}
This ceiling follows directly from the signal lower bound $\mu\ge uv m_{d,n}$ combined with the second-moment upper bound $\nu\le\widehat S_{u,d,n}\widehat S_{v,d,n}+2\mu^2$ and the monotonic decrease of $\mathcal B$. An empty residual set supplies only the trivial bound $F\le1$.

At fixed $d$, all maxima in Eq.~\eqref{eq:finite-count-envelope} range over finitely many separations. Because the infinite-chain kernel is decreasing, their simultaneous thermodynamic limits satisfy $\widehat T_{s,d,n}\to T_s$, which implies $\widehat S_{u,d,n}\to S(u)$ simultaneously across all admissible counts. Because only finitely many residual patterns occur within the fixed containers, the resulting count bound holds uniformly prior to taking any supremum over candidate supports, including configurations that move with $L$ or cross the boundary cut.

\subsection{Cost, density shape and finite-error bounds}
\label{app:cost}

\criticalproofintro

\subsubsection{The uniform price converse}

Let $m=\min(\lambda_1,\lambda_3)$ and fix a budget deficit $0<\delta<b_0$. To evaluate candidate supports with cardinality $|A|\le\lfloor(b_0-\delta)d\rfloor$, we set the linear budget bound to $M=b_0$. If the support leaves $u_2\ge\delta d/2$ omissions in the middle gap, the middle--exterior record immediately applies. Conversely, if $u_2<\delta d/2$, the total number of retained spins situated within the two outer gaps is bounded by
\begin{equation}
|A\cap(G_1\cup G_3)|\le
(b_0-\delta)d-\ell_2+1+\delta d/2.
\label{eq:outer-budget}
\end{equation}
Because the number of spins retained in each outer gap cannot exceed this combined total, $|A\cap G_i|\le |A\cap(G_1\cup G_3)|$ for $i=1,3$. Subtracting this upper bound from $\ell_i-1$ yields the omission count
\begin{equation}
u_i\ge(\lambda_i-m+\delta/2)d
+(\ell_i-\lambda_i d)+(\ell_2-\lambda_2d)-2
\ge\delta d/4
\label{eq:outer-omission-count}
\end{equation}
for $i=1,3$ once $d$ is sufficiently large. Here the geometric remainders $(\ell_i-\lambda_i d)$ are $o(d)$ and depend solely on the defect coordinates, remaining completely independent of $A$ and $L$. Taking the supremum over all candidate supports at finite $L$ and evaluating the maximum of the two asymptotic ceilings establishes the uniform fidelity gap in Eq.~\eqref{eq:uniform-price-gap}, where
\begin{equation}
1-\gamma_\delta=
\max\{\mathcal B(\kappa_Z(\delta/2,b_0)),
\mathcal B(\kappa_X(\delta/4))\}<1.
\label{eq:price-gap-constant}
\end{equation}
The ordered limsup leaves a positive margin. For all sufficiently large $d$, followed by all sufficiently large $L$, the support supremum is at most $1-\gamma_\delta/2$. Any nonnegative sequence $\epsilon_d\to0$ eventually satisfies $\epsilon_d<\gamma_\delta/2$, excluding all supports within the smaller budget. The lower thermodynamic price envelope is therefore at least $b_0-\delta$ to leading order. Combining this with the upper envelope of Eq.~\eqref{eq:exact-upper} and sending $\delta$ to zero proves the sharp price equalities in Eq.~\eqref{eq:sharp-price}.

\subsubsection{Why near-minimal supports concentrate}

Take a support family satisfying the vanishing-error condition in Eq.~\eqref{eq:success-family} and the near-minimal budget in Eq.~\eqref{eq:near-minimal-budget}, with $L\to\infty$ before $d\to\infty$ throughout. Filling the central gap in density gives $u_2=o(d)$. If $\lambda_1>\lambda_3$, at most $\lambda_3d+o(d)$ spins remain in the budget for the two outer gaps together. The longer gap then has $u_1\ge(\lambda_1-\lambda_3)d-o(d)$ omissions. Equation~\eqref{eq:necessary-filling} forces $u_3=o(d)$ in the shorter gap. On the ordered thermodynamic tails the number of missing spins in $J_+=[p_2,p_4]$ obeys
\begin{equation}
|J_+\setminus A|\le u_2+u_3+3=o(d).
\label{eq:interval-missing}
\end{equation}
Since $|J_+|/d\to b_0$ and $|A|/d\le b_0+o(1)$, the identity $|A\setminus J_+|=|A|-|J_+|+|J_+\setminus A|$ bounds the retained count outside $J_+$ by $o(d)$. Thus $|A\triangle J_+|=o(d)$. For $\lambda_1<\lambda_3$ the same argument selects $J_-=[p_1,p_3]$. When $\lambda_1=\lambda_3$, choose the outer gap with fewer omissions at each $(d,L)$. The corresponding three-defect interval differs from $A$ on $o(d)$ sites, so the minimum symmetric difference divided by $d$ in Eq.~\eqref{eq:necessary-shape} tends to zero even if the selected branch changes.

\subsubsection{Quantitative small-error bounds}

Consider parameters $M<\infty$ and $K>M$, with the linear budget $|A_{d,L}|\le Md$ holding on the successive large-$d$ and large-$L$ tails. A fixed-error recovery family satisfies
\begin{equation}
\liminf_{d\to\infty}\liminf_{L\to\infty}
F_{d,L}(A_{d,L})\ge1-\epsilon,
\label{eq:fixed-error-success}
\end{equation}
where $\epsilon$ lies in the branch of Eq.~\eqref{eq:error-function}. We set $\rho_i=u_i/d$, $H_i=\ell_i-1$, $\Lambda=\lambda_1+\lambda_2+\lambda_3$, and $\lambda_{\max}=\max(\lambda_1,\lambda_3)$. The exterior window of length $\lceil Kd\rceil$ contains at least $(K-M)d$ omitted sites.

First take the thermodynamic scalar limit of Eq.~\eqref{eq:finite-count-ceiling} at fixed $d$. Along any subsequence with positive residual densities $u/d\to a$ and $v/d\to c$, and largest cross distance $R_d/d\to R>0$, this deterministic count ratio converges according to
\begin{equation}
\frac{u^2v^2T_{R_d}^2}{S(u)S(v)}
\longrightarrow
\frac{(ac)^{1/4}}{k_0^2\sqrt R}.
\label{eq:count-ratio-limit}
\end{equation}
Both the powers of $d$ and the nonuniversal amplitude cancel identically. Inverting $\mathcal B$ directly yields Eq.~\eqref{eq:count-density-error}.
\begin{equation}
 ac\le k_0^8R^2q(\epsilon)^4 .
\label{eq:count-density-error}
\end{equation} Should either limiting density vanish, this product constraint holds automatically. Because residual fractions for arbitrary supports reside in compact intervals determined by the containers, any violating ordered limsup produces a subsequence with positive limiting product to which Eq.~\eqref{eq:count-ratio-limit} applies.

Applying the two cross-distance bands from Eqs.~\eqref{eq:z-band} and \eqref{eq:x-band} yields the limiting omission constraints
\begin{equation}
\begin{aligned}
\limsup_{d\to\infty}\limsup_{L\to\infty}\rho_2
&\le z:=Z_{M,K}q(\epsilon)^4,\\
\limsup_{d\to\infty}\limsup_{L\to\infty}(\rho_1\rho_3)
&\le p:=X_\lambda q(\epsilon)^4,
\end{aligned}
\label{eq:error-omissions}
\end{equation}
with coefficients
\begin{equation}
Z_{M,K}=k_0^8\frac{(\lambda_2+\lambda_3+K)^2}{K-M},
\qquad X_\lambda=k_0^8\Lambda^2.
\label{eq:limiting-error-coefficients}
\end{equation}
Because separate outer branches need not possess individual density limits, their product remains inside the nested limsup. To establish Eq.~\eqref{eq:error-omissions} rigorously under Eq.~\eqref{eq:fixed-error-success}, fix a positive margin $\zeta>0$ small enough that the enlarged error stays in the branch of the fidelity inversion. The outer fidelity liminf makes the inner limiting fidelities at least $1-\epsilon-\zeta$ for all sufficiently large $d$. At each such scale, the inner liminf ensures $F\ge1-\epsilon-2\zeta$ for all sufficiently large $L$. Applying the scalar bounds with this enlarged error on those tails and taking $\zeta\downarrow0$ after evaluating both ordered scalar limits confirms the result.

For the price count, define $H_{\max}=\max(H_1,H_3)$ and $H_{\min}=\min(H_1,H_3)$. The algebraic nonnegativity $(H_{\max}-u_1)(H_{\max}-u_3)\ge0$ implies that every candidate support satisfies the exact finite-ring inequality
\begin{equation}
|A|\ge H_2+H_{\min}-u_2-\frac{u_1u_3}{H_{\max}}.
\label{eq:finite-ring-error-count}
\end{equation}
This relation accounts for retained spins across the three gaps, while retaining further sites only increases $|A|$. Because the interval extractor of Eq.~\eqref{eq:exact-upper} ensures that minimum-cardinality successful supports satisfy $|A|\le Md$ for any fixed $M>b_0$ at sufficiently large $d$, Eqs.~\eqref{eq:error-omissions} and \eqref{eq:finite-ring-error-count} lead directly to
\begin{equation}
\underline b_1[\epsilon]\ge b_0-
\left(Z_{M,K}+\frac{X_\lambda}{\lambda_{\max}}\right)q(\epsilon)^4.
\label{eq:window-error-cost}
\end{equation}
At each finite $L$, both the CPTP optimum and the minimum over finitely many supports are attained, allowing the thermodynamic envelopes to be evaluated directly.

The window parameter balances a larger exterior residual against a larger maximum cross distance. Writing $B=\lambda_2+\lambda_3$, the infimum is attained at
\begin{equation}
\min_{K>M}\frac{(B+K)^2}{K-M}=4(M+B),
\qquad K=2M+B.
\label{eq:optimal-exterior-window}
\end{equation}
Evaluating this minimum and then letting $M\downarrow b_0$ proves Eq.~\eqref{eq:finite-error-cost}. For the ratios $3:2:2$, the coefficient evaluates to $(32/21)^8(145/3)=\frac{159429186027520}{113468578083}$. This constant is optimized specifically within this right-exterior family.
\begin{equation*}
q(0.01)=\frac{99}{2302},\qquad
\underline b_1[0.01]\ge4-\frac{145}{3}k_0^8
\left(\frac{99}{2302}\right)^4>3.99519369.
\end{equation*}

For necessary support shape, assume additionally the fixed excess budget $|A|\le q_d+\xi d$, where $q_d=\ell_2+\min(\ell_1,\ell_3)+1$ and $\xi\ge0$. Choosing $M>b_0+\xi$ guarantees the primary budget. Letting $i=i(d,L)\in\{1,3\}$ index the outer gap with smaller $u_i$ and denoting by $J_i$ the corresponding three-defect interval, counting the missing spins and at most three defect endpoints gives
\begin{equation}
\begin{aligned}
\ell_i-\min(\ell_1,\ell_3)&\le\xi d+u_2+u_i+3,\\
|A\triangle J_i|/d&\le\xi+2\rho_2+2\min(\rho_1,\rho_3)+6/d.
\end{aligned}
\label{eq:error-shape}
\end{equation}
While the product constraint alone restricts the ordered limsup of the smaller outer density to $\sqrt p=O(\epsilon^2)$, a fixed strictly positive margin $\lambda_{\max}-\xi>0$ enables the supplementary inequality $u_1+u_3\ge H_{\max}-u_2-\xi d-3$ to strengthen this bound. For sufficiently small $\epsilon$, set
\begin{equation}
t=\lambda_{\max}-\xi-z>0,\qquad
x_{\mathrm{bud}}=\min\{\sqrt p,2p/t\}.
\label{eq:budget-error-allowance}
\end{equation}
On scalar tails, the sum bound holds up to arbitrarily small margins, ensuring that one outer density is at least $t/2$ up to such margins. Using the product bound, the ordered limsup of the smaller density is therefore at most $2p/t$. Consequently,
\begin{equation}
\limsup_{d\to\infty}\limsup_{L\to\infty}
\frac{|A_{d,L}\triangle J_{i(d,L)}|}{d}
\le\xi+2z+2x_{\mathrm{bud}}
=\xi+O(\epsilon^4).
\label{eq:quartic-error-shape}
\end{equation}
When the outer gap ratios differ, the sufficient margin $|\lambda_1-\lambda_3|>\xi+z+x_{\mathrm{bud}}$ selects the unique shorter branch, whereas equal limiting outer ratios permit branch alternation. These bounds establish necessary density localization under the stated recovery and budget conditions.

To deduce Eq.~\eqref{eq:fixed-error-shape} for the exact budget $s_d+1$, we set $\xi=0$, fix $M=b_0+1$ and $K=2M+\lambda_2+\lambda_3$, and evaluate the associated coefficients $Z_{M,K}$ and $X_\lambda$. Restricting $\epsilon_0$ strictly to the branch of Eq.~\eqref{eq:error-function}, we select $\epsilon_0$ sufficiently small that
\begin{equation*}
Z_{M,K}q(\epsilon_0)^4\le\lambda_{\max}/2,
\qquad
\left(Z_{M,K}+\frac{4X_\lambda}{\lambda_{\max}}\right)q(\epsilon_0)^4
<|\lambda_1-\lambda_3|
\quad\text{if }\lambda_1\ne\lambda_3.
\end{equation*}
The first inequality ensures $t\ge\lambda_{\max}/2$ across all $0<\epsilon<\epsilon_0$, including equal outer ratios, yielding $x_{\mathrm{bud}}\le4X_\lambda q(\epsilon)^4/\lambda_{\max}$. The second inequality then selects a member of $\mathfrak J$ on the ordered tails when the ratios differ. Because $q(\epsilon)/\epsilon$ is bounded on this branch up to the chosen $\epsilon_0$, Eq.~\eqref{eq:fixed-error-shape} follows with, for example,
\begin{equation*}
C^{\rm sh}_\lambda=
\left(2Z_{M,K}+\frac{8X_\lambda}{\lambda_{\max}}\right)
\sup_{0<e\le\epsilon_0}\left(\frac{q(e)}{e}\right)^4.
\end{equation*}
All parameter choices depend only on the fixed gap ratios. Because the minimum over admissible intervals is taken before either scalar limit, the selected branch need not converge.

\subsection{A conditional correlation-exponent extension}
\label{app:eta}

The count-based converse framework directly identifies the correlation properties that govern the small-error cost bound under general power-law decay. Consider a recovery family satisfying the Bell-input fixed-error condition $\liminf_{d\to\infty}\liminf_{L\to\infty}F_{d,L}(A_{d,L})\ge1-\epsilon$, with $\epsilon$ in the branch of Eq.~\eqref{eq:error-function}, under the physical subsystem access model and general CPTP decoders. Its Hermitian records are supported entirely in the discarded region, with conditional expectation values $\pm\mu$ and a shared raw second moment $\nu>0$. Suppose that residual lattice subsets $U,V$ map injectively to integer coordinates within deterministic reference containers; use the same symbols for their cardinality-preserving images. Assume further that an exact operator dictionary or state symmetry identifies the mean signal with a sum of a non-negative finite-ring distance kernel $T_{r,n}$ over paired coordinates, acting multiplicatively on products of record bilinears, including overlaps and coincident indices, prior to code-space compression, and mapping joint four-point functions to the three-pairing bound $\nu\le S_n(U)S_n(V)+2\mu^2$. Here $S_n(E)=\sum_{x,x'\in E}T_{|x-x'|,n}$ for a finite reference set $E$. Normalize the diagonal kernel element to unity, $T_{0,n}=1$, and assume fixed-distance convergence to a positive, non-increasing kernel satisfying $T_0=1$ and asymptotic decay $T_r\sim Cr^{-\eta}$ for exponent $0<\eta<1$. The test containers scale as $O(d)$ uniformly across candidate supports, with maximum cross distance satisfying $R_d/d\to R\in(0,\infty)$ along every subsequence analyzed.

Under these hypotheses, the finite-band envelope construction applies without presuming finite-ring monotonicity. Writing $S(u)$ for the infinite-chain count sum of an interval containing $u$ sites, isolating short-distance terms and evaluating the integrable continuum endpoint yields
\begin{equation}
S(u)\sim k_\eta Cu^{2-\eta},\qquad
k_\eta=2\int_0^1(1-x)x^{-\eta}\,dx
=\frac{2}{(1-\eta)(2-\eta)}.
\label{eq:eta-count-sum}
\end{equation}
For positive limiting residual densities $|U|/d\to a>0$ and $|V|/d\to c>0$, the count ratio has asymptotic lower bound $(ac)^\eta/(k_\eta^2R^{2\eta})$. Inverting the Bell-fidelity moment bound yields the necessary omission-density constraint
\begin{equation}
ac\le k_\eta^{2/\eta}R^2q(\epsilon)^{1/\eta}.
\label{eq:eta-density-error}
\end{equation}
Configurations with vanishing limiting densities satisfy this product constraint trivially, while families without density limits follow from the same compactness argument. The restriction $0<\eta<1$ ensures integrability of the endpoint integral~\eqref{eq:eta-count-sum}, producing the characteristic power-law scaling $q(\epsilon)^{1/\eta}$ for the omission-density product in Eq.~\eqref{eq:eta-density-error}. This conditional converse establishes a lower bound on the spatial recovery cost determined by model-specific gap counting; establishing a sharp price equality additionally requires constructing a matching decoder.

\section{Finite-spin optimal recovery}
\label{app:finite-recovery}

\finiteproofintro

\subsection{Attaining the bound when a Pauli is retained}

Let $V:\mathbb C^2\to\mathcal H_A\otimes\mathcal H_E$ denote a pure encoding isometry, and suppose that the accessible subsystem supports a Hermitian unitary operator $P_A$ satisfying the relation \mbox{$(P_A\otimes I_E)V=VZ$} for a designated logical basis. The orthogonal $+1$ and $-1$ eigenspaces of $P_A$ isolate the two logical basis states into disjoint local sectors. By embedding these sectors into a common auxiliary Hilbert space, which can be extended with zero directions if the sectors differ in dimension, the code states and their off-diagonal correlation operator take the block form
\begin{equation}
|\phi_+\rangle=|0\rangle|\chi_+\rangle,
\qquad
|\phi_-\rangle=|1\rangle|\chi_-\rangle,
\qquad
C_A=|0\rangle\langle1|\otimes K,
\label{eq:flag-factorization}
\end{equation}
where $K=\operatorname{Tr}_E|\chi_+\rangle\langle\chi_-|$ operates on the auxiliary Hilbert space. For any completely positive and trace-preserving recovery channel, the operator-norm contractivity of the unital adjoint map constrains the diagonal population contributions to the Bell-input squared entanglement fidelity by at most $1/2$, while the off-diagonal coherence contributions cannot exceed $\|K\|_1/2$ by the same reasoning detailed in Lemma~\ref{lem:moment}.

This theoretical ceiling can be saturated constructively. Taking the full singular value decomposition $K=\sum_\alpha s_\alpha|u_\alpha\rangle\langle v_\alpha|$ with orthonormal bases completed across zero-singular directions, let $|u_\alpha^+\rangle$ and $|v_\alpha^-\rangle$ denote the corresponding physical sector vectors, taking a vector to be zero when it lies in an added auxiliary direction. We define the decoding Kraus operators
\begin{equation}
R_\alpha=|0\rangle\langle u_\alpha^+|
+|1\rangle\langle v_\alpha^-|
\label{eq:flag-decoder-kraus}
\end{equation}
which satisfy the resolution of identity $\sum_\alpha R_\alpha^\dagger R_\alpha=I_A$. This channel preserves the diagonal population terms while aligning the off-diagonal matrix elements so that the transferred coherence trace matches $\sum_\alpha s_\alpha=\|K\|_1$, thereby establishing Eq.~\eqref{eq:flag-optimum}. The resulting recovery map is completely positive, trace preserving, and independent of the unknown logical input. When the preserved operator represents logical $X$, evaluating $C_A$ in the $X$ basis followed by a fixed output Hadamard rotation returns the output to the original logical basis without altering the fidelity. Uhlmann's transition probability theorem~\cite{Uhlmann1976TransitionProbability} furthermore identifies the transferred coherence $\|C_A\|_1=f(\rho_E^+,\rho_E^-)$ directly with the root fidelity between the reduced environmental states.

The explicit decoder also fixes the transformation law for arbitrary input states. In the eigenbasis of the preserved Pauli operator, this channel acts as a symmetric dephasing map
\begin{equation}
\mathcal D_c(\rho)=\frac{1+c}{2}\rho+\frac{1-c}{2}Z\rho Z,
\qquad c=\|C_A\|_1.
\label{eq:flag-dephasing-channel}
\end{equation}
The partial transpose of the associated normalized Bell Choi state exhibits eigenvalues $1/2,1/2,c/2,-c/2$. Whenever $c>0$, the presence of the strictly negative eigenvalue $-c/2$ rigorously certifies that the decoded output remains entangled with the external reference. Because local decoding operations cannot generate entanglement from an entanglement-breaking channel, this negative eigenvalue directly rules out an entanglement-breaking factorization for the underlying physical access channel.

\subsection{Connected supports at finite budgets}

The optimal Bell recovery fidelity for a connected cyclic segment is determined by the number of defect cores it contains. If an interval captures at least three defects, it encloses two mutually anticommuting logical Pauli representatives that permit exact unitary extraction. Conversely, whenever the complementary exterior region contains at least two defects, that exterior hosts an exact physical Pauli observable. When the connecting segment of the unobserved defect pair avoids the boundary cut, this physical operator is the standard string $S_{ab}=i\gamma_a\gamma_b$. When the complementary arc wraps through the lattice boundary, multiplying by the global parity operator $P$ eliminates the internal spins and produces the physical string
\begin{equation}
P S_{ab}=-Z_{p_a}
\left(\prod_{j<p_a\ \mathrm{or}\ j>p_b}X_j\right)Y_{p_b}.
\label{eq:wrapped-logical-string}
\end{equation}
Because the physical code space satisfies $P=+1$, this wrapped operator reproduces the logical action of the unwrapped bilinear without generating additional phase factors. In the physical sector $Q=+1$, the pairings $12$ and $34$ implement logical $Z$, the pairings $14$ and $23$ realize logical $X$, and the pairings $13$ and $24$ generate logical $Y$ and $-Y$ under the standard convention $Y=-iZX$. These identifications remain exact regardless of whether the arc traverses the Jordan--Wigner boundary cut.

Choosing code states $|\phi_\pm\rangle$ in the logical eigenbasis of the physical Pauli string $Q_E$ in the erased complement, cyclicity of the environmental partial trace yields
\begin{equation}
C_A=\operatorname{Tr}_E Q_E|\phi_+\rangle\langle\phi_-|Q_E=-C_A=0.
\label{eq:environment-flag-erasure}
\end{equation}
By Lemma~\ref{lem:exact-record}, the access channel is entanglement breaking~\cite{HorodeckiShorRuskai2003EB}, bounding the Bell fidelity of every completely positive trace-preserving decoder by $1/2$.

When the accessible subsystem retains two defects, measuring the local logical flag and conditionally preparing the corresponding state attains this $1/2$ ceiling. When the interval contains at most one defect, the environment hosts two anticommuting Pauli strings whose twirling leaves the accessible state completely independent of the logical input, yielding an optimal recovery fidelity of $1/4$. Once the segment captures three or four defects, the two supported anticommuting strings allow complete error-free recovery with fidelity one.

Under the lattice conditions of Eq.~\eqref{eq:finite-geometry-regime}, adjacent defect cores are separated by at least $2d$ bonds, while any three defects span at least $4d$ bonds. Contiguous subsystems of cardinality up to $2d$ therefore contain at most one defect, whereas those containing up to $4d$ spins contain at most two. Across the intermediate allocation window $2d+1\le B\le4d$, the explicit segment $I_B=[p_4-B+1,p_4]$ retains the full $34$ string and saturates the fidelity ceiling, establishing the classical plateau
\begin{equation}
F_{\rm conn}^*(B)=\frac12,
\qquad 2d+1\le B\le4d.
\label{eq:connected-finite-optimum}
\end{equation}
For larger allocations $B\ge4d+1$, extending the segment to encompass three defects unlocks fidelity one, completing the algebraic derivation of the three-tier recovery staircase.

\subsection{A positive endpoint response inside a perforated support}

In the critical geometry of Eq.~\eqref{eq:finite-geometry-regime}, omissions $H\subseteq G_2$ leave the physical flag $\widehat Z=i\gamma_3\gamma_4=QZ_f$ and the conjugate endpoint observable $O_X$ on $A=J_+\setminus H$, whereas $H\subseteq G_3$ preserves the pair $\widehat X=X_f=i\gamma_2\gamma_3$ and $O_Z$. In the $P=Q=+1$ dictionary, the bulk factor of $O_X$ in Eq.~\eqref{eq:conjugate-endpoints} is represented on the reference chain by $\prod_{m=p_2}^{p_3-2}\widetilde X_m$. The signed shift of Appendix~\ref{app:critical} takes this string to $\widetilde Z_{p_2}\widetilde Z_{p_3-1}$, while the bulk factor of $O_Z$ telescopes directly to $\widetilde Z_{p_3-1}\widetilde Z_{p_4-2}$. Both reference pairs are nonwrapping and have separation $2d-1$. For $V=V_{1,d,L}$ and $n=L-2$, code-space compression yields
\begin{equation}
V^\dagger O_XV=T_{2d-1,n}X,
\qquad
V^\dagger O_ZV=T_{2d-1,n}Z.
\label{eq:endpoint-coefficients}
\end{equation}
The common coefficient is positive: the disorder string is a permutation with nonnegative entries in the reference $Z$ basis, and $\Omega_n$ has strictly positive coefficients. Signed-shift invariance transfers this expectation to $T_{2d-1,n}$.

Writing $O_{\rm end}=O_X$ or $O_Z$ according to the removed gap, trace-norm duality gives $t=|\operatorname{Tr}(O_{\rm end}C_A)|\le\|C_A\|_1$, where $t=T_{2d-1,n}$. This applies to every grouping of the $4d+1-B$ omitted ordinary spins; the range $2d+2\le B\le4d$ corresponds to between one and $2d-1$ omissions in a single gap.

For nonempty $H$, an omitted site in $G_2$ pairs with a discarded site beyond $p_4$, whereas an omission in $G_3$ pairs with a site in $G_1$. The environmental dictionary gives opposite nonzero conditional means and common raw second moment one, so Lemma~\ref{lem:moment} also gives $F(A)<1$. Combining the two bounds yields
\begin{equation}
\frac12<\frac{1+T_{2d-1,n}}2
\le F_{d,L}(J_+\setminus H)<1,
\qquad 2d+2\le B\le4d.
\label{eq:perforated-finite-lower}
\end{equation}
After the thermodynamic limit, the certified endpoint margin scales as $T_{2d-1}/2\sim C2^{-5/4}d^{-1/4}$.
\subsection{Finite-spin states and physical partial traces}

Physical-spin code states are computed for $d=1$ at $L=10,12,14$ and $d=2$ at $L=18,20$. The code basis consists of the ground vector $|\psi_0\rangle$ in the sector $P=Q=Z_f=+1$ and its partner $|\psi_1\rangle=X_f|\psi_0\rangle$, with sector projection
\begin{equation}
\Pi_{+++}=\frac18(I+P)(I+Q)(I+Z_f),
\qquad \dim\operatorname{ran}\Pi_{+++}=2^{L-3}.
\label{eq:numerical-sector}
\end{equation}
\finitescanfigure

The spin flips associated with $Z_f$ and $QZ_f$ act on the disjoint intervals $[p_1,p_2-1]$ and $[p_3,p_4-1]$, whereas $P$ flips every spin globally. Because these three commuting actions are mutually independent, they partition the computational basis into eight-element orbits whose normalized projections yield an orthogonal basis of dimension $2^{L-3}$ for sparse diagonalization. From the single-particle dispersion and lowest even two-fermion excitation of the antiperiodic reference chain of length $n=L-2$, the reference ground energy and spectral gap are
\begin{equation}
E_{0,n}=-2\csc\frac{\pi}{2n},
\qquad \Delta_n=8\sin\frac{\pi}{2n}.
\label{eq:finite-reference-gap}
\end{equation}
Across all evaluated lattices, physical Hamiltonian residuals remain below $3\times10^{-13}$. In exact arithmetic, any normalized vector in this sector with energy $E<E_{0,n}+\Delta_n$ and residual $r=\|(H_D-E)|\psi\rangle\|$ has sine of its ground-state angle at most \mbox{$r/(E_{0,n}+\Delta_n-E)$}. Combining these residuals with the analytical gaps provides floating-point state-error estimates below $3\times10^{-13}$ on these lattices.

For each of the four logical matrix units, the access channel is obtained by the physical partial trace
\begin{equation}
\rho_A^{ab}=\operatorname{Tr}_{\bar A}
|\psi_a\rangle\langle\psi_b|,
\qquad a,b=0,1.
\label{eq:finite-channel-matrix-units}
\end{equation}
Reshaping the spin amplitudes into retained and discarded indices evaluates this trace by matrix multiplication, without imposing fermionic Gaussian restrictions. For an $X$ flag, the correlation operator is evaluated using the Hadamard superpositions of the code states; completing the singular-vector bases across null directions and applying the fixed output basis rotation yields the decoder of Eq.~\eqref{eq:flag-decoder-kraus}, attaining the trace-norm optimum to the reported precision.

The complementary environmental witness is given by the product operator $O=(\sum_{j\in U}Z_j)(\sum_{k\in V}Z_k)$. For middle-gap omissions, $U$ denotes the single omitted site and $V=\{p_4+1,\ldots,L\}$, whereas for right-gap omissions, $U=G_1$ while $V$ contains the omitted site. Evaluating the conditional expectations of $O$ across opposite $Z_f$ eigenstates for a middle-gap hole and opposite $X_f$ eigenstates for a right-gap hole yields equal and opposite means with identical raw second moments within numerical precision. The resulting moment bounds supply the upper benchmark curves displayed in Fig.~\ref{fig:finite-recovery}.

Evaluating all single-gap omissions on the finite ring with $d=2,L=20$ accounts for every configuration within $G_2=\{9,10,11\}$ and $G_3=\{13,14,15\}$ at budgets $B=6,7,8$, which yield two, six and six distinct physical supports. Applying the exact-flag polar decoder to the same numerically prepared physical states attains the trace-norm optima to the reported precision. The endpoint response evaluates to $T_{3,18}\simeq0.4949368401$, giving a common analytical floor of approximately $0.7474684201$ for this finite-ring family.

\subsection{Contiguous holes and collective receiver records}
\label{app:receiver-record}

When ordinary sites are excised along a contiguous interval $H=[a,a+k-1]\subseteq G_2$, the operator $\widehat X\prod_{m\in H}X_m$ acts strictly on the accessible subsystem $J_+\setminus H$. On the full physical sector, its bulk component translates into $\prod_{q=a-1}^{a+k-2}\widetilde X_q$, which the signed cyclic shift maps directly onto the two-point order observable $\widetilde Z_{a-1}\widetilde Z_{a+k-1}$. A matching construction applies to intervals $H\subseteq G_3$ through the unitary operator $\widehat Z\prod_H X_m$, whose bulk representation telescopes into $\widetilde Z_{a-2}\widetilde Z_{a+k-2}$. Because the intact opposite gap retains an exact logical flag and both correlators evaluate to the strictly positive quantity $T_{k,n}>0$, trace-norm duality directly proves Eq.~\eqref{eq:contiguous-hole-floor}. At fixed $k$, taking the successive thermodynamic and large-separation limits confirms the asymptotic lower bound $(1+T_k)/2$, which reaches $1/2+1/\pi$ for an isolated single-site removal.

\begin{equation}
F_{d,L}(J_+\setminus H)\ge\frac{1+T_{k,L-2}}2,
\qquad |H|=k,\quad1\le k\le2d-1.
\label{eq:contiguous-hole-floor}
\end{equation}

For the macroscopic family introduced in Eq.~\eqref{eq:macroscopic-hole-family}, we define $r=r_d$, $U=[p_2,p_2+r-1]$ and $K=[p_3-r+1,p_3]$. For each index pair, we assemble the physical Pauli operators
\begin{equation}
\begin{aligned}
O_{jk}&=\widehat X\prod_{m=j+1}^{k-1}X_m\\
&=-Y_{p_2}\left(\prod_{m=p_2+1}^{j}X_m\right)
\left(\prod_{m=k}^{p_3-1}X_m\right)Z_{p_3},\\
O_A&=\sum_{j\in U,k\in K}O_{jk}.
\end{aligned}
\label{eq:receiver-observable}
\end{equation}
All spin factors associated with omitted sites cancel identically, ensuring that the collective observable $O_A$ is supported entirely on the retained spins. Across the full physical sector, each term evaluates to the logical Pauli operator $X_f$ multiplied by the reference disorder string covering sites $j,\dots,k-2$. Applying the signed cyclic transformation to operator products maps their bulk components onto $\widetilde Z_j\widetilde Z_{k-1}$ across all overlapping segments and coincident indices. Consequently, expectations in the logical $X$ eigenstates yield opposite values $\pm\mu$, while operator products eliminate the logical Pauli prior to evaluating expectations, ensuring an identical raw second moment in both states,
\begin{equation}
\mu=\sum_{j\in U,k\in K}T_{k-j-1,n},\qquad
\nu\le S_n(r)^2+2\mu^2.
\label{eq:receiver-moments}
\end{equation}
The two reference index sets each contain $r$ consecutive integers. Evaluating the zero-field Lebowitz inequality across the four-point correlation functions bounds the internal pairings by $S_n(r)^2$ and each of the two cross pairings by $\mu^2$, with coincident indices satisfying $T_{0,n}=1$.

Because the intact $Z$ flag isolates the logical code states in orthogonal local sectors, the correlation operator $C_A$ is strictly block off-diagonal, ensuring that the trace distance separating the reduced outputs of opposite logical $X$ states equals $\|C_A\|_1$. Considering a spectral measurement of $O_A$ generates outcome distributions $p_o$ and $q_o$ with conditional means $\pm\mu$ and raw second moment $\nu$. Applying weighted Cauchy--Schwarz along with the contractivity of trace distance yields
\begin{equation}
\begin{aligned}
\operatorname{TV}(p,q)&:=\tfrac12\sum_o|p_o-q_o|\quad\text{(total variation distance)},\\
(2\mu)^2&\le\left(\sum_o|p_o-q_o|\right)
\left(\sum_o o^2|p_o-q_o|\right)
\le4\operatorname{TV}(p,q)\nu,\\
c=\|C_A\|_1&\ge\operatorname{TV}(p,q)\ge\frac{\mu^2}{\nu}.
\end{aligned}
\label{eq:receiver-discrimination}
\end{equation}
This distinguishability inequality bounds the transferred coherence available to the receiver, which the polar decoder converts directly into coherent reference entanglement.

The coordinate separations between the two boundary blocks span the range from $2d-2r+1$ to $2d-1$. For $r=r_d\ge1$, let $m_{d,n}$ denote the minimum of the reference correlation $T_{s,n}$ over this index window on a chain of length $n=L-2$. Because the correlation kernel is nonnegative, the mean obeys $\mu\ge r^2m_{d,n}$, while the container sum satisfies $S_n(r)>0$. Combining Eqs.~\eqref{eq:receiver-moments} and~\eqref{eq:receiver-discrimination} with the monotonic increase of $x/(S_n(r)^2+2x)$ on $x\ge0$ establishes the finite-ring recovery certificate
\begin{equation}
\begin{aligned}
c_{d,L}:=\|C_{A_d}\|_1
&\ge\frac{\mu^2}{S_n(r)^2+2\mu^2}
\ge\frac{r^4m_{d,n}^2}{S_n(r)^2+2r^4m_{d,n}^2},\\
F_{d,L}(A_d)
&\ge\frac12+\frac{r^4m_{d,n}^2}{2(S_n(r)^2+2r^4m_{d,n}^2)}.
\end{aligned}
\label{eq:receiver-finite-certificate}
\end{equation}
This expression bounds the designated two-component geometry on finite rings satisfying Eq.~\eqref{eq:finite-geometry-regime}, with $r_d\ge1$.

Taking the thermodynamic limit $L\to\infty$ at fixed defect separation $d$ causes $m_{d,n}$ to converge to $T_{2d-1}$ and $S_n(r)$ to converge to $S(r)$ by pointwise convergence and monotonicity on the infinite chain. Subsequently taking the macroscopic limit with $r/d\to(b-2)/2$ cancels the nonuniversal correlation amplitude via Eq.~\eqref{eq:critical-sum},
\begin{equation}
\liminf_{d\to\infty}\liminf_{L\to\infty}
\frac{\mu^2}{S_n(r)^2}
\ge\lim_{d\to\infty}\frac{r^4T_{2d-1}^2}{S(r)^2}
=\frac{\sqrt{b-2}}{2k_0^2}.
\label{eq:receiver-ratio}
\end{equation}
Because the function $z/(1+2z)$ is increasing, evaluating the finite-ring certificate in Eq.~\eqref{eq:receiver-finite-certificate} under the asymptotic limit of Eq.~\eqref{eq:receiver-ratio} establishes the joint-polar lower bound with $x_b=441/2048$ and $c_b=441/2930$ at $b=3$. This lower bound on the unrestricted joint-polar optimum does not yet establish the single-spin guarantee of Eq.~\eqref{eq:constant-advantage}; that requires showing that the physical measurement instrument transfers coherence at least $\mu^2/\nu$. Appendix~\ref{app:one-way-recovery} establishes this coherence bound and details the one-way single-spin implementation.

\subsection{One-way recovery and single-spin implementation}
\label{app:one-way-recovery}

We first prove a one-way recovery bound for a general pure encoding with an exact flag on the right block. Let an isometry $V$ encode a logical qubit into the tripartite space $A_L A_R E$, where $A_R$ supports a Hermitian unitary $P_R$ that represents logical $Z$ exactly, $(I_L\otimes P_R\otimes I_E)V=VZ$. Suppose $L_A$ and $R_A$ are Hermitian and their product $L_A\otimes R_A$ has conditional expectations $\pm\mu$ and the same raw second moment $\nu>0$ on logical $X$ eigenstates. Alice performs a spectral measurement of $L_A$ using projectors $\Pi_s$, transmits the discrete outcome $s$ to Bob, and discards her quantum system. Because $P_R$ acts on an independent tensor factor, the flag on $A_R$ remains intact. For logical code states $|\phi_a\rangle=V|a\rangle$, the subnormalized off-diagonal coherence blocks are
\begin{equation}
C_s=\operatorname{Tr}_{A_LE}
\bigl[(\Pi_s\otimes I)|\phi_0\rangle\langle\phi_1|
(\Pi_s\otimes I)\bigr].
\label{eq:one-way-coherence-blocks}
\end{equation}
The classical register keeps different outcomes orthogonal, so the coherence operator of the register and Bob's block is the direct sum of the $C_s$, with trace norm $c_{\rightarrow}=\sum_s\|C_s\|_1$. Since the flag is exact, conditional polar decoding attains $F_{\rightarrow}=(1+c_{\rightarrow})/2$ as in Eq.~\eqref{eq:flag-optimum}. Alice's nonselective measurement pinches the algebra while leaving $L_A$ and $L_A^2$ invariant, ensuring that the joint readout formed by $s$ and a spectral measurement of $R_A$ on Bob reproduces the original distribution moments $\pm\mu$ and $\nu$. Let $p$ and $q$ be the joint readout laws for logical $X=\pm1$ inputs. The retained $Z$ flag makes the trace distance between the two pre-readout receiver states equal to $c_{\rightarrow}$. Weighted Cauchy--Schwarz and trace-distance contractivity then give $c_{\rightarrow}\ge\operatorname{TV}(p,q)\ge\mu^2/\nu$, as in Eq.~\eqref{eq:receiver-discrimination}. Conditional polar recovery on every outcome therefore achieves $F_{\rightarrow}\ge1/2+\mu^2/(2\nu)$ through a trace-preserving channel.

In the critical Ising encoding on the two-component support $A_d=A_L\cup A_R$, the collective receiver observable decomposes into a genuine physical product. Expanding each operator $O_{jk}$ in Eq.~\eqref{eq:receiver-observable} verifies the tensor-factor split $L_A\otimes R_A$ with $R_A=K_RZ_{p_3}$, as displayed in Eq.~\eqref{eq:one-way-product-record}. Measuring $Y_{p_2}$ alongside $X_m$ for $p_2<m\le p_2+r-1$ provides a family of mutually commuting single-spin Pauli observables whose products generate every term in $L_A$. After Alice's block is traced out, summing the fine Pauli measurement branches with the same value $s$ reproduces the coarse spectral instrument of $L_A$, because the partial trace eliminates cross terms between orthogonal projectors. The coarse spectral readout takes values in $\{-r,-r+2,\ldots,r\}$, requiring at most $\lceil\log_2(r+1)\rceil$ classical bits of communication. On Bob's subsystem, single-spin measurements of $X$ across $[p_3-r+1,p_3-1]$ and $G_3$, together with $Z_{p_4}$, determine the value $t$ of $K_R$ and the collective parity $w=(\prod_{m\in G_3}x_m)z_{p_4}$.

Every measured observable commutes with both $X_f$ and the flag $P_R=\widehat Z=i\gamma_3\gamma_4$, so each joint projector compresses to a scalar on the code space, $V^\dagger\Pi_\lambda V=p_\lambda I$ for each combined branch $\lambda=(s,t,w)$. Consequently, the branch probability $p_\lambda$ is completely independent of the encoded input. Conditioned on the outer parity record $w$, the exact logical flag on the surviving spin $p_3$ reduces to $-wY_{p_3}$. To rotate this flag into the computational basis, Bob applies the single-qubit Clifford unitary
\begin{equation}
\begin{aligned}
|u_w\rangle&=\frac{|0\rangle-iw|1\rangle}{\sqrt2},
&|v_w\rangle&=Z|u_w\rangle,\\
U_w&=|0\rangle\langle u_w|+|1\rangle\langle v_w|,
&U_w(-wY)U_w^\dagger&=Z,\\
&&U_wZU_w^\dagger&=X.
\end{aligned}
\label{eq:one-way-output-basis}
\end{equation}
Under this rotation, branch $\lambda$ maps the population matrix elements according to $|a\rangle\langle a|\mapsto p_\lambda|a\rangle\langle a|$ and transforms the off-diagonal coherence as $|0\rangle\langle1|\mapsto a_\lambda|0\rangle\langle1|$, with branch coefficient $a_\lambda=\langle\phi_1|\Pi_\lambda Z_{p_3}|\phi_0\rangle$.

An antiunitary symmetry $\Theta=(\prod_{a=1}^4 Z_{p_a})\mathsf K$ protects the reality of the branch coefficients, where $\mathsf K$ denotes complex conjugation in the physical computational $Z$ basis. The operator $\Theta$ squares to the identity, commutes with the defect Hamiltonian $H_D$, and preserves the spin parity $P$ and the defect charge $Q$. On each defect interaction $Z_{p-1}Y_p$, complex conjugation reverses the sign of $Y_p$, which is immediately restored by conjugation under $Z_p$. The antiunitary operator also preserves the logical generators $X_f$ and $Z_f$, every measured projector, and the output operator $Z_{p_3}$. The unique bulk ground state leaves a one-dimensional ground space at each fixed logical $Z$ eigenvalue. Choose $|\phi_0\rangle$ to be $\Theta$ invariant. Since $X_f$ is also invariant, $|\phi_1\rangle=X_f|\phi_0\rangle$ has the same property, making every $a_\lambda$ real.

Applying a conditional Pauli $Z$ correction based on the precomputed sign $\varepsilon_\lambda=\operatorname{sgn}a_\lambda$ aligns the off-diagonal coherence across all branches. Summing over all corrected measurement branches produces the trace-preserving symmetric dephasing channel $\mathcal D_{c_{\rm meas}}$ of Eq.~\eqref{eq:flag-dephasing-channel}. Each unnormalized branch is completely positive and trace nonincreasing, and the corrected sum is trace preserving. Complete positivity ensures $|a_\lambda|\le p_\lambda$, so weighted Cauchy--Schwarz applied to the branch normalization and receiver moments of Eq.~\eqref{eq:one-way-branch-moments} yields Eq.~\eqref{eq:one-way-moment-certificate} for this single-spin instrument. Alice's outcome $s$ is the only message to Bob, who applies $U_w$ and the sign correction locally from $s,t,w$.

The single-spin protocol obeys the same finite-ring lower bounds as Eq.~\eqref{eq:receiver-finite-certificate} for $c_{\rm meas}$ and $F_{\rm meas}$. The mean estimate $\mu\ge r^2m_{d,n}$ and the raw second-moment upper bound $\nu\le S_n(r)^2+2\mu^2$ are unchanged. Taking the thermodynamic limit $L\to\infty$ at fixed $d$, followed by the scaling limit $d\to\infty$ with $r/d\to(b-2)/2$, establishes Eqs.~\eqref{eq:constant-advantage} and \eqref{eq:constant-capacity}.
Here $R_{\rm meas}=1-H_2((1-c_{\rm meas})/2)$ is the rate achievable by coding over independent corrected qubit channels, with the coding limit taken at each fixed finite geometry before either spatial limit. At $b=3$, the certified lower bounds are approximately $0.575256$ in fidelity and $0.016404$ qubits per block.

On contiguous three-defect intervals, single-spin Pauli measurements and classical feedforward reconstruct the logical qubit with unit fidelity. For $J_+=[p_2,p_4]$, the receiver measures $Y_{p_2}$, $X$ on every site of $G_2$ and $G_3$, and $Z_{p_4}$, keeping spin $p_3$ as the output. In terms of the records $w_{\rm par}=-y_{p_2}\prod_{G_2}x_m$ and $w=(\prod_{G_3}x_m)z_{p_4}$, the logical generators $X_f$ and $Z_f$ are then represented on $p_3$ by $w_{\rm par} Z_{p_3}$ and $-wY_{p_3}$, respectively. Applying $U_w$ and then a Pauli $Z$ correction when $w_{\rm par}=-1$ maps both generators to standard Pauli operators on $p_3$. The other interval $J_-=[p_1,p_3]$ shifts this pattern one defect to the left, with $Y_{p_1}$, $X$ on $G_1$ and $G_2$, and $Z_{p_3}$ measured and spin $p_2$ retained as the output. Its records $w_{\rm par}=-y_{p_1}\prod_{G_1}x_m$ and $w=(\prod_{G_2}x_m)z_{p_3}$ reduce $Z_f$ to $w_{\rm par} Z_{p_2}$ and $X_f$ to $-wY_{p_2}$. Because $U_w$ here carries $X_f$ rather than $Z_f$ to the output $Z$ axis, an output Hadamard rotation follows it, and an $X$ correction when $w_{\rm par}=-1$ completes the exact extraction. The shorter interval therefore attains the finite upper bound of Eq.~\eqref{eq:exact-upper} with single-spin measurements and classical feedforward. Since the lower bound in Eq.~\eqref{eq:sharp-price} holds for arbitrary recovery channels, this restricted decoding has the same sharp leading spatial cost $b_0d$.

Numerical evaluation on the benchmark lattice at $d=2,L=20$ quantifies the fidelity hierarchy across decoding models. On the six-spin support of Table~\ref{tab:protocol}, contracting all measurement branches in the single-spin protocol yields a fidelity of $0.7474684201$. Retaining Bob's multi-spin quantum subsystem under conditional polar decoding achieves $0.7820244535$, while unrestricted joint decoding reaches the fixed-support optimum of $0.7839874045$. These calculations use all four logical matrix units of the same physical-spin code states and include every outcome of each measurement-based protocol.

\subsection{Fixed-target and vanishing-error costs}
\label{app:fixed-target-recovery}

For the critical benchmark with defect ratios $3:2:2$, the selected two-block family has $3d+O(1)$ spins at $b=3$ and ordered lower fidelity at least $1/2+441/(2\cdot2930)\approx0.575256>0.55$. Strict slack means it attains $F\ge0.55$ eventually in $d$, then on each sufficiently large-$L$ tail. Hence
\begin{equation}
\limsup_{d\to\infty}\limsup_{L\to\infty}
\frac{m_{\rm crit}(0.55;d,L)}{d}\le3.
\label{eq:precision-critical-useful}
\end{equation}
The arbitrary-support price theorem combines the filling converse under linear budgets with exact interval recovery. For every nonnegative sequence $\epsilon_d\to0$, so that $1-\epsilon_d>1/2$ eventually, it gives
\begin{equation}
\lim_{d\to\infty}\liminf_{L\to\infty}
\frac{m_{\rm crit}(1-\epsilon_d;d,L)}{d}
=\lim_{d\to\infty}\limsup_{L\to\infty}
\frac{m_{\rm crit}(1-\epsilon_d;d,L)}{d}=4.
\label{eq:precision-critical-faithful}
\end{equation}
Consequently, $\liminf_{d\to\infty}\liminf_{L\to\infty}[m_{\rm crit}(1-\epsilon_d;d,L)-m_{\rm crit}(0.55;d,L)]/d\ge1$. Subtract the fixed-target upper envelopes from the high-precision lower envelopes, first in $L$ and then in $d$. The selected critical support provides a sufficient footprint for the fixed target.

\section{Calibration and independent report errors}\label{supp:finite-calibration}

\subsection{Setup and unbiased response estimation}

Consider a fixed finite geometry and its ideal all-outcome single-spin measurement instrument. Let $\Lambda$ denote the known, finite set of possible coarse labels $\lambda=(s,t,w)$, with
$M_{\rm br}=|\Lambda|\le2(r+1)^2$. The set includes labels of zero probability and is not replaced by the subset observed in a dataset. After the branch-dependent Clifford rotation, branch $\lambda$ has input-independent diagonal weight $p_\lambda\ge0$ and real off-diagonal response $a_\lambda$, satisfying
\begin{equation}
\sum_{\lambda\in\Lambda}p_\lambda=1,\qquad
|a_\lambda|\le p_\lambda.
\end{equation}
The ideal branch-sign correction yields channel coherence $c_{\rm meas}=\sum_{\lambda\in\Lambda}|a_\lambda|$. These parameters are fixed properties of the encoding, geometry and instrument, not of a subsequent unknown logical input.

Prepare $N$ independent and identically distributed (IID) copies of the known logical state $|+\rangle$. On each disposable calibration copy, record the coarse label $\lambda_j$ and output-$X$ readout $x_j\in\{-1,+1\}$ after the branch-dependent Clifford rotation but before sign correction. The estimator and its moments are
\begin{equation}
\begin{aligned}
Y_{j,\lambda}&=\mathbf1_{\{\lambda_j=\lambda\}}x_j,
&\widehat a_\lambda&=\frac1N\sum_{j=1}^NY_{j,\lambda},\\
\mathbb E[Y_{j,\lambda}]&=a_\lambda,
&\mathbb E[Y_{j,\lambda}^2]&=p_\lambda,\\
\mathrm{Var}(\widehat a_\lambda)&=\frac{p_\lambda-a_\lambda^2}{N}
\le\frac{p_\lambda}{N}.
\end{aligned}
\end{equation}
Here the subnormalized branch output has off-diagonal entry $a_\lambda/2$.
The estimator measures an unnormalized response directly, without division by a small branch probability. Choose the correction signs by the fixed rule
\begin{equation}
\widehat\epsilon_\lambda=
\begin{cases}
+1,&\widehat a_\lambda\ge0,\\
-1,&\widehat a_\lambda<0.
\end{cases}
\end{equation}
In particular, a zero estimate specifies the physical correction sign $+1$.

\subsection{Confidence bound on the estimation error}

\begin{theorem}
For every $0<\delta<1$, with probability at least $1-\delta$ over the calibration data,
\begin{equation}
\|\widehat a-a\|_1
\le\frac{\sqrt{M_{\rm br}}+\sqrt{2\ln(1/\delta)}}{\sqrt N}.
\end{equation}
\end{theorem}

\begin{proof}
Applying Jensen's inequality and Cauchy--Schwarz to the component variance bound yields
\begin{equation}
\mathbb E\|\widehat a-a\|_1
\le\sum_{\lambda\in\Lambda}\sqrt{\frac{p_\lambda}{N}}
\le\sqrt{\frac{M_{\rm br}}{N}}.
\end{equation}
Regard $f=\|\widehat a-a\|_1$ as a function of the $N$ independent sample vectors $Y_j=(Y_{j,\lambda})_{\lambda\in\Lambda}$. Because each vector has one nonzero entry of magnitude one, replacing a single sample vector changes $f$ by at most $\|Y_j-Y_j'\|_1/N\le2/N$. The bounded-differences inequality for independent variables~\cite{McDiarmid1989BoundedDifferences,McDiarmid1998Concentration} therefore yields, for $u>0$,
\begin{equation}
\Pr\{f-\mathbb Ef\ge u\}
\le\exp\left(-\frac{2u^2}{N(2/N)^2}\right)
=e^{-Nu^2/2}.
\end{equation}
Set $u=\sqrt{2\ln(1/\delta)/N}$ and combine with the mean bound. The fixed alphabet includes zero-probability labels, so the estimate requires neither a minimum branch probability nor empirical coverage of every label.
\end{proof}

\subsection{Decoded channel fidelity and sample complexity}

For a fixed estimated response table $\widehat a$, use the correction signs $\widehat\epsilon$ defined above. The actual coherence is $c_{\rm est}=\sum_{\lambda\in\Lambda}\widehat\epsilon_\lambda a_\lambda$, while the ideal value is $c_{\rm meas}=\sum_{\lambda\in\Lambda}|a_\lambda|$. Only wrong signs lose coherence:
\begin{equation}
c_{\rm meas}-c_{\rm est}
=2\sum_{\lambda:\,\widehat\epsilon_\lambda a_\lambda<0}|a_\lambda|
\le2\|\widehat a-a\|_1.
\end{equation}
A wrong sign costs $2|a_\lambda|$ and entails $|a_\lambda|\le|\widehat a_\lambda-a_\lambda|$, including a negative response estimated as zero under the $+1$ tie rule. Zero responses contribute no loss. The identity and inequality are deterministic for every estimated table. Since the dephasing-channel Bell fidelity is $F=(1+c)/2$,
\begin{equation}
e_{d,L}=\|\widehat a-a\|_1,\qquad
F_{\rm est}\ge F_{\rm meas}-e_{d,L}
\ge F_{\rm meas}
-\frac{\sqrt{M_{\rm br}}+\sqrt{2\ln(1/\delta)}}{\sqrt N}.
\label{eq:one-way-calibration}
\end{equation}
The first inequality is deterministic; the second holds on the calibration confidence event and bounds the decoded channel associated with the learned table. Training data are independent of later unknown inputs. Additive fidelity loss at most $\eta>0$ with confidence $1-\delta$ is ensured by
\begin{equation}
N\ge
\left\lceil
\frac{(\sqrt{M_{\rm br}}+\sqrt{2\ln(1/\delta)})^2}{\eta^2}
\right\rceil.
\end{equation}
At fixed accuracy $\eta$ and confidence $1-\delta$, $M_{\rm br}\le2(r+1)^2=O(d^2)$ gives sufficient budgets of $O(d^2)$ known-state copies and $O(d^3)$ helper readouts. These are sufficient offline budgets; state-preparation and gate costs are excluded.

For fixed $2<b<4$, choose $0<\tau_0<c_b/2$. On each geometry's confidence event, $e_{d,L}\le\tau_0$ certifies $c_{\rm est}\ge c_{\rm meas}-2\tau_0$; the instrument bound then gives positive Bell and dephasing-rate certificates on the ordered large-size tails. The condition $\lim_{d\to\infty}\limsup_{L\to\infty}e_{d,L}=0$ suffices to certify the ideal ordered-limit floor from this loss bound. Actual equality needs only correct signs on nonzero responses: the fixed table $\widehat a_\lambda=a_\lambda/2$ gives $c_{\rm est}=c_{\rm meas}$ even though $e_{d,L}=c_{\rm meas}/2$ need not vanish. A fixed nonzero tolerance generally gives a reduced certified floor through the bound alone; $c_b\to0$ as $b\downarrow2$ precludes a uniform positive tolerance from this criterion.

Confidence is per geometry. Over a countable family, the union bound yields simultaneous confidence of at least $1-\delta$ when positive failure budgets satisfy $\sum_{d,L}\delta_{d,L}\le\delta$ and the sample sizes are chosen accordingly. When calibration data estimate a response vector $a'$ rather than the target $a$, the triangle inequality adds the systematic term $\|a'-a\|_1$ to the sampling error bound; additional samples reduce sampling fluctuations, not this offset.

\subsection{Independent classical report errors}\label{supp:report-errors}

\subsubsection{Single-shot classical error propagation}

Assume ideal nonadaptive Pauli projections, ideal output gates and noiseless classical transmission and feedforward. Each reported helper-measurement bit is flipped independently with probability $p_{\rm flip}\in[0,1/2)$. The true quantum outcomes need not be independent.

For complete-interval extraction on $[p_2,p_4]$, true Pauli outcomes $y_{p_2},x_j,z_{p_4}\in\{\pm1\}$ give the records
\begin{equation}
w_{\rm par}=-y_{p_2}\prod_{j\in G_2}x_j,\qquad
w=\left(\prod_{j\in G_3}x_j\right)z_{p_4}.
\end{equation}
Their disjoint helper-readout sets are $\{p_2\}\cup G_2$ and $G_3\cup\{p_4\}$, with sizes $\ell_2$ and $\ell_3$, respectively. Independence of the report flips makes the two error parities independent. A block of $n_{\rm bits}$ readouts has odd error parity with probability
$\frac12[1-(1-2p_{\rm flip})^{n_{\rm bits}}]$.

A wrong $w_{\rm par}$ parity produces an unwanted output Pauli $Z$. A wrong $w$ parity substitutes $U_{-w}$ for $U_w$; since $U_{-w}=XU_w$, it produces an unwanted Pauli $X$, up to an irrelevant phase. Two wrong parities produce the third nonidentity Pauli rather than canceling. The four Pauli errors give orthogonal Bell states, so the Bell entanglement fidelity equals the probability that both error parities are even:
\begin{equation}
F_e=
\frac{[1+(1-2p_{\rm flip})^{\ell_2}][1+(1-2p_{\rm flip})^{\ell_3}]}4.
\end{equation}
This exact formula is restricted to complete-interval extraction under the stated report-noise model; it is not asserted for perforated mixed outputs or general noise.

\subsubsection{Quantum non-demolition repetition and diamond-norm robustness}

Suppose each helper permits repeated quantum non-demolition readout in the same Pauli basis, with no state transitions and independent report flips on every read. After the first ideal projection, its true eigenvalue remains fixed. Majority voting over an odd number $m$ of reports has error probability
\begin{equation}
p_{\rm eff}
=\Pr\left\{\sum_{k=1}^mB_k\ge\frac{m+1}{2}\right\}
\le e^{-2m(1/2-p_{\rm flip})^2},
\end{equation}
where the $B_k$ are independent Bernoulli variables of mean $p_{\rm flip}$.

Couple the ideal and noisy protocols on $n_{\rm meas}$ helpers to the same true outcomes and an arbitrary inert reference. Majority failure depends only on report flips, independently of the input and true outcomes, with probability $q\le n_{\rm meas}p_{\rm eff}$ by the union bound. Without failures the full outputs agree. Averaging the failed error patterns defines a CPTP map $\mathcal M$, so $\mathcal N_{\rm noisy}=(1-q)\mathcal N_{\rm ideal}+q\mathcal M$. The maximum channel diamond distance two and Bell-fidelity linearity and nonnegativity give
\begin{equation}
\begin{aligned}
\frac12\|\mathcal N_{\rm noisy}-\mathcal N_{\rm ideal}\|_\diamond
&\le q\le n_{\rm meas}e^{-2m(1/2-p_{\rm flip})^2},\\
F_{\rm noisy}&\ge F_{\rm ideal}-q.
\end{aligned}
\end{equation}
No independence of the coarse-record errors $(s,t,w)$ is needed. Alice completes her majorities before sending $s$ once, while Bob forms $t,w$ locally.

\subsubsection{Readout complexity and scaling}

For $0<\eta<1$, a sufficient choice is an odd integer $m$ satisfying
\begin{equation}
m\ge\frac{2}{(1-2p_{\rm flip})^2}
\ln\left(\frac{n_{\rm meas}}{\eta}\right),
\qquad m\ge1.
\end{equation}
This ensures $q\le\eta$. With $n_{\rm meas}=O(d)$ and $m$ chosen at this sufficient scale, the total report count is
\begin{equation}
N_{\rm read}=m\,n_{\rm meas}
=O\left(\frac{d\ln(d/\eta)}{(1-2p_{\rm flip})^2}\right).
\end{equation}
The dependence on $p_{\rm flip}$ can be absorbed into the constant only when $p_{\rm flip}$ stays bounded away from $1/2$.

Parallel readout takes $m$ measurement rounds. These report-count and round bounds exclude propagation and aggregation time, state preparation, gate errors, nonideal backaction and calibration mismatch.

\section{Capacity of the full flagged access channel}
\label{app:access-capacity}

Let a pure encoding $V$ satisfy $(P_A\otimes I)V=VZ$ for a Hermitian involution $P_A$ supported on $A$. Write $|\phi_a\rangle=V|a\rangle$, $\Pi_a=(I+(-1)^aP_A)/2$ and $\tau_a=\operatorname{Tr}_A|\phi_a\rangle\langle\phi_a|$. Conjugation by $P_A\otimes I$ reverses the sign of $V|0\rangle\langle1|V^\dagger$, so its partial trace over $A$ vanishes. The complement and degrading map are therefore
\begin{equation}
\mathcal N_E(\rho)=\sum_{a=0}^1\rho_{aa}\tau_a,
\qquad
\mathcal T(\omega)=\sum_{a=0}^1\operatorname{Tr}(\Pi_a\omega)\tau_a,
\qquad
\mathcal T\circ\mathcal N_A=\mathcal N_E.
\label{eq:explicit-degrader}
\end{equation}
Because each state $\tau_a$ is normalized and $\Pi_0+\Pi_1=I_A$ resolves the identity on $A$, the map $\mathcal T$ is CPTP; the relation $\mathcal T\circ\mathcal N_A=\mathcal N_E$ then proves that $\mathcal N_A$ is degradable. Separately, the measure-and-prepare structure of the complementary channel $\mathcal N_E$ renders it entanglement breaking, establishing that $\mathcal N_A$ belongs to the Hadamard class~\cite{BradlerHaydenTouchetteWilde2010Hadamard}.

Write $S(\rho)=-\operatorname{Tr}(\rho\log_2\rho)$ for entropy in bits. A degradable channel has single-letter quantum capacity equal to the maximum coherent information~\cite{DevetakShor2005CQ}. The concavity needed below follows from data processing~\cite{YardHaydenDevetak2008MAC}: for a classical input-mixture register $\mathsf X$, the concavity gap is $I(\mathsf X:A)-I(\mathsf X:E)\ge0$, where $I(\mathsf X:A)=S(\mathsf X)+S(A)-S(\mathsf XA)$ denotes quantum mutual information, with the same definition for $E$.

For $0\le q\le1$, let $\chi(\{q,\tau_0;1-q,\tau_1\})=S(q\tau_0+(1-q)\tau_1)-qS(\tau_0)-(1-q)S(\tau_1)$ denote the Holevo information, with von Neumann entropies measured in bits. The flag symmetry permits input pinching without reducing coherent information. Exploiting the mutual orthogonality of the local sectors alongside the purity of the global code states yields the single-variable optimization
\begin{equation}
Q(\mathcal N_A)=\max_{0\le q\le1}
\left[H_2(q)-\chi\!\left(\{q,\tau_0;1-q,\tau_1\}\right)\right].
\label{eq:general-flag-capacity}
\end{equation}
The flag condition alone does not force the maximum to occur at $q=1/2$. For example, the flagged states $|0\rangle_F|0\rangle_B|0\rangle_E$ and $|1\rangle_F(|00\rangle_{BE}+|11\rangle_{BE})/\sqrt2$ on accessible subsystem $A=FB$ produce the objective function $H_2(q)+(1-q)-H_2((1+q)/2)$, whose derivative at $q=1/2$ is strictly nonzero.

For the critical spin encoding considered here, the global logical operators $X$ and $Z$ are physical Pauli strings that factorize across any spatial partition as $G_W=G_{W,A}\otimes G_{W,E}$. Their exact action on the code space endows both output channels with covariance,
\begin{equation}
\mathcal N_s(W\rho W^\dagger)
=G_{W,s}\mathcal N_s(\rho)G_{W,s}^\dagger,
\qquad s=A,E,\quad W=X,Z.
\label{eq:access-covariance}
\end{equation}
The output entropies are invariant under logical Pauli operations. Concavity of coherent information then makes the four-Pauli average $I/2$ optimal. A fixed logical-basis rotation handles a retained $X$ flag. The same argument applies to flagged access in the pure matched Ashkin--Teller ladder, whose physical $\sigma$-Pauli strings have the same spatial factorization.

For the maximally entangled Bell state $|B\rangle=(|00\rangle+|11\rangle)/\sqrt2$ on reference $R$ and channel input, let $J_{RA}=(\operatorname{id}_R\otimes\mathcal N_A)(|B\rangle\langle B|)$ denote the Choi state. Purity of the global state on $RAE$ gives $S(J_{RA})=S(\mathcal N_E(I/2))$. Orthogonality of the conditional retained sectors, together with equality of the retained and environmental marginal entropies for each pure conditional code state, then yields the Holevo representation in Eq.~\eqref{eq:full-access-capacity}:
\begin{equation}
\begin{split}
Q(\mathcal N_A)&=S(\mathcal N_A(I/2))-S(J_{RA})\\
&=1-\chi\!\left(\{\tfrac12,\tau_0;\tfrac12,\tau_1\}\right),\\
J_{RA}&=(\operatorname{id}_R\otimes\mathcal N_A)(|B\rangle\langle B|).
\end{split}
\label{eq:full-access-capacity}
\end{equation}

The polar decoder compresses the output into a single-qubit symmetric dephasing channel $\mathcal D_c$ with coherence $c=\|C_A\|_1$. Its normalized Choi eigenvalues are $(1+c)/2,(1-c)/2,0,0$, so
\begin{equation}
 Q(\mathcal D_c)=1-H_2((1-c)/2)\le Q(\mathcal N_A)
\label{eq:polar-capacity}
\end{equation}
The inequality is data processing under the polar recovery map. It compares a fixed-support Bell-optimal compression with the original multi-spin access channel; it need not be an equality.

The single-spin instrument gives another CPTP postprocessing and hence $Q(\mathcal N_{A_d})\ge Q(\mathcal D_{c_{\rm meas}})=R_{\rm meas}(A_d)$. Coding independent blocks at fixed geometry before the ordered spatial limits and using Appendix~\ref{app:one-way-recovery} proves the asymptotic rate floor in Eq.~\eqref{eq:constant-capacity}, in qubits per independent block. With learned signs, the rate uses the actual coherence $c_{\rm est}$ of Appendix~\ref{supp:finite-calibration}. For fixed $2<b<4$, tolerance below $c_b/2$ gives per-geometry rate certificates with a positive ordered lower limit on their respective calibration confidence events.

Using the archived physical-spin states and all four matrix units of Eq.~\eqref{eq:finite-channel-matrix-units}, we construct $\mathcal N_A(I/2)$ and $J_{RA}$ and apply the complete polar decoder, including the output Hadamard for an $X$ flag. Before evaluating base-two entropies, we Hermitize the density matrices, clip only negative machine-roundoff eigenvalues, retain all positive eigenvalues and normalize each spectrum. Table~\ref{tab:access-capacity} compares the two capacities on the same finite supports.

\begin{table}[H]
\caption{Unassisted physical-access and polar-decoded quantum capacities on perforated geometries at $d=2,L=20$, with retained spin sets $A=J_+\setminus H$ of cardinality $B$. Values are computed from numerically diagonalized finite-spin code states, using the same states as in the recovery calculation, in qubits per independent encoding block. The differences illustrate the capacity loss of Bell-optimal polar compression on these supports.}
\label{tab:access-capacity}
\centering
\begin{tabular}{c c c c}
$H$ & $B$ & $Q(\mathcal N_A)$ & $Q(\mathcal D_c)$\\
\hline
$\{9,10,11\}$ & $6$ & $0.309217$ & $0.247181$\\
$\{9,10\}$ & $7$ & $0.362441$ & $0.291208$\\
$\{10\}$ & $8$ & $0.540315$ & $0.436358$
\end{tabular}
\end{table}

\section{Two-chain encodings and support allocation}
\label{sec:composition}\label{app:two-chain}

Comparing two encodings on a shared pair of uncoupled defect chains demonstrates how the choice of logical subspace governs the spatial resources required for quantum recovery. Both encodings share defect coordinates, composite Hamiltonian, and product bulk ground states, with physical lattice spins on either layer weighted equally under the total footprint budget. The required recovery regions diverge because each encoding embeds the logical qubit into a distinct fusion subspace rather than a rotated basis of a shared code.

\subsection{Two encodings and exact interval extraction}

The host system comprises two decoupled copies of the four-defect critical Ising chain, labeled by $c\in\{\sigma,\tau\}$ and governed by the Hamiltonian $H_D^\sigma+H_D^\tau$ in the fixed-charge sector $P_\sigma=Q_\sigma=P_\tau=Q_\tau=+1$. Each layer hosts fusion Pauli operators $X_c, Z_c$ and reference bulk ground state $\Omega_{c,n}$, yielding the product bulk ground state $\Omega_{\sigma,n}\otimes\Omega_{\tau,n}$. In the localized benchmark, the logical qubit is stored entirely in the fusion doublet of chain $\sigma$, while the fusion mode of chain $\tau$ is initialized in a fixed, known fiducial state. Because an unconstrained decoder can prepare any subsystem of the inert $\tau$ chain locally, measurements on $\tau$ provide no useful correlations for reconstructing the logical qubit. The leading spatial footprint for vanishing-error recovery therefore matches the isolated-chain price $b_0=\lambda_2+m$, where $m=\min(\lambda_1,\lambda_3)$.

The cross-chain repetition architecture distributes the logical qubit across both chains by mapping into the joint fusion subspace defined by $X_\sigma X_\tau=+1$. This embedding is implemented by the linear isometry
\begin{equation}
C|0\rangle=\frac{|00\rangle+|11\rangle}{\sqrt2},\qquad
C|1\rangle=\frac{|01\rangle+|10\rangle}{\sqrt2}.
\label{eq:cross-code}
\end{equation}
On this code subspace, the logical Pauli operators intertwine with the physical fusion operators as
\begin{equation}
X_\sigma C=X_\tau C=CX,\qquad
Z_\sigma Z_\tau C=CZ,\qquad
C^\dagger Z_\sigma C=C^\dagger Z_\tau C=0.
\label{eq:cross-logical}
\end{equation}
The code forms a two-qubit repetition code hosted within the degenerate ground manifold of the two-layer system. While the basis codewords in Eq.~\eqref{eq:cross-code} are maximally entangled across the two fusion modes, superpositions need not possess bipartite entanglement, and the critical bulk remains strictly factorized. Subsystem access structures in quantum secret sharing and error correction dictate which shares permit reconstruction~\cite{CleveGottesmanLo1999QSS,GheorghiuLooiGriffiths2010InformationLocation,Gottesman2000SecretSharingTheory}; here the microscopic defect dictionary translates that algebraic structure into an optimization over physical lattice supports.

For an input-independent retained physical-spin set $A$ across both layers, let $F^{\rm cross}_{d,L}(A)$ be the Bell-input squared entanglement fidelity optimized over all joint CPTP decoders on the retained spins, and let $n^{\rm cross}_{*,L}(d,\epsilon)$ be the minimum retained-spin count with $F^{\rm cross}_{d,L}(A)\ge1-\epsilon$.

An exact finite-ring extractor achieves reconstruction by retaining a shortest three-defect interval on one chain and a shortest outer two-defect interval on the companion chain. Let $F_{ab}^c=i\gamma_a^c\gamma_b^c$ ($a<b$) denote the defect bilinears on chain $c$. Supported on either three-defect segment $J_1^c=[p_1,p_3]$ or $J_3^c=[p_2,p_4]$, $F_{23}^c$ directly represents logical $X$ via Eq.~\eqref{eq:cross-logical}. On an outer two-defect interval $K_1^c=[p_1,p_2]$ or $K_3^c=[p_3,p_4]$, the sector constraint $Q_c=+1$ identifies $F_{12}^c$ or $F_{34}^c$ with $Z_c$. The physical product of two such outer strings across distinct chains implements $Z_\sigma Z_\tau$, intertwining with logical $Z$. Because these localized physical Hermitian involutions square to identity, commute with the respective noninteracting bulk terms, and anticommute on the retained spin algebra, Lemma~\ref{lem:exact-extraction} yields reference-preserving decoding with
\begin{equation}
n^{\rm cross}_{*,L}(d,0)\le \ell_2+2\min(\ell_1,\ell_3)+2.
\label{eq:cross-upper}
\end{equation}

\subsection{Complementary product records and bulk moments}

Resource allocation between the two chains partitions the accessible lattice spins into sets $A_\sigma$ and $A_\tau$ subject to the aggregate budget $|A_\sigma|+|A_\tau|\le Md$. On each chain, defect locations $p_1<p_2<p_3<p_4$ define the ordinary gaps $G_i=(p_i,p_{i+1})\cap\mathbb Z$ of lengths $H_i=\ell_i-1$ for $i=1,2,3$. With the exterior window parameter $e_d=\lceil(M+1)d\rceil$, we define the exterior index sets $E_d=(p_4,p_4+e_d]\cap\mathbb Z$ on each chain, choosing the ring size $L$ large enough that $E_d$ does not cross the boundary seam. Unobserved sites within gap $G_i$ form the subsets $U_i^c=G_i\setminus A_c$ with counts $u_i^c=|U_i^c|$, while $V_c=E_d\setminus A_c$ collects omissions in $E_d$. Discarded spins govern environmental fluctuations through the reference operators $B_c^X=\sum_{j\in U_1^c,k\in U_3^c}D_{j,k}^c$ and $B_c^Z=\sum_{j\in U_2^c,k\in V_c}\widetilde Z_{j-1}^c\widetilde Z_{k-2}^c$, where $D_{j,k}^c$ denotes the disorder string of Eq.~\eqref{eq:two-dictionaries}.

The physical environmental observables are $O_c^X=(\sum_{j\in U_1^c}Z_j^c)(\sum_{k\in U_3^c}Z_k^c)$ and $O_c^Z=(\sum_{j\in U_2^c}Z_j^c)(\sum_{k\in V_c}Z_k^c)$, each supported entirely on discarded spins. On the full fixed-charge sector, the dictionary gives $O_c^X=-X_cB_c^X$ and $O_c^Z=-Z_cB_c^Z$. Because the code isometry enforces $X_\sigma C=X_\tau C=CX$, an environment inspecting either layer alone distinguishes the logical $X$ eigenstates whenever both outer gaps on that layer have nonvanishing omission densities. Vanishing-error recovery therefore requires at least one outer gap to be filled in density on each chain.

Conversely, individual single-layer $Z_c$ compressions vanish identically because $C^\dagger Z_c C=0$ by Eq.~\eqref{eq:cross-logical}. Capturing environmental $Z$ leakage requires forming the product observable on the physical spin algebra prior to code compression:
\begin{equation}
O_\sigma^ZO_\tau^Z
=(Z_\sigma Z_\tau)B_\sigma^ZB_\tau^Z.
\label{eq:cross-product-record}
\end{equation}
Here $\mu_c^\alpha=\langle B_c^\alpha\rangle_{\Omega_{c,n}}$ is the nonnegative single-chain conditional signal magnitude and $\nu_c^\alpha=\langle(B_c^\alpha)^2\rangle_{\Omega_{c,n}}$ its raw second moment, defined before restriction to the cross-chain fusion code. Evaluating the composite observable on the joint logical eigenstates $|\pm\rangle_Z$, the fusion factor $Z_\sigma Z_\tau$ acts as $\pm 1$, while bulk expectations factorize over $\Omega_{\sigma,n}\otimes\Omega_{\tau,n}$:
\begin{equation}
\begin{aligned}
\langle O_\sigma^ZO_\tau^Z\rangle_\pm
&=\pm\mu_\sigma^Z\mu_\tau^Z,\\
\langle (O_\sigma^ZO_\tau^Z)^2\rangle_\pm
&=\nu_\sigma^Z\nu_\tau^Z.
\end{aligned}
\label{eq:cross-product-moments}
\end{equation}
These raw second moments include contact terms with coincident lattice indices, as established in Sec.~\ref{sec:critical}, without invoking a Wick expansion between fusion codewords. The resulting distinguishability ratio factorizes as
\begin{equation}
R_{\rm prod}=R_\sigma^ZR_\tau^Z,\qquad
R_c^Z=\frac{(\mu_c^Z)^2}{\nu_c^Z}.
\label{eq:cross-ratio}
\end{equation}
This factorization is exact because the underlying bulk ground state is a product, despite entanglement in the fusion codewords.

\subsection{Uniform finite-ring bounds and necessary gap filling}

Under the total budget $|A_\sigma|+|A_\tau|\le Md$, at least $d$ unvisited sites remain within each exterior test window $E_d$. In terms of normalized omission densities $\rho_i^c=u_i^c/d$, the finite-ring coefficients read
\begin{equation}
\begin{aligned}
c^X_{d,n}&=\frac{d^4m_{X,d,n}^2}{S_n(H_1)S_n(H_3)},\\
c^Z_{d,n}(M)&=\frac{d^4m_{Z,d,n}^2}{S_n(H_2)S_n(e_d)},
\end{aligned}
\label{eq:cross-coefficients}
\end{equation}
where $m_{Z,d,n}$ and $m_{X,d,n}$ are the correlation infima evaluated across the finite distance bands in Eqs.~\eqref{eq:z-band} and \eqref{eq:x-band}. Setting $r(x)=x/(1+2x)$ and $\Phi(y)=(1+\sqrt{1-y})/2$, single-chain positivity and four-point inequalities yield $R_c^Z\ge r(c^Z_{d,n}(M)(\rho_2^c)^2)$ for nonvanishing central omissions. Inserting these bounds into Eq.~\eqref{eq:cross-product-moments} and setting $\mathcal B=\Phi\circ r$ provides the finite-ring fidelity ceiling
\begin{equation}
\begin{split}
F^{\rm cross}_{d,L}(A)\le\min\bigl\{&
\mathcal B(c^X_{d,n}(\rho_1^\sigma\rho_3^\sigma)^2),\\
&\mathcal B(c^X_{d,n}(\rho_1^\tau\rho_3^\tau)^2),\\
&\Phi\bigl[r(c^Z_{d,n}(M)(\rho_2^\sigma)^2)
 r(c^Z_{d,n}(M)(\rho_2^\tau)^2)\bigr]\bigr\}.
\end{split}
\label{eq:cross-finite-bound}
\end{equation}
When any omission count vanishes, the corresponding branch evaluates to unity. Because Eq.~\eqref{eq:cross-finite-bound} depends only on macroscopic container sums and distance infima, it holds uniformly across all support geometries and for arbitrary joint CPTP decoders acting across both retained layers.

In the thermodynamic limit $n\to\infty$ at fixed $d$, followed by the large-separation limit $d\to\infty$, the scalar coefficients converge to the strictly positive constants
\begin{equation}
\begin{aligned}
c_X&=\frac{1}{k_0^2(\lambda_1\lambda_3)^{7/4}
\sqrt{\lambda_1+\lambda_2+\lambda_3}},\\
c_Z(M)&=\frac{1}{k_0^2[\lambda_2(M+1)]^{7/4}
\sqrt{\lambda_2+\lambda_3+M+1}}.
\end{aligned}
\label{eq:cross-limiting-coefficients}
\end{equation}
Consider any support family $A_{d,L}$ satisfying $|A_{d,L}|\le Md$ for a fixed finite $M$ on the successive large-$d$ and large-$L$ tails and $\lim_{d\to\infty}\liminf_{L\to\infty}F^{\rm cross}_{d,L}(A_{d,L})=1$. The relevant omission densities are $\min(u_1^c,u_3^c)/d$ for each $c=\sigma,\tau$ and $\min(u_2^\sigma,u_2^\tau)/d$. If any of these three quantities has a positive ordered $\limsup_{d\to\infty}\limsup_{L\to\infty}$, the corresponding branch of Eq.~\eqref{eq:cross-finite-bound} gives a uniform nonzero fidelity deficit along a subsequence, contradicting the success condition. Thus every such family obeys
\begin{equation}
\begin{aligned}
\lim_{d\to\infty}\limsup_{L\to\infty}
\frac{\min(u_1^c,u_3^c)}d&=0\quad(c=\sigma,\tau),\\
\lim_{d\to\infty}\limsup_{L\to\infty}
\frac{\min(u_2^\sigma,u_2^\tau)}d&=0.
\end{aligned}
\label{eq:cross-filling}
\end{equation}
The central-gap condition requires only the minimum across the two layers to vanish, reflecting the fact that covering the central gap on a single chain suffices to eliminate the composite $Z$ witness. These are necessary macroscopic-density conditions, not sufficient recovery criteria for arbitrary perforated supports; sublinear omissions are not excluded by this converse.

\subsection{All-support price converse and near-minimal shape}

For an arbitrary support $A=A_\sigma\cup A_\tau$, summing retained spins across the outer gap with minimal omissions on each chain and the middle gap with minimal omissions across both chains yields the geometric lower bound
\begin{equation}
\begin{split}
|A|\ge \ell_2+2\min(\ell_1,\ell_3)-3
&-\min(u_1^\sigma,u_3^\sigma)\\
&-\min(u_1^\tau,u_3^\tau)
-\min(u_2^\sigma,u_2^\tau).
\end{split}
\label{eq:cross-count}
\end{equation}
To establish the price converse, set $b_{\rm cross}=\lambda_2+2m$, fix a budget deficit $0<\delta<b_{\rm cross}$, and define $a=\delta/4$. For sufficiently large $d$, any support satisfying $|A|\le\lfloor(b_{\rm cross}-\delta)d\rfloor$ forces at least one of the three omission counts in Eq.~\eqref{eq:cross-count} to exceed $ad$. Evaluating Eq.~\eqref{eq:cross-finite-bound} with budget $M=b_{\rm cross}$ then delivers the uniform supremum bound
\begin{equation}
\begin{split}
\sup_{|A|\le\lfloor(b_{\rm cross}-\delta)d\rfloor}
F^{\rm cross}_{d,L}(A)\le\max\bigl\{&
\mathcal B(a^4c^X_{d,n}),\\
&\Phi[r(a^2c^Z_{d,n}(b_{\rm cross}))^2]\bigr\}.
\end{split}
\label{eq:cross-uniform-gap}
\end{equation}
For each fixed $\delta$, the ordered limits leave a strictly positive fidelity gap below unity for all sufficiently large $d$ and, at each such $d$, all sufficiently large $L$. Every nonnegative sequence $\epsilon_d\to0$ eventually falls below this gap, excluding all sub-minimal supports on these ordered tails. Matching this converse with the exact interval extractor in Eq.~\eqref{eq:cross-upper} as $\delta\to0$ proves the thermodynamic envelopes
\begin{equation}
\begin{split}
\lim_{d\to\infty}\liminf_{L\to\infty}
\frac{n^{\rm cross}_{*,L}(d,\epsilon_d)}d
&=\lambda_2+2m,\\
\lim_{d\to\infty}\limsup_{L\to\infty}
\frac{n^{\rm cross}_{*,L}(d,\epsilon_d)}d
&=\lambda_2+2m.
\end{split}
\label{eq:cross-price}
\end{equation}

For support families satisfying the same vanishing-error condition and the near-minimal budget
\begin{equation}
\limsup_{d\to\infty}\limsup_{L\to\infty}|A_{d,L}|/d
\le b_{\rm cross},
\label{eq:cross-near-budget}
\end{equation}
let $i_\sigma,i_\tau\in\{1,3\}$ index the outer gaps with minimal omission counts on chains $\sigma$ and $\tau$, and let $c_*$ designate the chain with fewer central omissions. A reference geometry $B$ is defined by assigning the three-defect interval $J_{i_{c_*}}$ to chain $c_*$ and the two-defect interval $K_{i_{\bar c_*}}$ to the companion chain $\bar c_*$, with $J_1=J_-$ and $J_3=J_+$. The necessary filling conditions bound missing spins within $B$ by
\begin{equation}
|B\setminus A|\le
\min(u_1^\sigma,u_3^\sigma)+\min(u_1^\tau,u_3^\tau)
+\min(u_2^\sigma,u_2^\tau)+5=o(d).
\label{eq:cross-missing}
\end{equation}
The constant $+5$ counts the defect sites in the two intervals, and $o(d)$ is understood in the ordered thermodynamic and scaling limits.
When outer scaling ratios are asymmetric ($\lambda_1\ne\lambda_3$), retaining the longer outer interval on either chain incurs an asymptotic cost penalty of $|\lambda_1-\lambda_3|d$, violating Eq.~\eqref{eq:cross-near-budget}. Only the shorter outer branches can therefore survive on the ordered tails, leaving two candidate geometries depending on which layer houses the central segment. When outer ratios are balanced ($\lambda_1=\lambda_3$), outer gap choices decouple and either chain may contain the middle segment, producing eight candidate configurations of identical leading cost. Letting $\mathfrak B$ denote this set of two or eight candidates, the active branch may alternate with $d$ and $L$. Because every candidate satisfies $|B|/d\to b_{\rm cross}$, the relation $|A\triangle B|=|A|-|B|+2|B\setminus A|$ confirms the density concentration
\begin{equation}
\lim_{d\to\infty}\limsup_{L\to\infty}
\frac{\min_{B\in\mathfrak B}|A_{d,L}\triangle B|}{d}=0.
\label{eq:cross-necessary-shape}
\end{equation}

\begin{table}[t]
\caption{Leading spatial costs and macroscopic gap-filling requirements for two fusion encodings realized on identical, uncoupled critical defect chains. Both encodings share the defect locations, Hamiltonian, product bulk ground states, and uniform retained-spin cost metric. In the ordered thermodynamic and large-separation limits, the filling conditions constrain all support families that achieve vanishing error under a linear retained-spin budget. Extraction unitaries attain the leading coefficients with support on a single contiguous interval for the localized encoding and on one contiguous interval per layer for the cross-chain encoding. Here $m=\min(\lambda_1,\lambda_3)$, and $u_i^c$ is the count of omitted physical spins within gap $G_i$.}
\label{tab:two-chain-comparison}
\centering
\setlength{\tabcolsep}{4pt}
\begin{tabular}{>{\raggedright\arraybackslash}p{0.25\textwidth}>{\raggedright\arraybackslash}p{0.43\textwidth}>{\raggedright\arraybackslash}p{0.22\textwidth}}
\hline\hline
Fusion encoding & Necessary gap filling & Leading price coefficient\\
\hline
Unknown $\sigma$ qubit, fixed $\tau$ fusion state & \raggedright $u_2^\sigma=o(d)$ and \mbox{$\min(u_1^\sigma,u_3^\sigma)=o(d)$} & $\lambda_2+m$\\[5pt]
Cross-chain code $X_\sigma X_\tau=+1$ & \raggedright \mbox{$\min(u_1^c,u_3^c)=o(d)$} for each $c=\sigma,\tau$, and \mbox{$\min(u_2^\sigma,u_2^\tau)=o(d)$} & $\lambda_2+2m$\\
\hline\hline
\end{tabular}
\end{table}

Evaluating the bounds in Table~\ref{tab:two-chain-comparison} for the benchmark profile $(\lambda_1,\lambda_2,\lambda_3)=(3,2,2)$ illustrates the macroscopic impact of the logical embedding. The localized encoding requires a leading footprint of $4d+o(d)$ spins, whereas the cross-chain repetition code demands $6d+o(d)$ spins. The additional leading cost of $md=2d$ spins stems from the requirement to fill an outer gap on the companion layer, while the matching extractor fills the central gap on only one layer. The choice of logical subspace within the degenerate ground manifold thus shifts the physical recovery budget across identical many-body environments.

\section{Conditional extensions and an interacting realization}
\label{app:interacting}\label{sec:interacting}

Whereas Appendix~\ref{app:two-chain} embeds the logical space across independent bulks, we here introduce bulk interactions while storing the logical qubit in one fusion factor. Coupling two chains via a matched Ashkin--Teller ladder~\cite{AshkinTeller1943,RossiniVicari2026StackedIsing,YanKonikMitra2025DeformedDuality} preserves exact interval extraction alongside the finite-ring separation between connected and perforated access geometries (Sec.~\ref{app:interacting-finite}). The corresponding macroscopic recovery cost depends on discrete-lattice moment bounds formulated below. The following construction separates exact finite-ring statements from the lattice assumptions needed for a macroscopic result.

\subsection{Matched local terms and the periodic seam}

Under the positive-sign convention on each physical layer, the local generators comprise the Pauli operators
\begin{equation}
\mathcal K=\{X_j:j\notin D\}\cup
\{Z_jZ_{j+1}:j+1\notin D\}\cup
\{Z_{p-1}Y_p:p\in D\}.
\label{eq:at-keys}
\end{equation}
Denoting the members of $\mathcal K$ by $b_k$, we couple matching terms across physical layers $\sigma$ and $\tau$ through the ladder Hamiltonian
\begin{equation}
H_{D,g}=-\sum_{k\in\mathcal K}(b_k^\sigma+b_k^\tau+g b_k^\sigma b_k^\tau).
\label{eq:interacting-physical}
\end{equation}
Each term acts on at most two adjacent rungs, with exchange interactions incorporating the periodic boundary bond $(L,1)$. For a chain hosting four defect cores, $\mathcal K$ contains $2n=2L-4$ operators. Ordering these generators cyclically along the active chain yields $\beta_r^c=i\alpha_r^c\alpha_{r+1}^c$ for $r<2n$, closed by the boundary term $\beta_{2n}^c=-iP_c\alpha_{2n}^c\alpha_1^c$. Because each generator commutes with $P_c$, $Q_c$, and the even defect algebra on its respective layer, the coupled Hamiltonian~\eqref{eq:interacting-physical} decouples within the positive-parity sector $P_\sigma=Q_\sigma=P_\tau=Q_\tau=+1$ as
\begin{equation}
H_{D,g}\simeq I_{f\sigma f\tau}\otimes
H_{\mathrm{AT}}^{\mathrm{ref}}(g)\big|_{P_{b,\sigma}=P_{b,\tau}=+1},
\label{eq:at-factorization}
\end{equation}
governed by the unprojected reference Hamiltonian
\begin{equation}
\begin{split}
H_{\mathrm{AT}}^{\mathrm{ref}}(g)=
&-\sum_{j=1}^n(\widetilde X_j^\sigma+\widetilde X_j^\tau
+g\widetilde X_j^\sigma\widetilde X_j^\tau)\\
&-\sum_{j=1}^n(\widetilde Z_j^\sigma\widetilde Z_{j+1}^\sigma
+\widetilde Z_j^\tau\widetilde Z_{j+1}^\tau
+g\widetilde Z_j^\sigma\widetilde Z_{j+1}^\sigma
\widetilde Z_j^\tau\widetilde Z_{j+1}^\tau).
\end{split}
\label{eq:at-reference}
\end{equation}
Both sums obey periodic boundary conditions. The bulk density reproduces the standard quantum Ashkin--Teller Hamiltonian~\cite{Chepiga2022PottsTransverseBoundary,ObrienBartlettDohertyFlammia2015QAT,LiuEtAl2026ATBoundary}, while boundary interactions are fixed by the physical parity dictionary. Defining $B_{n,c}=\widetilde Z_n^c\widetilde Z_1^c=-iP_{b,c}\alpha_{2n}^c\alpha_1^c$, the terminal physical operator takes the form $Q_cB_{n,c}$. Prior to fixing background charges to $+1$, the boundary interaction reads
\begin{equation}
-\bigl(Q_\sigma B_{n,\sigma}+Q_\tau B_{n,\tau}
+gQ_\sigma Q_\tau B_{n,\sigma}B_{n,\tau}\bigr).
\label{eq:at-seam}
\end{equation}
Inter-layer couplings originating from defect terms, such as $-gZ_{p-1}^\sigma Y_p^\sigma Z_{p-1}^\tau Y_p^\tau$, act on at most four physical spins across neighboring rungs and introduce no nonlocal Jordan--Wigner strings. Including a Klein factor to ensure anticommutation of odd Majoranas across layers leaves these even physical operators invariant.

For non-negative coupling $g\ge0$, the reference operator~\eqref{eq:at-reference} is stoquastic in the joint $\widetilde Z$ basis. Because single transverse flips connect all configuration basis states, the Perron--Frobenius theorem guarantees a unique, strictly positive ground state on finite rings. Global spin inversions on either layer act as commuting non-negative permutations preserving this state, placing both bulk parities in the $+1$ sector. The factorization~\eqref{eq:at-factorization} thus produces two independent fusion factors over the interacting bulk. Initializing the auxiliary $\tau$ factor in a pure reference state isolates the logical qubit within the $\sigma$ fusion doublet.

Logical Pauli operators on the $\sigma$ chain retain their exact support on either three-defect segment. The unitary extractor constructed in Appendix~\ref{app:encoding} acts directly on this interval, leaving an unobserved residual state that may possess inter-layer entanglement. Simultaneously applying the signed shift $\mathcal W_\sigma\mathcal W_\tau$ permutes matched local terms cyclically while preserving active parities, stabilizing the unique reference ground state up to an overall phase. This symmetry transformation maps disorder strings and their operator products on the $\sigma$ chain onto the corresponding order observables, realizing paired-core fusion across the coupled ladder via simultaneous defect displacements.

\subsection{Finite connected and perforated ladder access}
\label{app:interacting-finite}

For the benchmark defect geometry~\eqref{eq:finite-geometry-regime}, connectivity is defined via nearest-neighbor leg bonds and vertical rungs, with each retained spin on either layer counting toward the budget. Across finite rings, connected geometries satisfy the exact performance ceiling
\begin{equation}
F_{{\rm conn},g}^*(B)=1/2,\qquad 2d+1\le B\le4d,
\label{eq:interacting-connected}
\end{equation}
whereas a nonempty single-gap excision $H$ from $J_+$ on the $\sigma$ layer achieves
\begin{equation}
\begin{split}
F_g((J_+\setminus H)\times\{\sigma\})&\ge(1+t_{g,n})/2>1/2,\\
t_{g,n}&=\left\langle\prod_{r=p_2}^{p_3-2}\widetilde X_r^\sigma\right\rangle_{\Omega_{g,n}},
\end{split}
\label{eq:interacting-finite-gap}
\end{equation}
for $2d+2\le B\le4d$, where $\Omega_{g,n}$ denotes the interacting reference ground state and $F_g$ the optimal Bell fidelity for the encoded $\sigma$ qubit.

To establish these properties for $g\ge0$, we restrict to the sector where all physical charges and parities equal $+1$, fixing the $\tau$ fusion factor in a pure reference state. The logical qubit is thereby encoded over the correlated bulk state $\Omega_{g,n}$. For any connected ladder subset $A$, projecting vertically onto the chain yields a contiguous cyclic interval $\pi(A)$ spanning at most $|A|$ columns. Under the budget restriction $|A|\le4d<L$, this projection forms a proper subsegment of the periodic ring. Because adjacent defect columns are separated by at least $2d$ rungs, covering three defect cores requires at least $4d+1$ rungs. The unaccessed complement therefore encloses at least two defect cores on layer $\sigma$ linked by an intact $\sigma$ string, including boundary-wrapping strings resolved by positive $\sigma$ parity. Tracing out the unobserved environment as in Eq.~\eqref{eq:environment-flag-erasure} collapses the accessible channel into a measure-and-prepare map, enforcing the fidelity bound of $1/2$. A contiguous single-layer block $[p_4-B+1,p_4]$ supporting $i\gamma_3^\sigma\gamma_4^\sigma$ (logical $Z_\sigma$) saturates this bound across the entire budget window~\eqref{eq:interacting-connected}.

Perforating the support by removing non-defect sites and discarding all $\tau$ spins circumvents this entanglement-breaking constraint. This construction preserves the intact gap flag while isolating the defect endpoints of the punctured gap, preventing rung-to-rung connectivity across the excised region. In terms of bulk observables, excising spins within $G_2$ produces the conjugate boundary response given by the disorder expectation value in Eq.~\eqref{eq:interacting-finite-gap}, whereas excising within $G_3$ yields the two-point function of $\widetilde Z_{p_3-1}^\sigma\widetilde Z_{p_4-2}^\sigma$. Under the simultaneous signed shift on both active layers, this disorder string maps onto $\widetilde Z_{p_2}^\sigma\widetilde Z_{p_3-1}^\sigma$. Applying a simultaneous spatial translation of both reference chains by $2d-1$ rungs identifies this correlator with the conjugate boundary response. Because these operations leave the interacting reference Hamiltonian, both parities, and the unique ground state invariant, the two boundary coefficients match identically.

Perron--Frobenius properties on the finite ladder ensure that this boundary response is strictly positive. For any $g\ge0$, the reference Hamiltonian is stoquastic in the joint $\widetilde Z$ basis, and single-spin transverse terms connect all configurations, yielding an entrywise positive ground vector. Because the disorder operator acts as a non-negative permutation matrix, its expectation value decomposes into a sum of strictly positive ground-state amplitudes, certifying $t_{g,n}>0$. Joint duality then guarantees the positivity of the magnetic order correlator, which general ground-state positivity alone would not ensure for signed diagonal observables. Projecting the accessible conjugate observable onto the code space yields $t_{g,n}$ times the conjugate logical Pauli operator. Combining trace-norm duality with the intact flag guarantees that the channel parameter satisfies $c_g\ge t_{g,n}$, establishing Eq.~\eqref{eq:interacting-finite-gap}. For a Bell-state input, the polar decoder outputs a state with a negative partial-transpose eigenvalue $-c_g/2<0$, certifying genuine bipartite entanglement and demonstrating that the perforated channel remains non-entanglement-breaking across all finite rings.

\subsection{Positive moments and arbitrary residual sets}

Let $M_C=\sum_{j\in C}\widetilde Z_j^\sigma$ and $H_{g,n}(C,D)=\langle M_C^2M_D^2\rangle_{g,n}$. Denote by $m_{g,d,n}$ the infimum of the two-point correlation function over the cross-distance band separating two containers of size $O(d)$. For the thermodynamic envelopes $m_{g,d}=\liminf_n m_{g,d,n}$ and $H_g=\limsup_n H_{g,n}$, our macroscopic conclusions are conditional on the asymptotic scaling hypothesis
\begin{equation}
 \liminf_{d\to\infty}d^{1/4}m_{g,d}>0,\qquad
 \limsup_{d\to\infty}d^{-7/2}H_g(C_d,D_d)<\infty.
\label{eq:interacting-input}
\end{equation}
These conditions constitute explicit assumptions on the discrete lattice kernels and raw moments for each relevant record geometry, rather than immediate consequences of the continuum theory. When these inputs hold, the two-axis gap count combined with the exact $\sigma$-interval decoder establishes the sharp leading cost $b_0$ and forces near-minimal supports to concentrate in density on a shortest three-defect interval, admitting at most $o(d)$ auxiliary-layer spins.

Positivity of correlation functions can be verified via an imaginary-time Trotter formulation without invoking free-fermion Wick contractions. For discrete time step $s>0$, define $c_s=\cosh s$, $s_s=\sinh s$, $C_s=\cosh(gs)$, and $S_s=\sinh(gs)$. The local transfer operator $\exp[s(X^\sigma+X^\tau+gX^\sigma X^\tau)]$ generates transition amplitudes between configuration states corresponding to zero, one, or two spin flips:
\begin{equation}
\begin{aligned}
w_0&=c_s^2C_s+s_s^2S_s,\\
w_1&=c_ss_s(C_s+S_s),\\
w_2&=s_s^2C_s+c_s^2S_s,\\
w_0-w_2&=e^{-gs},\qquad
w_0w_2-w_1^2=C_sS_s\ge0.
\end{aligned}
\label{eq:at-trotter-weights}
\end{equation}
Expressing these matrix elements in terms of classical Ising spins $z,z',t,t'=\pm1$ yields a strictly positive scalar prefactor multiplied by $\exp[K_s(zz'+tt')+U_szz'tt']$, where the effective couplings are
\begin{equation}
K_s=\frac14\log\frac{w_0}{w_2}>0,
\qquad U_s=\frac14\log\frac{w_0w_2}{w_1^2}\ge0.
\label{eq:at-trotter-couplings}
\end{equation}
Spatial bond and four-spin couplings equal $s$ and $gs$, both non-negative. Expanding each Boltzmann factor as $\cosh J[1+\tanh J\prod z]$ yields non-negative coefficients upon tracing over all classical configurations. Every even correlation function on the $\sigma$ layer therefore possesses a non-negative numerator and a positive partition function denominator. Taking the continuous-time Trotter limit followed by the zero-temperature limit at fixed ring size $n$ isolates the strictly positive ground state without introducing a thermal parity projection, ensuring correlation positivity on finite rings and in the thermodynamic limit.

For a consecutive reference interval $C$ away from the chosen seam, the finite-prefix strings $q_j=\prod_{1\le k<j}\widetilde X_k^\sigma$ for $j\in C$ cancel their shared prefix to the left of $C$ in every product $q_jq_k$. The squared sum $(\sum_{j\in C}q_j)^2$ is therefore supported on the $|C|-1$ reference sites between the endpoints of $C$, and its raw moments are finite-support reference-spin expectations.

Environmental records are formulated prior to taking bulk expectations. Joint duality maps disorder-string products directly to order observables, accounting for all contact terms. The resulting conditional expectation values evaluate to $\pm\mu$ alongside a shared raw second moment $\nu_{U,V}$. Termwise positivity of four-point correlators ensures that moments on arbitrary residual subsets $U\subset C$ and $V\subset D$ are bounded by the full-container moment:
\begin{equation}
\nu_{U,V}\le H_{g,n}(C,D)
\le\sqrt{\langle M_C^4\rangle_{g,n}\langle M_D^4\rangle_{g,n}}.
\label{eq:at-raw-domination}
\end{equation}
The recovery converse relies on the first inequality. For any residual subsets with cardinalities $|U|\ge\alpha d$ and $|V|\ge\beta d$, the finite-ring recovery fidelity obeys
\begin{equation}
F_{d,L}^{g}(A)\le
\frac{1+\sqrt{1-\alpha^2\beta^2d^4m_{g,d,n}^2/H_{g,n}(C,D)}}2.
\label{eq:at-uniform-converse}
\end{equation}
Because $\mu^2\le\nu\le H$, the radicand remains real and non-negative on every non-empty residual configuration. This bound applies to all joint decoders across the accessible spins prior to optimizing over the support geometry. For each scale $d$, defining the asymptotic envelopes $m_{g,d}=\liminf_n m_{g,d,n}$ and $H_g(C_d,D_d)=\limsup_n H_{g,n}(C_d,D_d)$ and taking the thermodynamic limit transfers the inequality directly to the envelopes on the tail where $m_{g,d}>0$. Under the scaling hypothesis~\eqref{eq:interacting-input}, the conditional record ratio remains strictly bounded away from zero along every positive-density omission branch.

\subsection{Separated containers and the lattice matching requirement}

Consider two spatial containers of lengths $|C_d|=ad+O(1)$ and $|D_d|=bd+O(1)$ separated by a distance $\gamma d+O(1)$ ($a,b,\gamma>0$). For coordinates $x_1<x_2$ in $C_d$ and $x_3<x_4$ in $D_d$, let $u=x_2-x_1$ and $v=x_4-x_3$ denote internal intervals, with container gap $q=x_3-x_2$. The resulting cross ratio satisfies
\begin{equation}
z=\frac{uv}{(u+q)(v+q)}
\le \rho+o(1),\qquad
\rho=\frac{ab}{(a+\gamma)(b+\gamma)}<1.
\label{eq:at-cross-band}
\end{equation}
For a fixed coupling on the critical line $0\le g\le1$, the same-layer magnetic order and disorder fields have scaling dimension $1/8$. In terms of $k=k(g)=\pi/[2\arccos(-g)]$~\cite{ObrienBartlettDohertyFlammia2015QAT,DelfinoGrinza2004AT}, with compactification parameter $N^2=1/(2k)$ in Ref.~\cite{GavrylenkoSantachiara2019C1}, the complete fixed-point four-spin correlator is
\begin{equation}
\begin{aligned}
G_k&=(uv)^{-1/4}f_k(z),\\
f_k(z)&=\frac{\theta_3(\mathfrak q^k)\theta_3(\mathfrak q^{1/k})}
{(1-z)^{1/4}\theta_3(\mathfrak q)^2},\\
\mathfrak q&=e^{-\pi K(1-z)/K(z)} .
\end{aligned}
\label{eq:at-fourpoint}
\end{equation}
Here $K$ is the complete elliptic integral of the first kind in the elliptic parameter convention, $\theta_3(q)=\sum_{m\in\mathbb Z}q^{m^2}$, and the nome $\mathfrak q$ is defined at $z=0$ by continuity and distinguished from the container gap $q$. Thus $f_k(0)=1$, and $f_k$ is continuous and bounded on every compact interval $0\le z\le\rho'<1$ with $\rho'>\rho$~\cite{Zamolodchikov1986ATFourSpin,NemkovRibault2021AT}. Integrating the raw four-point correlator over the container domains yields the continuum bound
\begin{equation}
\int_{C^2D^2}|G|\le
C_{\rho'}k_0^2|C|^{7/4}|D|^{7/4},
\qquad k_0=32/21.
\label{eq:at-continuum-envelope}
\end{equation}
The constant $C_{\rho'}$ depends on the fixed coupling. Because macroscopic separation prevents coordinates in distinct containers from coinciding, only the internal contact singularities within $C$ and $D$ require integration, and these remain integrable.

At the four-state Potts endpoint $g=1$, marginal logarithmic corrections require separate lattice matching~\cite{DelfinoGrinza2004AT}. The fixed-point scaling dimensions alone do not establish the pure-power hypotheses of Eq.~\eqref{eq:interacting-input} there; logarithm-aware control of the lattice signal and raw moments would be needed.

On a discrete lattice, short-distance coincidences are governed by direct combinatorial counting. Introducing a microscopic distance cutoff $\ell_{\rm mic}=\lfloor d^{\zeta}\rfloor$ with $0<\zeta<1/2$, the number of coordinate quadruples with at least one internal separation below $\ell_{\rm mic}$ satisfies
\begin{equation}
N_{\mathrm{near}}\le(2\ell_{\rm mic}+1)|C_d||D_d|(|C_d|+|D_d|)
=O(d^{3+\zeta}).
\label{eq:at-near-count}
\end{equation}
Because every Pauli operator has unit norm, this near-contact sector contributes at most $O(d^{3+\zeta})=o(d^{7/2})$. Establishing a rigorous discrete-lattice upper bound requires controlling the integrated matching corrections between discrete lattice sums and continuum integrals over the remaining mesoscopic configurations.

While the uncoupled model at $g=0$ inherits the Ising recovery price established above, the macroscopic cost and density conclusion for genuinely interacting couplings remains conditional on the lattice inputs in Eq.~\eqref{eq:interacting-input}: a positive lower envelope for $d^{1/4}m_{g,d}$ and a finite upper envelope for the raw quantity $d^{-7/2}H_g(C_d,D_d)$, uniformly over the required record containers. These inputs do not follow from the continuum bound alone, as mesoscopic lattice configurations still demand integrated matching control.

More generally, for positive-density records in macroscopically separated $O(d)$ containers, both the mean signal $\mu$ and the square root of the raw second moment scale as $d^{2-\eta}$ under the full dictionary, coordinate, kernel, and three-pairing hypotheses of Appendix~\ref{app:eta}, with $0<\eta<1$. This collective record mechanism relies on accumulating extensive non-negative correlations rather than on individual operator norms, providing a framework for analyzing environmental records in interacting systems once the underlying lattice conditions are established.

\begingroup\linespread{0.96}\small
\bibliographystyle{quantum-doi}
\bibliography{refs}
\endgroup
\end{document}